\documentclass[preprint,11pt]{aastex}

\newcommand{\oate}{Q~0835+580}
\newcommand{\onine}{Q~0952+179}
\newcommand{\eleven}{Q~1126+101}
\newcommand{\twelve}{Q~1258+404}
\newcommand{\twone}{Q~2149+212}
\newcommand{\twoth}{Q~2345+061}
\newcommand{\ha}{H$\alpha$}

\newcommand{\chinu}{$\chi_{\nu}^2$}
\newcommand{\chisq}{$\chi^2$}

\newcommand{\ebv}{$E(B-V)$}
\newcommand{\etal}{{\it et~al.}}

\newcommand{\ho}{$H_0$}

\newcommand{\jk}{$J-K$}

\newcommand{\ks}{$K_s$}

\newcommand{\mgii}{Mg\,{\sc ii}}

\newcommand{\OIII}{[O\,{\sc iii}]}

\newcommand{\qo}{$q_0$}

\newcommand{\rk}{$r-K$}

\newcommand{\z}{$z$}
\newcommand{\zph}{$z_{ph}$}

\shortauthors{HALL ET AL.}
\shorttitle{GALAXIES AROUND Z=1.5 RLQS} 
\slugcomment{Accepted to The Astronomical Journal December 27, 2000}

\begin{document}

\title{Galaxies in the Fields of $z\sim1.5$ Radio-Loud Quasars}

\author{
Patrick B. Hall,\altaffilmark{1,7,8,9,12}
Marcin Sawicki,\altaffilmark{2}
Paul Martini,\altaffilmark{3,11}
Rose A. Finn,\altaffilmark{4}
C. J. Pritchet,\altaffilmark{5,8}
Patrick S. Osmer,\altaffilmark{3}
Donald W. McCarthy,\altaffilmark{4}
Aaron S. Evans,\altaffilmark{2,9,10}
Huan Lin,\altaffilmark{4,6,8}
and F. D. A. Hartwick\altaffilmark{5,8}
\altaffiltext{1}{Department of Astronomy, University of Toronto, 60 St. George Street, Toronto, ON M5S~3H8, Canada}
\altaffiltext{2}{California Institute of Technology, Mail Stop 320-47, Pasadena, CA 91125}
\altaffiltext{3}{Astronomy Department, The Ohio State University, 140 West 18th Avenue, Columbus, OH 43210-1173}
\altaffiltext{4}{Steward Observatory, The University of Arizona, Tucson, AZ 85721}
\altaffiltext{5}{Department of Physics and Astronomy, University of Victoria, Victoria, BC V8W~3P6, Canada}
\altaffiltext{6}{Hubble Fellow}
\altaffiltext{7}{Visiting Astronomer, Infrared Telescope Facility,
operated by the University of Hawaii under contract to the 
National Aeronautics and Space Administration.}
\altaffiltext{8}{Visiting Astronomer, Canada-France-Hawaii Telescope,
operated by the National Research Council of Canada, the Centre National 
de la Recherche Scientifique de France and the University of Hawaii.}
\altaffiltext{9}{Visiting Astronomer, James Clerk Maxwell Telescope,
operated by The Joint Astronomy Centre on behalf of the Particle Physics and
Astronomy Research Council of the United Kingdom, the Netherlands Organisation
for Scientific Research, and the National Research Council of Canada.}
\altaffiltext{10}{Current Address: Department of Physics and Astronomy, 
State University of New York at Stony Brook, Stony Brook, NY 11794-3800}
\altaffiltext{11}{Current Address: Carnegie Observatories, 813 Santa Barbara
St., Pasadena, CA 91101}
\altaffiltext{12}{Current Affiliations: 
Princeton University Observatory, Princeton, NJ 08544-1001 and
Pontificia Universidad Cat\'{o}lica de Chile, Departamento de Astronom\'{\i}a
y Astrof\'{\i}sica, Facultad de F\'{\i}sica, Casilla 306, Santiago 22, Chile;
E-mail: phall@astro.puc.cl}
}


\begin{abstract}

\small 

We have previously identified an excess population of predominantly red
galaxies around a sample of 31 radio-loud quasars (RLQs) at $1<z<2$.  
Here we show that these fields have a surface density of extremely red objects
(EROs, with $R-K>6$) 2.7 times higher than the general field.  
Assuming these EROs are passively evolved galaxies at the quasar redshifts,
they have characteristic luminosities of only $\sim L^*$.
Only one of four RLQ fields has an excess of $J-K$ selected
EROs with $J-K>2.5$; thus, those objects are mostly unrelated to the quasars.

We also present new multiwavelength data and analyses on the fields of
four of these quasars at $z_q\sim1.54$, obtained to build more detailed pictures
of the environments of these quasars and the galaxies within them.
First, wide-field $J$ and \ks\ data shows that the galaxy excess around \oate\ 
is of Abell richness 2$\pm$1 and extends to 140\arcsec, 
and that the galaxy excess around \eleven\ extends to only 50\arcsec\ even 
though the overall counts in the field are higher than the literature average.
Second, in three fields we present the deepest narrow-band
redshifted H$\alpha$ observations yet published.
We detect five candidate galaxies at the quasar redshifts, a surface density
2.5 times higher than in the only existing random-field survey of similar depth.
However, photometric spectral energy distribution (SED) fitting of one 
candidate suggests it is an \OIII\ detection background to the quasar.
Third, SCUBA submillimeter observations of three fields detect two of the
quasars and two galaxies with SEDs best fit as highly reddened galaxies
at the quasar redshifts.
Fourth, $H$-band adaptive optics (AO) imaging is used to estimate 
redshifts for two moderately red bulge-dominated galaxies in the \oate\ field
using the Kormendy relation between central surface brightness and half-light
radius.  Both have structural redshifts consistent with early-type galaxies 
foreground to the quasar at $z\lesssim0.2$ or $1 \lesssim z \lesssim 1.35$.
Photometric redshifts do not confirm these structural redshifts, however,  
possibly because our optical photometry for these objects is corrupted by
scattered light from the nearby bright AO guidestar.
Finally, quantitative SED fits are presented for numerous galaxies of interest
in two fields and are used to constrain their photometric redshifts $z_{ph}$.
Most galaxies in the spatially compact group around \oate\ are consistent with
being at the quasar redshift $z_q$.  
One of these is a candidate very old galaxy without ongoing star formation,
while the others appear to have ongoing or recent star formation.
Many very and extremely red objects across both fields have $z_{ph} \simeq z_q$,
and significant dust is required to fit most of them, including about half
of the objects whose fits also require relatively old stellar populations.
Large reddenings of \ebv~$\simeq0.6\pm0.3$ are also required to fit four 
$J-K$ selected EROs in the \eleven\ field, though all but one of them have 
best-fit redshifts $z_{ph}>z_q$.
These objects may represent a population of dusty high-redshift galaxies 
underrepresented in optically selected samples.

Taken together, these observations reinforce the claim that 
radio-loud quasars at $z_q>1$ can be found in galaxy overdensities.
Ongoing star formation with moderate amounts of dust seems to be common
among all but the very reddest galaxies in these overdensities.

\end{abstract}

\small 


\section{Introduction}	\label{Intro}

It is of considerable current interest to efficiently identify clusters and 
other concentrations of galaxies at $z>1$ in order to study the evolution of 
both galaxies and large scale structure.  Radio-loud quasars (RLQs) are
among the obvious signposts around which to search for clusters at $z>1$.
This is because RLQs almost exclusively reside in giant elliptical galaxies 
(a strongly clustered population)
and some RLQs have been spectroscopically confirmed to inhabit clusters at
$z\lesssim0.7$ \citep{ye93}.
In \cite{hgc98}, hereafter HGC98, and \cite{hg98}, hereafter HG98, 
we presented optical ($r\lesssim25.5$) and near-IR ($K\lesssim21$) 
imaging of the fields of 31 RLQs at $1<z<2$
which revealed an excess of predominantly red galaxies at $K>19$.
The excess has two spatial components --- one within 40\arcsec\ of
the quasars and one extending out to at least $\sim 2\arcmin$ ---
consistent with the quasars often residing in galaxy groups or poor clusters
which are themselves found within richer filaments or sheets of galaxies.
The images of HGC98 were at most 3\arcmin$\times$3\arcmin\ in size, and so 
could only place a lower limit on the size of the large-scale galaxy excess
(relative to blank-field galaxy counts at these magnitudes) around the quasars.
The magnitude and color distributions of the excess galaxies are consistent 
with a population of predominantly early-type galaxies at the quasar redshifts.
However, a possible exception is some or all of the $J$-band dropouts
($J-K>2.5$) found in five fields for which we have $J$ imaging.  
Such very red \jk\ colors require that the galaxies are reddened by dust, are
at $z\gtrsim2.5$ so that the 4000~\AA\ break lies beyond the $J$ band, or both.

Spectroscopic confirmation of overdensities at the quasar redshifts in these
RLQ fields is still lacking, as are spectroscopic redshifts for the $J$-band 
dropouts, due to the difficulty of obtaining spectroscopic redshifts for 
galaxies at $1.5 \lesssim z \lesssim 2.5$, and extremely red galaxies at all
$z\gtrsim1$.
To provide a more comprehensive picture of these systems in the absence of
spectroscopy, in this paper we present further analyses of the excess galaxy
population as well as various new observations of four of these RLQ fields.
The results reported here supersede earlier results reported as work in
progress in \citet{hal99tp3a} and \citet{hal99tp3b}.
In \S\ref{wide-field} we present new wide-field near-IR imaging
	and study the extent of the galaxy overdensities.
In \S\ref{surf} we discuss the surface density of extremely red objects (EROs)
	in these RLQ fields compared to field surveys.
In \S\ref{irtf} we present narrow-band imaging 
	sensitive to \ha\ emission from galaxies at the quasar redshifts.
In \S\ref{submm} we present sub-millimeter mapping
	sensitive to dust in luminous star-forming galaxies.
In \S\ref{aob} we present adaptive-optics imaging of two galaxies.
In \S\ref{photz} we present multicolor SED fits used to derive photometric 
	redshifts, reddenings, and age estimates for selected galaxies.
We summarize our results in \S\ref{concl}.  We adopt a $\Lambda=0$ cosmology
with $H_{\rm 0}$=75 km~s$^{-1}$~Mpc$^{-1}$ and $q_{\rm 0}$=0.1
(for a projected scale of 7.4~$h_{75}$ kpc/arcsec at $z=1.54$).
except as noted for comparison with the literature.

\section{Wide-Field Near-IR Imaging} \label{wide-field}

We obtained $J$ and $K_s$ data over wider fields around two RLQs from HGC98
to help verify the reality of the large-scale galaxy excess and to search for 
additional $J$-band dropouts. 
Information about these RLQs, the two other RLQs studied in this paper, and
the observations obtained in the field of each is given in Table~\ref{t_obs}.
The near-IR imaging data were reduced using 
PHIIRS.\footnote{PHIIRS is Pat Hall's Infrared Imaging Reduction Software
\citep{hgc98}, a package of IRAF routines available from the first author 
or the IRAF contributed software website.  The Image Reduction and
Analysis Facility (IRAF) is distributed by the National Optical Astronomy
Observatories, which is operated by AURA, Inc., under contract to the National
Science Foundation.}
Standard infrared observing and reduction procedures were used, including
flattening using either domeflats or ``running skyflats''
created from the science images themselves.  
Object detection and photometry using FOCAS \citep{val82a} followed the
procedure of HGC98.

\subsection{Q~0835+580 Field Observations and Reductions} \label{wf0835}

We obtained $J$ and $K_s$ data over a $\sim$9\arcmin\ diameter unvignetted 
field of view at platescale 0\farcs5/pixel
at the Steward Observatory 90$''$ telescope
with PISCES \citep{mcc98,mea00}, which incorporates a 1024$^2$ HgCdTe detector.
PISCES has significant optical distortions in both $J$ and $K_s$ which must be
removed before coadding.
Observations of M~67 \citep{gir89} were used to calculate distortion solutions
for the April 1999 run, yielding images oriented to true North and East
(as defined by the USNOA-2.0 catalog) to within $\leq$0.12\arcdeg\ accuracy.
%
Calibration used ARNICA \citep{hun98} and NICMOS \citep{per98} standards
and assumed extinction coefficients of 0.08 in $J$ and 0.07 in $K_s$.
Both object detection and aperture definition were done on the summed $J+K_s$ 
image (total area 52.561~arcmin$^2$), and FOCAS total magnitudes were produced 
using those same apertures on the $J$ and $K_s$ images.
%


\subsection{Q~1126+101 Field Observations and Reductions} \label{wf1126}

We imaged \eleven\ with the near-infrared imager/spectrograph TIFKAM
\citep[a.k.a. ONIS at Kitt Peak;][]{pog98}
at the 2.4m Hiltner telescope of the MDM Observatory.  
TIFKAM has a $512 \times 1024$ InSb array with a platescale of 0\farcs3/pixel
at the 2.4m, corresponding to a $2.5' \times 5'$ field of view.  
We observed a field centered $80''$ W of the quasar (which includes the
quasar near the eastern edge of the array) in $J$ and $K_s$ in 
nonphotometric conditions of $\sim$1\farcs5 seeing.
The coadded images were trimmed to the region of maximum exposure time
(12.051 arcmin$^2$).
The final $J$ and $K_s$ images completely overlap the $r-$band data of HGC98
and partially overlap the $J$ and $K_s$ data of HGC98.
We rotated and magnified our new near-infrared observations to match those
images and then catalogued all the objects in the sum of the $r$ image and
our new $J$ and $K_s$ images.
Finally, we calibrated our $J$ and $K_s$ images using the objects in common
with HGC98.

\subsection{Wide-Field Near-IR Imaging Results} \label{wfdisc}

The $K<20.5$ galaxy surface density is 17.5$\pm$0.6 arcmin$^{-2}$ in the
\oate\ field and 20.7$\pm$1.3 arcmin$^{-2}$ in the \eleven\ field.
These values are respectively $2.4\sigma$ and $3.5\sigma$ above the literature
compilation value of 13.7$\pm$1.5 (including RMS field-to-field scatter)
from HG98.  
Both values agree within $1.3\sigma$ with those measured in HGC98 
using smaller ($\sim$3\arcmin$\times$3\arcmin) images of these fields.
The $J<21.5$ galaxy surface density is 11.8$\pm$0.5 arcmin$^{-2}$ in the
\oate\ field and 14.2$\pm$1.1 arcmin$^{-2}$ in the \eleven\ field.  These
values are slightly higher than the $J$-band field galaxy counts of \cite{tmm99}
and $>3\sigma$ higher than those of \cite{sar99}.
Thus these new wide-field observations confirm that the surface density of
faint $J$- and $K$-selected galaxies is higher in these RLQ fields than in
random fields.

\begin{figure} 
\plottwo{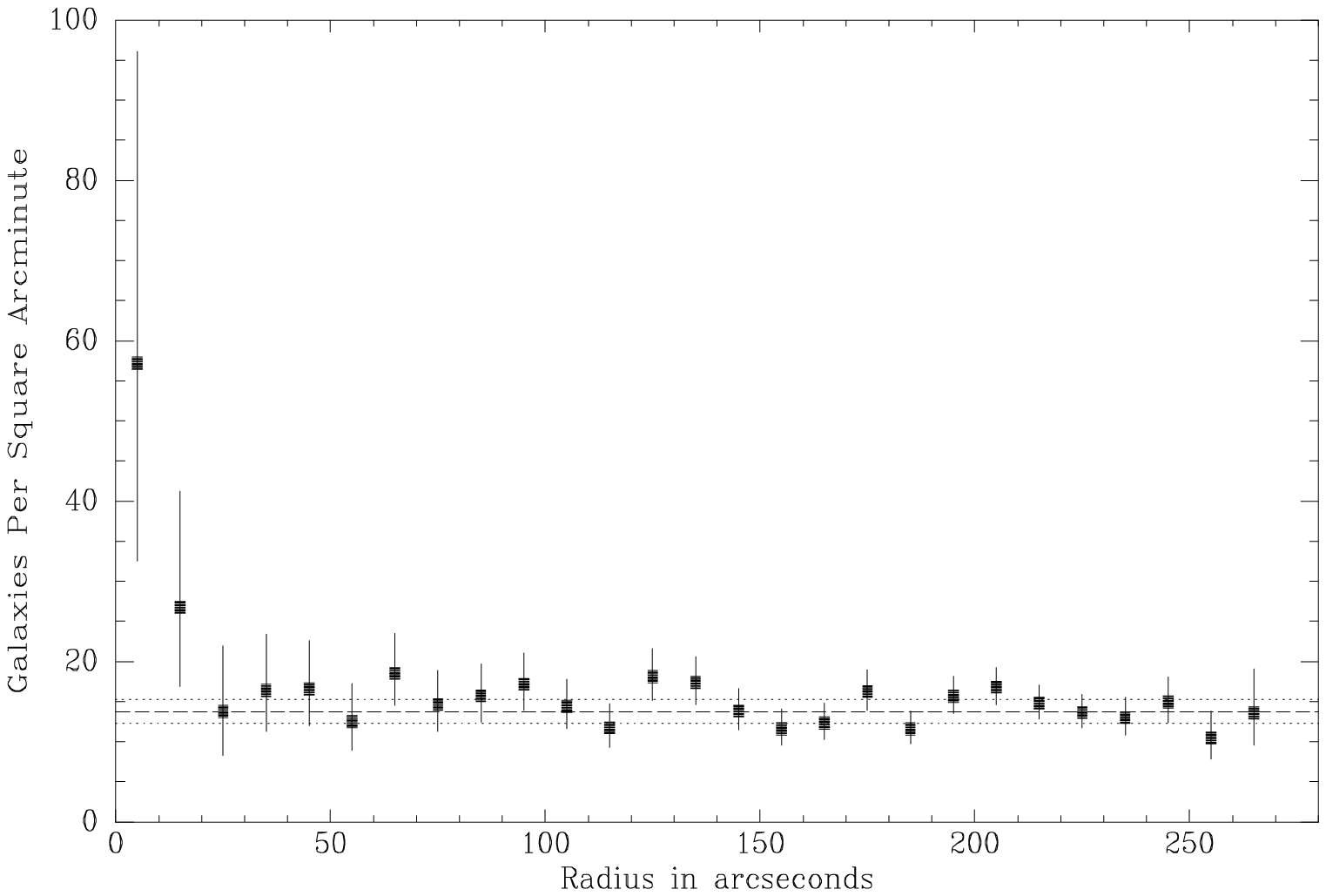}{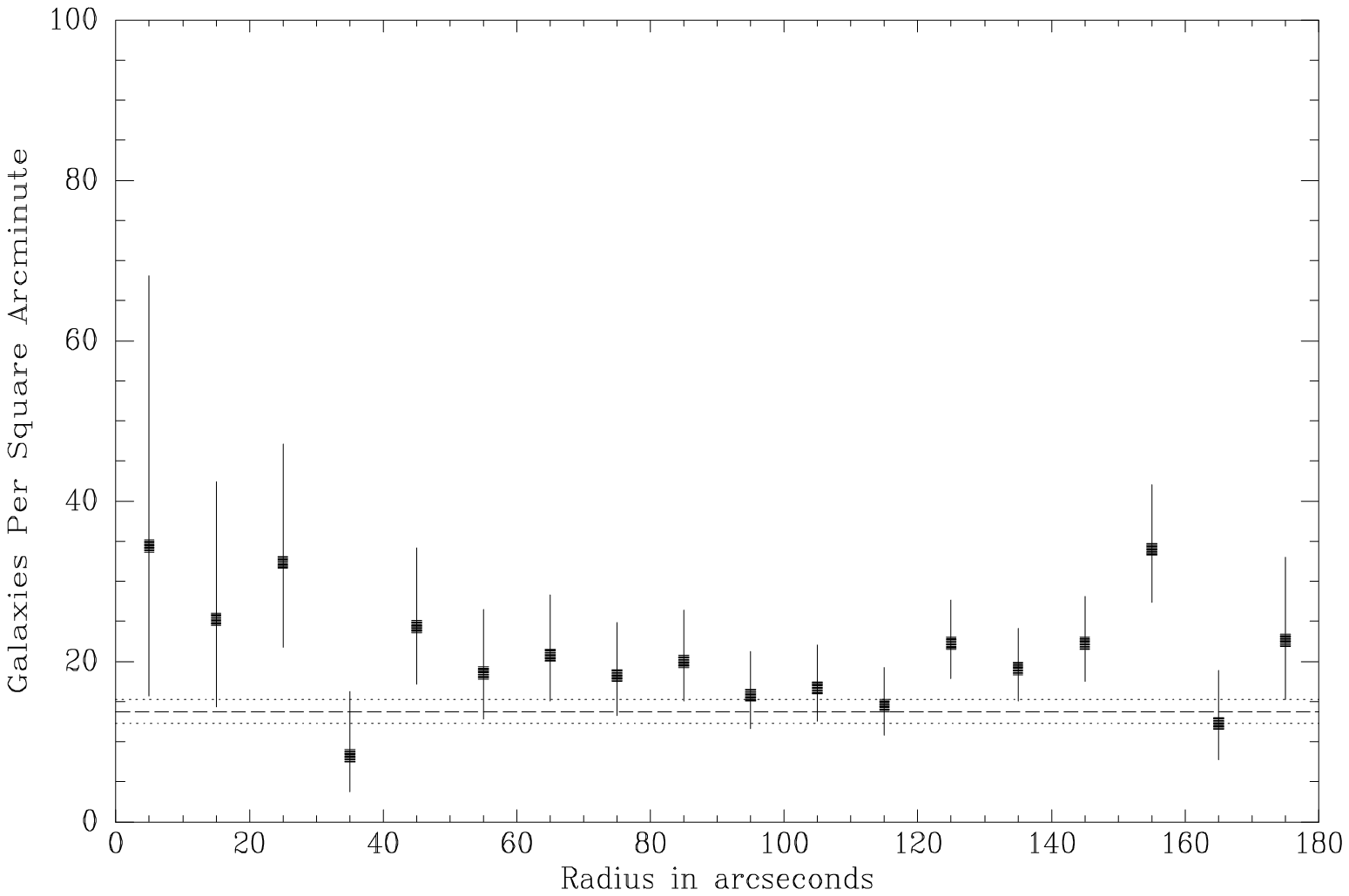}
\caption[]{
\singlespace 
The radial distributions of all $K<20.5$ galaxies relative to the two quasars
with new wide-field $JK$ imaging are plotted as points with error bars from
Gehrels (1986).
The average $K<20.5$ counts and $\pm1\sigma$ RMS scatter from the literature
data compiled in Hall \& Green (1998) are plotted as dashed and dotted lines,
respectively.
(a) Q~0835+580 field.
There is a clear galaxy excess at $<$20\arcsec.
The overdensity on larger scales appears to extend out to $\sim$140\arcsec\ at
$2.35\sigma$ significance.
(b) Q~1126+101 field.  
There is an excess on all scales relative to the literature, but only a weak
excess at $\lesssim$50\arcsec\ around the quasar relative to the counts in this
field.
The spike at $\sim$150\arcsec\ is due to a galaxy grouping on the sky
which by chance is aligned tangentially to the quasar.
}\label{fig_rp}
\end{figure}

The radial distribution of all $K<20.5$ galaxies relative to \oate\ 
(Figure~\ref{fig_rp}a) shows a clear excess at $<$20\arcsec.
There also appears to be an overdensity extending out to $\sim$140\arcsec\ 
($\sim2h_{75}^{-1}$~Mpc).  From the surface density at $>$140\arcsec, 
we would expect 245 galaxies within 140\arcsec\ of the quasar.
We observe 285, which is a $2.35\sigma$ excess.
The radial distribution of all $K<20.5$ galaxies relative to \eleven\ 
(Figure~\ref{fig_rp}b)
shows a constant surface density within the uncertainties to 180\arcsec, 
apart from a weak excess at $\lesssim50$\arcsec.
%
Given this relatively flat radial distribution, part of the excess population
of faint $J$- and $K$-selected galaxies in the \eleven\ field is probably
unrelated to the quasar.  
This is consistent with the $J-K$ and $r-K$ color distributions, which show an
excess of blue galaxies as well as of the red galaxies we expect to find
at the quasar redshifts.
Another possibility is a structure at the quasar redshift of size
$\geq6h_{75}^{-1}$~Mpc \citep[cf.][]{sgs99},
which is less probable but cannot be ruled out 
without spectroscopy or data on even wider fields.

As a measure of the richness of the structure around \oate, 
we computed $N_{0.5}$ \citep{hl91}, the excess number of galaxies 
within 0.5~Mpc of the quasar and no more than 3 magnitudes fainter
than a brightest cluster galaxy at the quasar redshift.
$N_{0.5}$ was computed as in HG98 except that galaxies $>$140\arcsec\ from the
quasar were used as the background, rather than the average literature counts.
We find $N_{0.5}$=27$\pm$11 (corresponding to Abell richness 2$\pm$1)
compared to the value of 16$\pm$13 found in HG98.
This newer, more robust measurement illustrates the value of wide-field data
in determining the background counts locally rather than taking them from the
literature or from small radii which may still be inside the
large-scale structure being studied.
In the case of \eleven, even though the overall counts in the field are higher
than the literature average, wide-field data suggests that the
quasar is only embedded in a small-scale overdensity.

\section{Surface Density of EROs in $1<z<2$ RLQ Fields} \label{surf}

Since the appearance of infrared array detectors, numerous discoveries of 
extremely red objects (EROs) have been noted in the literature 
\citep[for recent summaries see][]{dea99,dad00}.
EROs remain a poorly understood population of galaxies
about which several different interpretations are plausible.
Their observed 
colors of $R-K \ge 6$ are so red that they seem to be explainable only by
old stellar populations at $z\gtrsim1$ or by heavily dust-reddened galaxies
or AGN preferentially located at $z\gtrsim1$.
This is confirmed by the handful of spectroscopic redshifts for
EROs \citep{dea99,soi99,liu00} and very red objects (VROs),
as defined by \cite[][; cf. \S\ref{photboth}]{coh99b}. 
Previous narrow-field optical and near-IR surveys have hinted
that EROs are more common along lines of sight to distant AGN 
\citep[for recent summaries see][]{sma99,cim00}.
Our $1<z<2$ RLQ fields can be used to test this hypothesis more carefully, using
recent estimates of the surface density of field EROs from wide-field surveys.

\subsection{\rk\ selected EROs} \label{rkeros}

\cite{teb99} find 29 $K'\leq19$ EROs with $R-K'\geq6$ in four fields of
$\sim$150~arcmin$^2$ each, 
for a surface density of 0.048$_{-0.009}^{+0.011}$ arcmin$^{-2}$ (we use the
methods of \cite{geh86} to find all $1\sigma$ Poissonian uncertainties where
small number statistics apply).  
The details for one field have been published in 
Thompson \etal\ (1999a; hereafter T99).
The T99 $K'$ observations were calibrated using UKIRT standards \citep{ch92},
as were the $K_s$ observations of HG98.  We refer to both magnitudes
simply as $K$ hereafter.  
We also adopt $R-K\geq6$ as the definition of a ERO, or
$r-K\geq6.322$ for Gunn $r$ imaging such as ours (HG98, Appendix A),
and $K=19$ as the division between bright and faint EROs.
However, we conservatively adopt 
a $3\sigma$ detection limit in the optical instead of $2\sigma$ as in T99.
This yields a difference of $\sim$0\fm44 in the lower limit on the $r-K$ 
color of undetected objects.

We select EROs from all 22 RLQ fields of HGC98
with $3\sigma$ limits of $r\ge25.322$ and $5\sigma$ limits of $K\ge19$ and
from the new larger area around \eleven\ imaged with TIFKAM (\S\ref{wf1126}).
The total area is 160.95~arcmin$^2$, 
almost identical to that of T99,
and the depth of our imaging is also similar.
We find 24 EROs, including two presumed stars 
unresolved in archival HST WFPC2 snapshots 
and one presumed star identified 
via its blue \jk\ color.  Four of the remaining 21 EROs are compact enough 
that they could be stars given the seeing, but the rest are extended.
T99 found two of their eight bright EROs to be unresolved in 1$\farcs$1
seeing, and both were subsequently confirmed to be stars.  
There is no reason to expect more stellar contamination in our ERO sample
than in the T99 sample, since both are from high Galactic latitude fields
of almost exactly the same total areas.
Therefore we assume that only three of our 24 EROs are stars.
The remaining 21 \rk\ selected extragalactic EROs yield
a surface density of 0.130$_{-0.029}^{+0.035}$ arcmin$^{-2}$ to $K=19$, 
which is 2.7 times higher than the field ERO counts of \cite{teb99} and a
$2.64\sigma$ excess over the number expected in this area based on those counts.
%
This is consistent with the excesses found by the smaller surveys of
\cite{cmp00} at $K<20$ around 4 radio galaxies with $z\sim1$
and \cite{cim00} at $K\lesssim19$
in 40~arcmin$^2$ around 14 radio-loud AGN with $z>1.5$.

\subsection{$J-K$ selected EROs} \label{jkeros}

\cite{dro99} find a surface density of 
0.098$\pm$0.009~arcmin$^{-2}$ for $K\leq19.5$ EROs with $J-K\geq2.5$ from 
124 objects over 1260~arcmin$^2$ 
from the random-field MUNICS survey \citep{mdo98}.
Stellar contamination of these $J-K$ selected EROs should be negligible.
A literature search revealed no predicted colors of $J-K>2.2$
for unreddened stars of any type from brown dwarf to supergiant
and very few observations of objects with $J-K>2.5$ which were not galaxies:
two objects presumed to be stars since they were unresolved on
$HST$ WFPC2 images \citep{dep99} and a small number of known AGN
\citep{cut00,bh00}.  
Thus objects with $J-K>2.5$ are almost certainly not stars, are unlikely to 
be AGN, and are probably galaxies.

In four RLQ fields from HGC98 with deep $J$ data (\oate, \onine, \eleven, and 
\twelve), we find six such $J-K$ selected EROs.  
We also find three $J-K$ selected EROs in the new area around \oate\ imaged 
with PISCES, but none in the new area around \eleven\ imaged with TIFKAM.
Five of the EROs 
are from the \eleven\ field, which as discussed in HG98 has a considerable
excess of these $J$-band dropouts.\footnote{Two of these EROs 
in the \eleven\ field have inconsistent measurements 
(at $\geq3\sigma$) between the old HGC98 and new TIFKAM datasets.
Examination of the images suggests that their $J$ fluxes were probably
underestimated in HGC98.  The HGC98 KPNO 4m IRIM data required `destriping'
to remove the large scale scattered light pattern present in some fields
(see \S3.2.7 of HGC98).  
The new $J$ data shows that this procedure worked quite well overall,
but in a few cases the masking of objects was insufficient to prevent 
removal of some real flux.} 
%
%
Excluding the \eleven\ field because of its excess number of these objects,
we find four \jk\ selected EROs in 70.373~arcmin$^2$ for a surface density of 
0.057$_{-0.024}^{+0.045}$~arcmin$^{-2}$,
within $1\sigma$ of the field measurement of \cite{dro99}.
In the \eleven\ field we find 3 EROs in 15.549~arcmin$^2$
for a surface density of 0.19$_{-0.09}^{+0.19}$~arcmin$^{-2}$,
3.3 times higher than in random fields but only significant at 1.54$\sigma$
due to small number statistics.

We also consider the surface density of $J$-band dropouts in the \eleven\ 
field to 
$K=20$, since all our data reaches deep
enough to select EROs with $J-K>2.5$ to at least this magnitude.  In the new
TIFKAM \eleven\ data we find 6 such EROs (0.50$^{+0.30}_{-0.20}$ arcmin$^{-2}$).
In the HGC98 \eleven\ data we find 8 such EROs (0.93$^{+0.46}_{-0.32}$ arcmin$^{-2}$).
In the other three fields with deep $J$ data from HGC98 plus the PISCES
\oate\ data we find 11 such EROs (0.16$^{+0.06}_{-0.05}$ arcmin$^{-2}$).
The enhancement in the \eleven\ field is still a factor of three or more,
but remains only $\sim2\sigma$ significance due to small number statistics.
The same holds true down to $K=20.5$ in the HGC98 data alone.
Deeper or wider-field data are unlikely to dramatically increase the statistical
significance of the overdensity of $J$-band dropouts, but their clustered
distribution within the \eleven\ field (Figure~24 of HG98)
and the overdensity of galaxies with less extreme colors suggests that 
this is a physical concentration and not just a random fluctuation.
However, some recent work suggests that such random field-to-field fluctuations
can be quite substantial even to $K=20$ and beyond 
\citep[cf. ][]{dad00,eis00,ss00}.

\subsection{Extremely Red Objects:  Discussion} \label{disceros}

We find bright \rk\ selected EROs with $K\leq19$ to be 2.7 times more common 
in the fields of 22 RLQs with $1<z_q<2$ than in the general field.  
We find \jk\ selected EROs with $K\leq19.5$ to be from 1 to 3.3 times as common
in the fields of four RLQs with $1.5<z_q<1.75$, depending on whether the 
\eleven\ field is included or not.
These quantitative measurements show that the excess ERO surface density
around distant AGN is considerably smaller than previous estimates 
of a factor of 10 to 100 excess \citep{dsd95}.  

One outstanding question regarding EROs is their intrinsic luminosities.
The presence of the quasars in these fields enables us to estimate the
luminosities of the \rk\ selected EROs by assuming that the EROs are
at the quasar redshifts.  
This is a reasonable assumption since there is no evidence for strong or weak
gravitational lensing in these fields and since the excess galaxy population
is concentrated around the quasars but not correlated with the presence of
intervening \mgii\ absorption systems (see \S2 and \S4.3 of HG98 for details
and discussion, and \S\ref{photboth} for further evidence of association).

Assuming the EROs in each field are at the redshift of the quasar in that field
($<$$z_q$$>$=1.61) and applying elliptical $k$-corrections only from 
\cite{frv97}, we find $M^{ERO}_K$=--25.4$\pm$0.5 for the \rk\ selected EROs
in our standard 
cosmology.
Applying 
evolutionary corrections ($e$-corrections) from \cite{pog97} as well, 
we find $M^{ERO}_K$=--24.1$\pm$0.4.  
This is consistent with 
the \cite{gar97} measurement of $M^*_K$=--24.2$\pm$0.2 for field galaxies
at $<$$z$$>$=0.14 
(see HG98 \S5.1)
and 
with the \cite{dep98} measurement of $M^*_K$=--23.9 in the Coma cluster.

Thus, these EROs have absolute magnitudes consistent with those expected for
passively evolving luminous elliptical galaxies at the quasar redshifts,
and need not be an extraordinarily luminous galaxy population.
The higher density of EROs in RLQ fields can be easily understood
if the RLQs are located in overdense regions.
Our measured ERO luminosities are consistent with the estimated luminosity 
evolution in $M_K^*$ for excess galaxies to $K\geq20$ in these fields 
\citep[\S5 of HG98; see also ][]{kaj00b}.  The faint end of our ERO 
distribution is set by our $K\leq19$ selection criterion, however, 
so we can say nothing further about the ERO luminosity function.

Using $k$- and $e$- corrections for bluer galaxies of later spectral types 
yields higher estimated luminosities for the EROs, but the red colors of 
these EROs strongly suggest that such corrections are not appropriate.  
On the other hand, if dust reddening is important for these EROs at $z>1$ but
not for their descendants at $z=0$, strong $k$- and $e$- corrections would be
required.  If dust were important we might expect considerable overlap between 
\rk\ and \jk\ selected EROs.
\cite{teb99} find that only 4 of 29 \rk\ selected EROs are \jk\ selected EROs,
and of our four objects with $r-K>6.322$ and $J$ data, only one has $J-K>2.5$.  
Our SED fitting (\S\ref{photboth}) yields a best-fit 
\ebv=0.75$\pm$0.50 for this object, \ebv=0.45$\pm$0.20 for one of the
$J-K<2.5$ objects, and \ebv=0 for the remaining two $J-K<2.5$ objects.
If the same fractions hold among the \rk\ selected EROs of \cite{teb99},
up to $\sim$12 of 29 ($\sim$40\%) could have strong dust reddening.
Multicolor data on larger \rk\ selected ERO samples reaching fainter magnitudes
will be needed to determine how frequent and strong dust reddening really is 
among EROs, but for now we make no correction for dust reddening. 


\section{Narrow-band H$\alpha$ Imaging}	\label{irtf}

H$\alpha$ is a good tracer of the instantaneous star formation rate (SFR)
in galaxies \citep{ken98}.
Narrow-band surveys until recently yielded only limits on the
space density of H$\alpha$ emitters at $z>1$, even in the fields of quasars
and radio galaxies (Pahre \& Djorgovski 1995).  
However, in the past few years detections 
have been made of objects located in random fields
and objects associated with damped Ly$\alpha$ absorbers,
with strong intervening metal-line absorbers,
or with quasars or radio galaxies \citep[for a recent overview see][]{tmm99}.
Given the excess galaxy population seen in our RLQ fields, they make promising
targets for narrow-band observations 
to search for H$\alpha$ emission at the quasar redshifts.
In particular, we would hope to detect galaxies whose SEDs suggest they are
dust-reddened and thus possibly actively star-forming and we would expect
not to detect galaxies whose SEDs suggest they are old and dust-free.

The fields of Q~0835+580 and Q~2345+061 were imaged in $H$ at
0$\farcs$300/pixel and in a narrow band within $H$ at 
0$\farcs$148/pixel using NSFCAM \citep{ld96} at the NASA Infra-Red Telescope
Facility (IRTF).  
Relevant details of the observations are given in Table~\ref{t_obs}.
Narrow-band imaging utilized a circularly variable filter
(CVF) with resolution $R=90$ 
tuned to the wavelength of H$\alpha$ at the quasar reshifts.
%
Photometric calibration 
was done using UKIRT standards \citep{ch92}.
Domeflats were used in $H$ but not for CVF imaging since the fringing across
the array in CVF mode was different between domeflats and the sky.
Photometry was performed in matched 4.2$''$ (14 pixel) diameter 
apertures on the $H$ images and on H$\alpha$ images resampled to the $H$-band
pixel scale.
Additional $H$- and narrow-band observations of the
fields of Q~2345+061 and Q~2149+212 were obtained by 
C. Pritchet \& F. Hartwick (Table~\ref{t_obs}).
These data were taken 
with CFHT and REDEYE (0\farcs5/pixel) and a custom-made narrow-band filter
62~\AA\ wide centered at 1.6662~\micron.
The data were calibrated using bright objects in common with the IRTF 
\twoth\ data,
and magnitudes were measured in 4\farcs0 (8 pixel) diameter apertures.
To quantify the significance of the excess narrow-band flux we follow
\cite{bun95} and use 
$\Sigma$, the number of standard deviations between the observed broad-band
counts and the number expected based on the narrow-band counts.  

\begin{figure} 
\plottwo{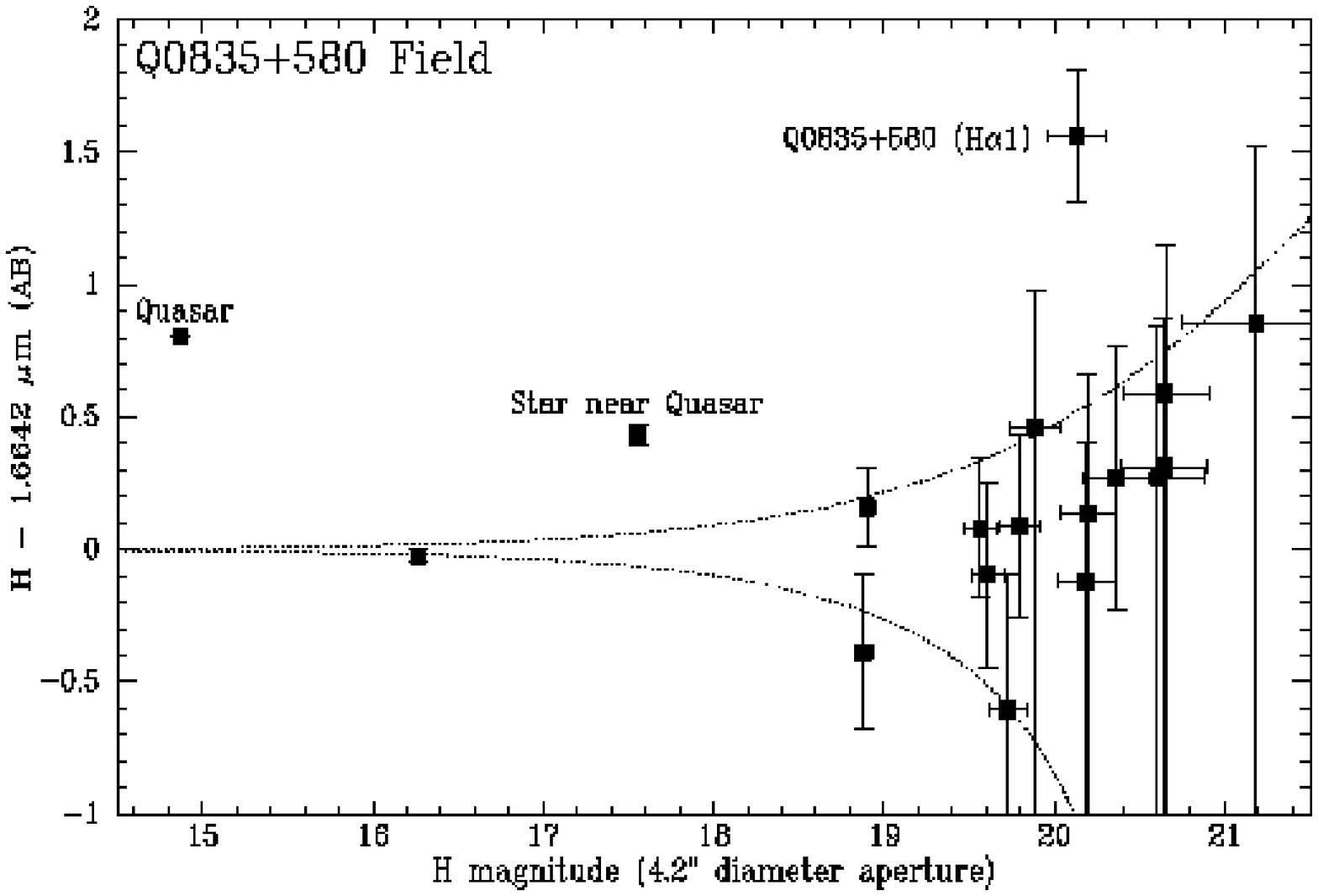}{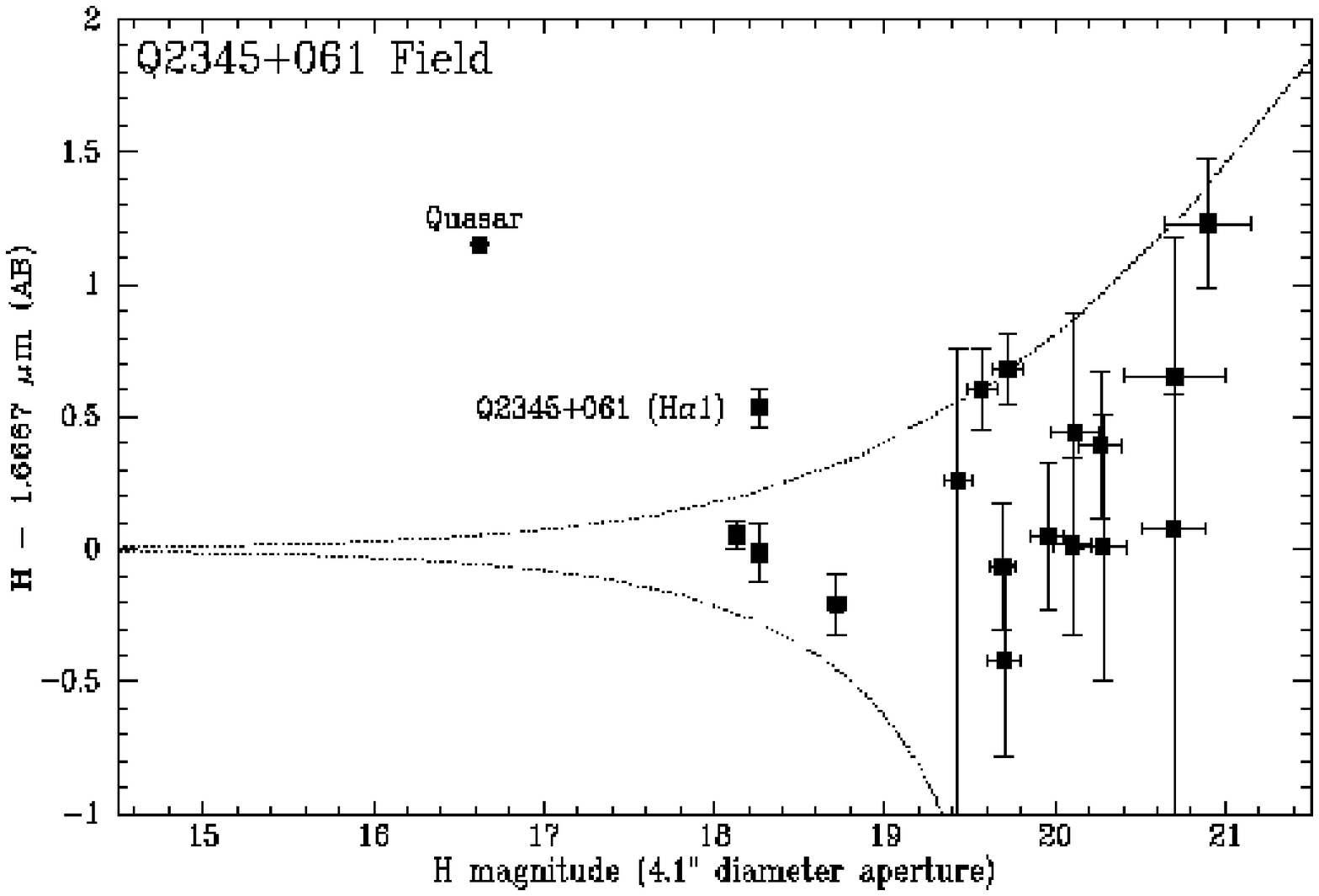}        
\caption[]{
\singlespace 
Color-magnitude diagrams from IRTF $H$- and narrow-band imaging.
Dotted lines enclose the expected range of objects with narrow-band excesses
of $<3\sigma$ significance.
(a) Q~0835+580 field.
(b) Q~2345+061 field.
}\label{fig_ha_irtf}
\end{figure}

There is one $>3\sigma$ detection in each field observed with IRTF
(see Table~\ref{t_ha_objs}).  Figure~\ref{fig_ha_irtf}a shows the
color-magnitude diagram for the \oate\ field.
The \ha\ emitter --- hereafter \oate~(H$\alpha$1) --- 
is an unremarkable faint blue galaxy with 
SFR$_{{\rm H}\alpha}$=14.7$\pm$2.5 $h_{75}^{-2} M_{\odot} {\rm yr}^{-1}$
using the \cite{ken83} relation between SFR$_{{\rm H}\alpha}$ and $L$(\ha),
which has an additional intrinsic dispersion of $\pm$50\%,
and assuming the detection is indeed \ha\ at the quasar redshift.
With this assumption, we can also estimate SFR$_{FUV}$ following \cite{ste96}
using $U$-band data which has been obtained for this field (see \S\ref{photz}), 
since $U$ samples rest-frame 1500~\AA\ at the quasar redshift almost exactly.
\oate~(H$\alpha$1) has $U$=23.052$\pm$0.075, which corresponds to
SFR$_{FUV}$=5.3$\pm$0.3 $h_{75}^{-2} M_{\odot} {\rm yr}^{-1}$ 
for a Salpeter IMF with an upper mass cutoff of 80~$M_{\odot}$.
The agreement with SFR$_{{\rm H}\alpha}$ is good given the various 
uncertainties (e.g. no correction for dust has been made to SFR$_{FUV}$).
Note that none of the dozen or so galaxies with $r-K\gtrsim5$
immediately surrounding \oate\ were detected in \ha.
If they are at the quasar redshift, they must have star formation rates of
$<1.8 h_{75}^{-2} M_{\odot} {\rm yr}^{-1}$ (3$\sigma$), 
consistent with them being red due to age and/or metallicity rather
than being extremely dust-reddened starburst galaxies (see \S\ref{photz}).

\begin{figure}
\plottwo{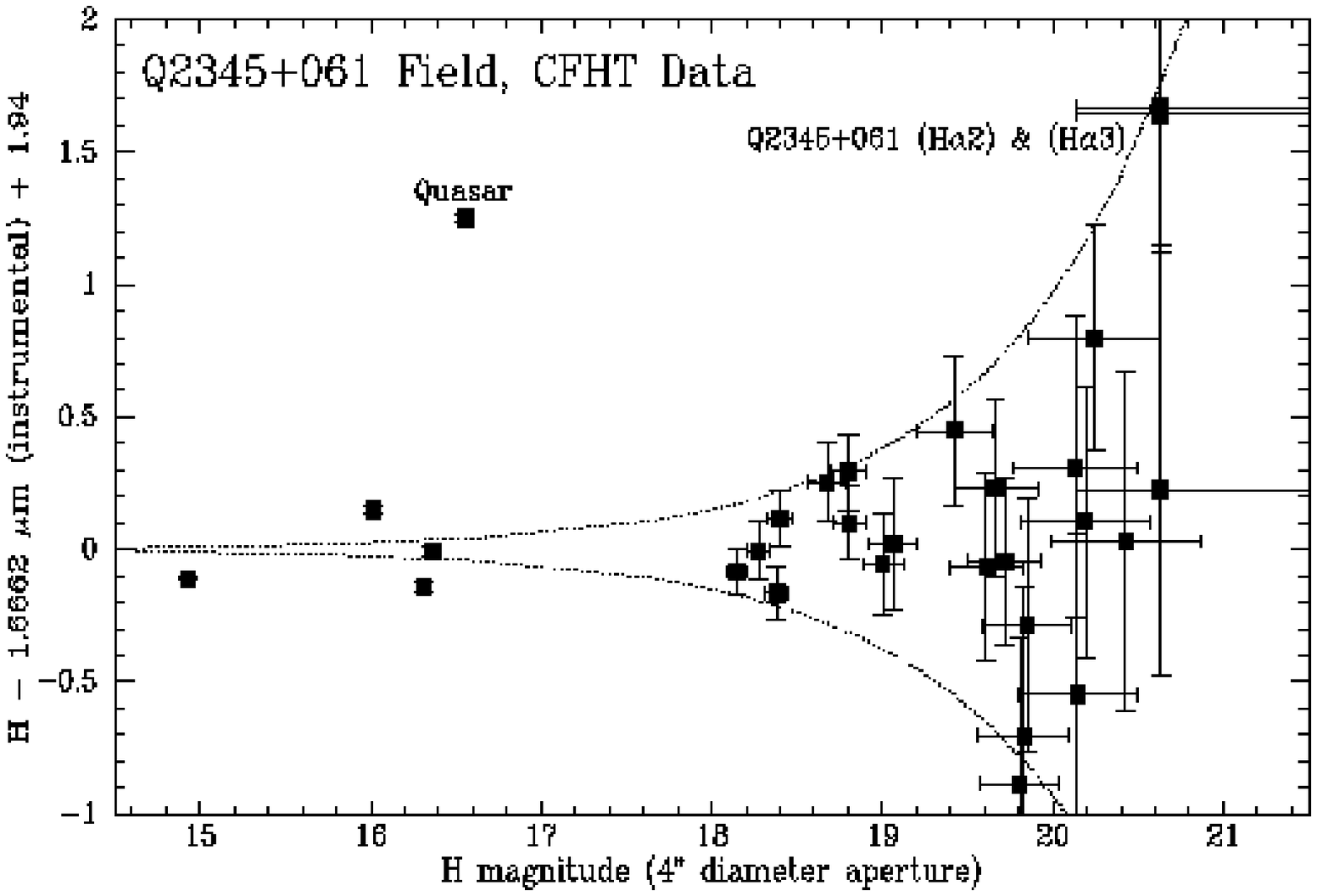}{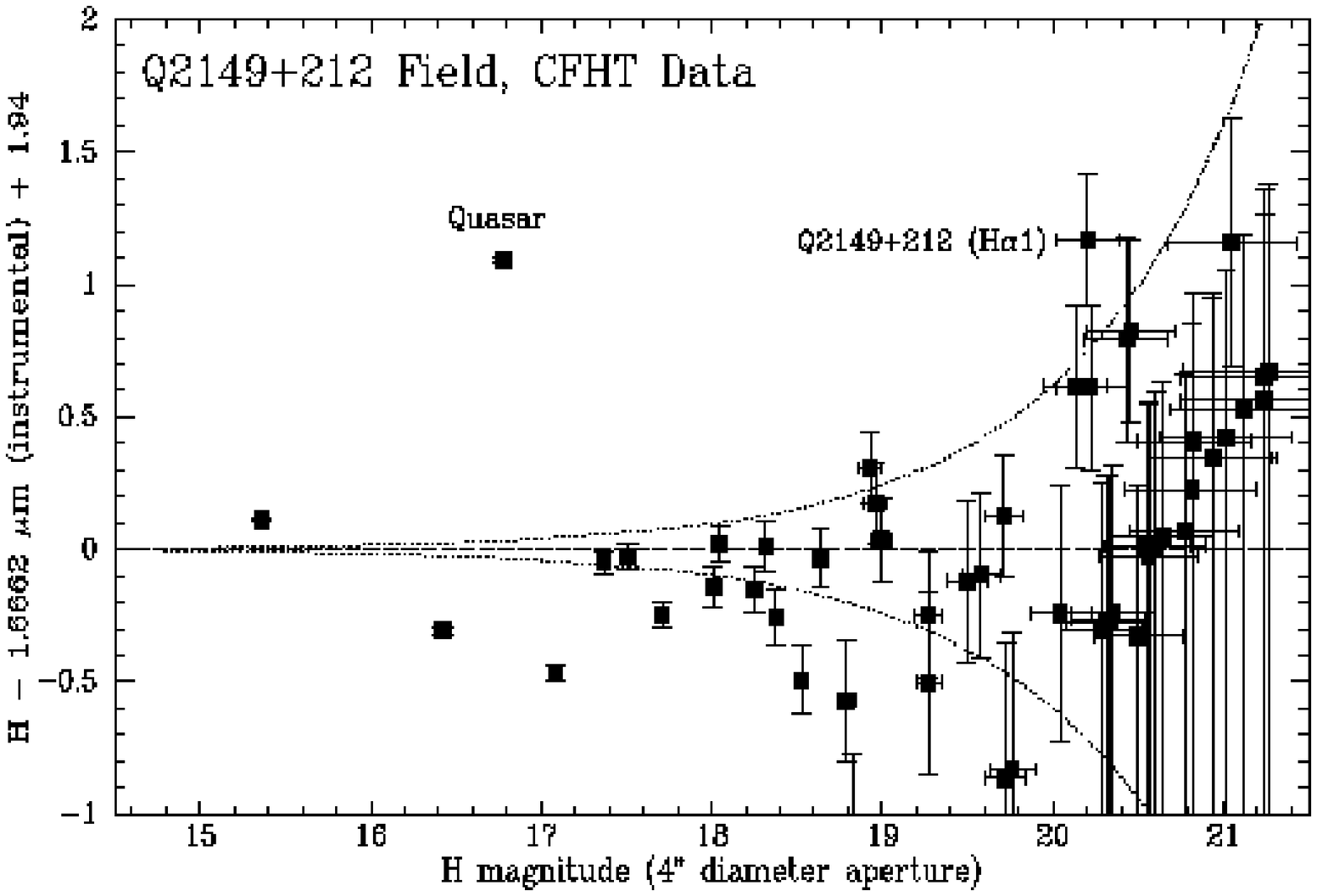} 
\caption[]{
\singlespace 
Color-magnitude diagrams from CFHT $H$- and narrow-band imaging.  
Dotted lines enclose the expected range of objects with narrow-band 
excesses of $<3\sigma$ significance.  
(a) Q~2345+061 field.
The two candidate $3\sigma$ H$\alpha$ emitters have 
$H$$\geq$20.6 and $H-H\alpha\gtrsim$1.7.
The three bright objects outside the $3\sigma$ significance range
range are galaxies larger than the 2\farcs0 radius photometric aperture.  
(b) Q~2149+212 field.
The candidate H$\alpha$ emitter has $H \sim 20.2$ and $H-H\alpha\sim1.2$.
}\label{fig_ha_cfht}
\end{figure}

Figure~\ref{fig_ha_irtf}b shows the IRTF data color-magnitude diagram of the
\twoth\ field, and Figure~\ref{fig_ha_cfht}a the same diagram for the CFHT data.
Figure~\ref{fig_ha_img}a shows the 
\ha\ image with the H$\alpha$ emitters marked.
The candidate \ha\ emitter seen with IRTF --- \twoth~(H$\alpha$1) ---
is not confirmed with CFHT, but the CFHT narrow-band filter is much 
narrower than the IRTF CVF 
and the line could 
lie outside the wavelength range of the CFHT filter.
There are also two objects visible in the narrow-band CFHT image 
but not on the $H$-band CFHT image.  
They both have H$\alpha$ excesses of $3\sigma$ significance with
$H-H\alpha\gtrsim1.7$ (Figure~\ref{fig_ha_cfht}a and Table~\ref{t_ha_objs}).
\twoth~(H$\alpha$2) is detected at $H$=20.3$\pm$0.1 and 
$H-H\alpha$=0.4$\pm$0.3 in the IRTF data
%
%
while \twoth~(H$\alpha$3) is not detected in the IRTF \ha\ image or in the deep \ks\
data of HGC98 to \ks$>$21.7.  It may be detected at $H$$\sim$21.5$\pm$0.5
in the IRTF data and at $r$$\sim$25.1$\pm$0.5 in the $r$ data of HGC98.
%
In any case, given the photometric uncertainties, the CFHT and IRTF \ha\ measurements
are consistent for both \twoth~(H$\alpha$2) and \twoth~(H$\alpha$3).

Figure~\ref{fig_ha_cfht}b
shows the color-magnitude diagram for the CFHT data on the \twone\ field.
There is one candidate H$\alpha$ emitter, with $H \sim 20.2$ 
(Figure~\ref{fig_ha_img}b and Table~\ref{t_ha_objs}).
It is a fairly red galaxy with $r-K=5$ and $H-K\sim1$.
This object is confirmed as an H$\alpha$ emitter in additional data taken the 
same night but not used due to calibration problems,
whereas the object at $H \sim 19$ and $H-H\alpha\sim0.3$ is not confirmed.

Our H$\alpha$ imaging reaches flux limits of 
1 to $2 \times 10^{-17}$ erg cm$^{-2}$ s$^{-1}$ ($3 \sigma$),
deeper than any previously published survey due to the very long IRTF CVF
exposures and the very low background in the CFHT narrow-band images.
Five detections in 10.156~arcmin$^2$ (counting both observations of the
\twoth\ field since different filters were used) gives a surface density of
0.5$_{-0.2}^{+0.3}$~arcmin$^{-2}$.
For comparison, no H$\alpha$ sources were found in two random fields surveyed to
$0.8 \times 10^{-17}$ erg cm$^{-2}$ s$^{-1}$ ($3 \sigma$) with the same CFHT
setup as in this work, for a $1\sigma$ upper limit of 0.2 sources per arcmin$^2$
(Pritchet \etal, in preparation).  
In the deepest published survey, \cite{mea99} find 0.57 sources per arcmin$^2$
to 4.1~$10^{-17}$ erg cm$^{-2}$ s$^{-1}$ ($3 \sigma$) from slitless spectroscopy
of random fields with NICMOS.  
We find only 0.2 sources per arcmin$^2$ to the same limit, but 
they sample a redshift range of $\Delta z=1.2$ 
compared to our average $\Delta z=0.013$.  
Our surface density is approximately 32 times higher, although that number is
biased high because we targeted a redshift where we had reason to believe a 
galaxy excess existed.
In surveys at the redshifts of quasars and radio galaxies,
\cite{pd95}, \cite{vdw97} and \cite{tmm99}
found 4 sources in 242 arcmin$^2$ to limits of
9 to 19~$10^{-17}$ erg cm$^{-2}$ s$^{-1}$ ($3 \sigma$).
We find at most one source in 10.156 arcmin$^2$ to these depths,
which is consistent with the previous surveys due to small number statistics
even though it is formally up to a factor of six excess.

In summary, our fields represent a factor of 2.5 
excess compared to the only existing random-field survey to equal depth, a
factor of $\leq$32 excess compared to the deepest published random-field survey,
and a factor of $\leq$6 excess compared to previous AGN-field surveys.  
The fields studied in this paper are among the best candidates for $1<z<2$
RLQs in clusters from a survey of 31 \citep{hgc98},
so this excess is not unexpected.
These H$\alpha$ detections constitute more evidence that the
galaxy overdensities in these RLQ fields are real, and that
at least some of the excess galaxies are at the quasar redshifts.
Although the number of member galaxies in the candidate host clusters of
\twone\ and \twoth\ is unknown, the deep CFHT $H\alpha$ images show
that there are only three such galaxies with star formation rates of
$\gtrsim 1.5 h_{75}^{-1} M_{\odot}~{\rm yr}^{-1}$
within fields $0.95 h_{75}^{-1}$~Mpc wide centered on these quasars.
This is a lower limit which neglects extinction 
and the velocity dispersion of the clusters
(e.g. \twoth~(H$\alpha$1) is not detected in the narrow CFHT filter), but it
is still a stringent limit which illustrates the potential of deep wide-field
narrow-band data in studying star formation rates in high redshift clusters.

\section{Sub-millimeter Mapping} \label{submm}

In the last few years it has become clear that much of the star formation
activity in the universe is obscured by dust and can be easily detected only
at far-IR and sub-mm wavelengths \citep[e.g.][]{dwe98,sma98}.
The presence of a number of galaxies with SEDs strongly indicative 
of substantial dust reddening in our RLQ fields suggested that
they might be detectable sub-mm sources.
Thus we obtained continuum observations at 450 and 850\micron\ 
with SCUBA, the Sub-mm Common-User Bolometer Array \citep{hea98},
on the James Clerk Maxwell Telescope.
Details of the observations are given in Table~\ref{t_obs}.
All observations used standard 64-point jiggle maps to fully sample both arrays.

Data on Q~2345+061 were obtained (and reduced) as a service observing program.
Azimuthal chopping and nodding was used.  
No sources are seen at 450\micron, but at 850\micron\ \twoth\ is detected
at $2.8\sigma$ (Table~\ref{t_scuba_objs}).  
No other potential sources (e.g. the H$\alpha$ emitters) are seen, but 
we are only sensitive to
hyperluminous IR galaxies ($L_{FIR} > 10^{13} h_{50}^{-2}\ L_{\odot}$).
The $2\sigma$ limit at 850\micron\ corresponds to 
$\sim1.3 \times 10^{13} h_{50}^{-2}\ L_{\odot}$ for the quasar redshift and 
a bolometric correction calculated using a dust emissivity spectral index of
1.0 and the Arp~220 dust temperature $T=47$~K, following \cite{bar99}.

Observations of Q~0835+580 and Q~1126+101 were made 
using a fixed chop throw of 40\arcsec\ 
and 12\farcs36 offsets N or S every other measurement.
Conditions ranged from fair to excellent during the run.
The data were
reduced and calibrated in the standard manner using SURF \citep{jl98}.
Figures~\ref{fig_scuba_img1} and \ref{fig_scuba_img2} show 850\micron\ contours
superimposed on $K$ images of the \oate\ and \eleven\ fields, respectively.
The 450\micron\ maps are not
shown as no objects were detected at that wavelength in either field.
To help reject spurious sources, we cross-correlated the maps with the beam
as given by a calibration observation of an unresolved source on the same night.
Each pixel's value in the cross-correlation map measures how closely
the sub-mm map centered on that pixel resembles the beam, providing a
measure of the correlation coefficient (CC) for each tentative source.
Real high-redshift objects should resemble the beam quite closely,
since objects more extended than a point source cannot be at high redshift 
given the large 
JCMT beam and objects more compact than a point source might be just noise.

There are two $\geq 3.5 \sigma$ sources in the \oate\ field
which coincide with peaks in the cross-correlation map, 
but the bright source (SM1; 10.1~mJy) has only CC=0.37
and the faint source (SM2; 7.5~mJy) only CC=0.29.
There are nearly equal numbers of negative and positive peaks at these flux and
CC values, so we regard both these sources as tentative sub-mm detections.
However, they both have possible optical/IR IDs:
a $K=19.8$ moderately blue galaxy (\oate~(117), \rk=3.9) 
3\farcs2 from the position of SM1,
and a $K=17.7$ red galaxy (\oate~(458), \rk=5.4) 
2\farcs1 from the position of SM2.
The ID numbers of these objects refer to our optical/NIR catalogs of these 
fields, available on request from the first author.
We include information on these objects in Table~\ref{t_scuba_objs}
(and Table~\ref{t_photz}),
but we emphasize that we do not consider them unambiguous sub-mm detections.
Both objects are faint enough in the optical and near-infrared
that they could be at the quasar redshift;
\oate~(SM2) would be 0\fm4 brighter than the magnitude of a brightest cluster
galaxy (as estimated in HG98) if it was at the quasar redshift,
but this is within the observed
scatter in such objects' magnitudes at $z\simeq1$ \citep{tas00}.
We discuss these objects further in \S\ref{phot0835submm}.
Note that \oate~(H$\alpha$1) is not detected, although the $3 \sigma$ limit
of SFR$_{FIR} < 2090\pm550$ $h_{75}^{-2} M_{\odot}~{\rm yr}^{-1}$
(calculated following \cite{dea99} using a Salpeter IMF and assuming it is at
  the quasar redshift) 
is two orders of magnitude larger than SFR$_{UV}$ and SFR$_{H\alpha}$
for this object (\S\ref{irtf}).

In the \eleven\ field, the quasar and several other sources are present
with $\geq 3 \sigma$ and CC$>$0.33.
There are twice as many fluctuations above sky as below sky
at these flux and CC values in this field; thus,
some of the detections are probably real but some are probably spurious.
%
Overall the most reliable detections are the quasar and SM1. 
Despite their low S/N, they were clearly detected on both nights,
have high correlation coefficients, and have FWHM$\sim$14\arcsec\ as expected
for the JCMT beam at 850\micron.
The possible counterpart 1\farcs1 to the N of SM1, 
object \eleven~(384), is discussed further in \S\ref{phot1126submm}.
The two other sources listed in Table~\ref{t_scuba_objs} are less certain,
despite their higher S/N on the raw map.  They have FWHM$\sim$9\arcsec,
suggesting they might be noise spikes rather than actual objects.
%
%
Neither \eleven~(SM2) nor (SM3) have probable IDs in the optical/IR data.

%

\section{CFHT Adaptive-Optics Imaging} \label{aob}

Little information is available on the detailed morphologies of EROs,
although our knowledge is growing rapidly
\citep{dea99,tea99,ben99,mcd00,st00}.  
To improve this situation, $H$-band
adaptive-optics data were obtained 
with the CFHT Adaptive Optics Bonnette (known as $Pue'o$) 
and the KIR camera (1024$^2$ HgCdTe detector, 0\farcs0348/pixel).
The $V$=10.3 star BD+58~1138 was used to guide on a field 
$\sim$35\arcsec\ from the star and $\sim$90\arcsec\ from \oate.
Standard infrared observing and reduction procedures 
were used, including domeflats for flattening.
Photometry was done using UKIRT standards 
and assuming a typical Mauna Kea $H$-band extinction coefficient of
0.051$\pm$0.012 \citep{kri87}.
Natural seeing conditions during the observations were poor, 
resulting in FWHM after correction of $\sim0\farcs42$,
as measured from the most compact object on the image.

The KIR field contains two objects bright enough for image profile analysis.
The fainter, redder object is designated \oate~(106) since it is \#106 in our
catalog of this field 
and has \ks=17.7, \rk=5 and \jk=2.
The other object, \oate~(112), has \ks=17.3 and \rk=2.4.
Radial profile analysis shows that both galaxies are much
better fit by an $r^{1/4}$-law profile than by an exponential disk profile,
with $r_e$=0\farcs97$\pm$0\farcs14 for \oate~(106) 
and $r_e$=0\farcs49$\pm$0\farcs07 for \oate~(112).

The resolution was not high enough for further morphological analysis, 
but if we assume the objects are early-type galaxies as indicated by their
radial profiles, we can use the Kormendy relation between effective radius and
average effective surface brightness \citep{kd89}
to determine what redshifts are consistent
with the objects' observed sizes and surface brightnesses \citep{eis96}.
To perform this analysis we use the elliptical colors and evolutionary and
$k$-corrections of \cite{pog97} and adopt their \ho=50, \qo=0.225 cosmology
for consistency.
For each redshift at which \cite{pog97} calculate $e$- and $k$-corrections,
we use the CFHT data to compute the rest-frame $B$-band surface brightness
within $r_e$ (accounting for (1+$z$)$^4$ surface brightness dimming)
and the physical size corresponding to the observed $r_e$.
The tracks for both objects are shown in Figure~\ref{fig_remue}
along with data for local ellipticals from \cite{sp90}.
Both objects intersect the blue surface brightness-effective radius
relation for ellipticals both at low and high redshift.
\oate~(112) could be at $0.04<z<0.24$ or $1<z<1.4$,
and \oate~(106) could be at $0.02<z<0.10$ or $1<z<1.3$.
We compare these estimates with those from photometric redshifts in
\S\ref{phot0835ao}.  

\section{Photometric Redshifts and Other Constraints from SED Fits}\label{photz}

Estimating photometric redshifts $z_{ph}$ and spectral types from multicolor
photometry can provide insights into the nature of faint, high-$z$ objects 
which are otherwise too faint or too numerous for efficient spectroscopy 
\citep[for a recent review see][]{yee98}.  In our RLQ fields we wish to
quantitatively confirm or refute the existence of excess galaxies at
or beyond the quasar redshifts and to investigate the SEDs --- and
hence the evolutionary state --- of such galaxies, particularly EROs.
A longer-term goal is to provide an efficient way to reject contaminants
foreground to the quasars in future spectroscopic followup studies.
Currently, we have sufficient multicolor imaging data for such
fitting in only the \oate\ and \eleven\ fields.

An established technique for measuring photometric redshifts is to compare
the photometry of objects of interest with that of an {\it empirical}
template set.  This technique works well for galaxies with normal SEDs,
where local template spectra \citep[e.g. the set of][]{cww80}
provide good results even to high redshifts \citep[e.g.][]{sly97}.  
However, because EROs are atypical objects, 
and potentially very dusty, it is inadvisible to
fit their photometry with templates that are representative only of
normal galaxies.  Instead, it is important to allow for the
possibility of heavy dust obscuration using synthetic
templates reddened with a range of dust values.

We thus constructed a grid of models spanning a range
in redshift, age, and reddening.  The basis of this grid is
the \citet{bc96} version of the spectral synthesis models of 
\citet{bc93}, spanning 221 age steps $0 \leq t \leq 2 \times 10^{10}$ yr.
We considered instantaneous burst and continuous star formation models,
which bracket the extremes of monotonically decreasing star formation
histories.
We assumed a metallicity of $0.2 Z_\odot$, though solar metallicity models 
were also considered for a subset of objects.  Our synthetic spectra
were attenuated using the \citet{cal97} starburst galaxy extinction recipe,
for a total of 67 values of $E(B-V)$ spanning $0 \leq E(B-V) \leq 2$.
The reddened spectra were then redshifted onto a grid of 101 redshifts
spanning $0 \leq z \leq 5$, after which they were convolved with
filter transmission curves to compute the final grid of galaxy colors
as a function of redshift, reddening, and age.  

The observed photometry for each object was compared with the model grids
and the best-fitting position in $z_{ph}$, age, $E(B-V)$ space was found 
by means of a maximum likelihood test.  Almost all the input data used in 
these calculations was reported in HGC98.  The exception is five hours of
$U$ imaging of the Q~0835+580 field obtained with the Steward
Observatory 90$''$ on UT 1997 December 31 and reduced in the same
manner as the data of HGC98.  FOCAS total magnitudes were measured 
in all filters using apertures from the summed $r + J + K_s$ image
and converted to fluxes using the conversions in Appendix
A of HGC98 plus that of \cite{fsi95} for the $U$ data.

Our procedure is thus similar to that used by \citet{sy98} in
their investigation of the spectral energy distributions of Lyman
break galaxies.  However, unlike in \citet{sy98} where the
redshifts of objects of interest were known a priori, here we must
find the best fitting model in the three-dimensional space that covers
redshift in addition to age and dust attenuation values.

For each object fit, \chinu\ (reduced \chisq) contour plots with projected 
90\% confidence error contours in the $z$/\ebv\ and age/\ebv\ planes were
constructed for each SFR scenario 
(Figures~\ref{fig_photz0835.339}-\ref{fig_photz425.1126}
  and Table~\ref{t_photz}).
All our quoted uncertainties and
plotted error bars are these projected 90\% confidence limits.  
They are thus somewhat more conservative that the standard $1\sigma$ 
(68\% confidence) gaussian error bars.  It should also be kept in mind 
that the 90\% confidence regions are typically not just elliptical
gaussians, but are elongated because of degeneracies between parameters
such as age and \ebv.  Plotting simple error bars, while convenient,
thus tends to exaggerate the apparent uncertainties. 
%
We define a successful fit as one where at most one of the three
variables of interest --- \zph, \ebv\ and age --- is unconstrained
at 90\% confidence (i.e., that variable has 90\% confidence
limits equal to its range of permitted values).
Note that a successful fit can still be a poor fit with a high \chinu\ value,
and vice versa.  There is a continuum of properties across our dividing line 
between successful and unsuccessful fits, of course, 
but objects not successfully fit typically have large photometric errors
which prevent their SEDs from usefully constraining the fitted parameters.
However, unsuccessful fits can also be caused by star formation histories 
which deviate strongly from the instantaneous burst or continuous histories 
we consider,
such as multiple strong bursts or slowly declining star formation rates.

In the following sections we present our photometric redshift results.
In addition to fitting a handful of specific objects of interest, we examine a
compact clump of predominantly red galaxies around \oate\ to confirm or refute
the visual impression that these galaxies are associated with the quasar.
We also examine the population of very red galaxies in these fields regardless
of their distance from the RLQ, because these galaxies may be part of extended
structures at the quasar redshift.  Furthermore, these red objects may be dusty
starbursting galaxies or high-redshift galaxies with old stellar populations,
both of which are of great interest to the study of galaxy evolution.

\subsection{\oate\ Field}	\label{phot0835}

The field of \oate\ 
has data in seven filters ($UrizJHK$) for most objects.
As discussed above, we do not compute \zph\ values for all galaxies in the
field, but instead focus in the following sections
on specific galaxies or galaxy populations.

\subsubsection{The Spatially Compact Group Around \oate} \label{phot0835rp}
%
There are 18 objects within 21\farcs5 of \oate\ to $K=21.24$ 
(the $5\sigma$ limit of our data) which form a spatially compact group
in the terminology of \citet[][]{sea97}.
This count excludes the quasar,
one bright morphologically classified star, and one probable star which
is unresolved on an {\em HST}\ WFPC2 snapshot image.  As discussed in HG98 
(see their Figure~23), the compact spatial distribution of these galaxies
and the red color of many of them makes it very likely that they are at the
redshift of the quasar ($z_q$=1.5358) or the intervening \mgii\ systems
(\z$\simeq$1.4368).  

To test this hypothesis, we have fitted the broadband magnitudes of these 18
objects.  They are denoted `SCG' in the Notes column of Table~\ref{t_photz},
		 which summarizes the fit results.
All objects had successful fits, as defined in the previous section.
Figure~\ref{fig_photz0835zebv}a shows \ebv\ vs. \zph\ 
(at the best fit model age) for the twelve objects with $K\leq20$ 
(large symbols) and the six with $K>20$ (small symbols).
Figure~\ref{fig_photz0835zebv}b shows the best fit model age vs. \zph.
Instantaneous bursts (squares) are preferred over
constant star formation histories (triangles)
in nine of twelve cases at $K\leq20$ but only two of six cases at $K>20$.
Figure~\ref{fig_photz0835zebv}b shows the best fit model age vs. \zph.
The dotted vertical line in both figures shows the quasar redshift.
Recall that the simple error bars we plot for convenience
do exaggerate the uncertainties somewhat.

Twelve of the 18 objects are consistent with being at the quasar redshift
at 90\% confidence.  Another which misses being consistent by $\Delta z = 0.06$
is coincident with the southern radio hotspot of the quasar 
(see \S7.2 of HG98), so we consider it to be at the quasar redshift as well.
Thus there is good evidence from photometric redshifts for 
galaxy overdensities at the quasar redshifts.
For these 13 galaxies the typical \ebv\ is $\sim$0.3 and the typical 
fitted age is $\sim$1~Gyr.
Both quantities have large uncertainties, and there is a large scatter 
in the ages.
Nonetheless, overall the plots are consistent with current or recent 
star formation in the majority of the galaxies at the quasar redshifts,
as denoted by a best-fit constant star formation history or a best-fit age of
$<1$~Gyr.

The notable exception to this is \oate~(339), which is a bright ERO with 
$r-K>7$ best fit by an instantaneous burst with $z_{ph}=1.65_{-0.30}^{+0.20}$,
\ebv~$=0.00_{-0.00}^{+0.20}$, and age $19.75_{-13.00}^{+0.25}$~Gyr
(Figure~\ref{fig_photz0835.339}a).
	Four other objects have similar best-fit ages which are older
	than the universe at the quasar redshift for all reasonable cosmologies,
	but this is the only one for which even the 90\% confidence lower limit
	on the age is older than the universe.
For reference, note that for the current best-fit flat, low-$\Omega_M$,
non-zero $\Omega_\Lambda$ cosmology, at $z=1.5$ the age of the Universe was
$(4.7_{-0.8}^{+1.7})h_{75}^{-1}$~Gyr ($\Omega_M = 0.2\mp0.1$, 
					$\Omega_\Lambda = 1 - \Omega_M$).
To see if we have overestimated this object's age by underestimating its
metallicity, we refitted it using solar metallicity models.
Both solar metallicity fits have lower \chinu\ than 20\% solar fits, though
the instantaneous burst is again preferred (Figure~\ref{fig_photz0835.339}b),
with slightly lower values of all three main parameters:
$z_{ph}=1.40_{-0.20}^{+0.20}$,
\ebv~$=0.00_{-0.00}^{+0.04}$, and age $12.5_{-8.0}^{+7.5}$~Gyr.
The 90\% confidence lower limit on the age for even this solar metallicity fit
is still older than the age of the universe at the quasar redshift,
except for flat $\Lambda$-dominated cosmologies with $\Omega_M \leq 0.2$.  
Thus, even though higher metallicities may further reduce the best-fit age,
\oate~(339) is clearly worthy of spectroscopic study as a candidate
very old galaxy at high redshift.

Another of the objects within 21\farcs5 of the quasar, \oate~(352),
is known to be at $z=0.236$ \citep{bur90}, which provides a useful check
of the accuracy of our photometric redshifts.  The object is confused with 
a separate red galaxy (Figure 23 of HG98) which contributes negligible flux
in $Uriz$ but $\sim$0\fm5 in $J$ and $H$ and $\sim$0\fm25 in $K$.
Our fit to its corrected photometry yields $z_{ph}=0.35_{-0.05}^{+0.20}$
and \ebv=0.16$\pm$0.06.  
This redshift is only $2.3\sigma$ above the known redshift,
and this \ebv\ value is close to the average value of $\simeq$0.1 estimated for
field galaxies at $<$$z$$>$=0.3 \citep{lin01}.
We thus consider our fit to be acceptable.

\subsubsection{\oate\ (H$\alpha$1)}	\label{phot0835ha}
%
The object \oate~(398) is also known as \oate~(H$\alpha$1),
a candidate H$\alpha$ emitter at the quasar redshift (\S\ref{irtf}).
However, it is best fit as a young $z_{ph}\simeq2.15\pm0.25$ galaxy with
\ebv$\simeq$0.35$\pm$0.20 under either the single stellar population or 
constant SFR models, although a similarly reddened object at $z_{ph}\sim0.25$
is also possible at 90\% confidence.
We also fit this object using solar metallicity models and found that the
best-fit redshift did not change significantly, though the best-fit reddening
was somewhat lower, \ebv$\simeq$0.20$\pm$0.20, and only the constant SFR model
still permitted a low-redshift ($z_{ph}\sim0.2$) solution at 90\% confidence.
It is possible that the narrow-band filter detected 
\OIII\,$\lambda$5007 at $z=2.3338\pm0.0186$ instead of H$\alpha$ at $z=1.5358$,
but the $z_{ph}\sim0.25$ option would be impossible to reconcile with the
narrow-band detection.
Spectroscopy is needed in this case to verify the photometric redshift.

\subsubsection{\oate\ Field Adaptive Optics Targets}	\label{phot0835ao}
%
Photometric redshifts were computed for \oate~(106) and (112) for comparison
with the structural redshifts calculated for these objects from adaptive optics
images (\S\ref{aob}).  
The direct SED fits do not agree very well with the structural redshifts.
Figure~\ref{fig_photz106.0835}a shows that \oate~(106) is best fit 
as a $z_{ph}=4.50\pm0.15$ galaxy with \ebv~$\simeq0.23_{-0.18}^{+0.25}$.
The structural redshift of $z\simeq0.8$ 
is excluded at high confidence for the models considered.  This result
supersedes the earlier modelling attempt reported in \citet{hal99tp3b}.
Figure~\ref{fig_photz106.0835}b shows that \oate~(112)
is not well fit by any of our simple models (\chinu~$\simeq20$).
A composite stellar population may improve the fit, but that may cast
some doubt on its morphological identification as an early-type galaxy.
From comparison of Figures~\ref{fig_photz106.0835}-\ref{fig_photz425.1126},
as well as similar plots for other objects (not shown), it appears that the
$riz$ fluxes for these two objects, especially \oate~(112), are unusually
bright relative to the $U$ and $JHK$ fluxes, even compared to blue objects.
Thus we believe the reason for the poor SED fits to these objects is that
their $riz$ photometry is affected by scattered light
from the nearby adaptive optics guidestar, despite attempts to account for
such light through masking, subtracting, and using local backgrounds.
Multiwavelength adaptive optics imaging may be necessary to obtain accurate
photometry for further study of these objects.

\subsubsection{\oate\ Field Possible Sub-mm Counterparts} \label{phot0835submm}
%
\oate~(117) and (458) were chosen for SED fitting
since they are possible counterparts of tentative sub-mm detections (\S\ref{submm}).
Figure~\ref{fig_photz117.0835}a shows that \oate~(117) is best fit as an
instantaneous burst with \ebv$=0_{-0.00}^{+0.12}$ at a redshift consistent with
that of the quasar ($z_{ph}=1.60\pm0.15$).
A constant star formation history requires \ebv$=0.20_{-0.16}^{+0.14}$ and is
also consistent with the quasar redshift, though with larger uncertainties.
Neither fit provides evidence for extreme dust reddening as might be expected
for a sub-mm-detected galaxy, 
and neither does a fit using solar metallicity spectral templates.
Thus the identification of \oate~(117) with \oate~(SM1) is doubtful.
However, such simple positional identifications of optical counterparts to 
sub-mm sources are not expected to be extremely robust given the $14\arcsec$
FWHM of the SCUBA beam at 850\micron.

Figure~\ref{fig_photz117.0835}b shows that \oate~(458) 
is best fit as an instantaneous burst at $z_{ph}=1.2_{-0.1}^{+0.2}$ with 
\ebv$=0.55_{-0.05}^{+0.15}$,
though a constant star formation history at $z_{ph}=1.35_{-0.15}^{+0.10}$
with the same reddening is also acceptable.  
This galaxy's photometric redshift, which places it
slightly foreground to the quasar, is consistent with it being slightly
brighter than expected for a brightest cluster galaxy at the quasar redshift.
Its high \ebv\ is consistent with it being the counterpart of \oate~(SM2).
This possible luminous sub-mm-detected starburst at (or nearly at) the quasar
redshift clearly warrants spectroscopic followup.

\subsection{\eleven\ Field}	\label{phot1126}

The field of \eleven\ ($z_q=1.5173$)
has data in five filters ($rzJHK$) for most objects,
and we can fit only objects with data (including upper limits) in all filters.
There is no distinct spatially compact group of galaxies around \eleven\ as 
there is around \oate.  However, there is a nearby clump of $J-K$ selected EROs
which may be galaxies background to the quasar or very dusty galaxies at
the quasar redshift (discussed in \S\ref{phot1126jdrops}), as well as a 
population of $R-K$ selected EROs which may be old galaxies in an overdensity 
at the quasar redshift (discussed in \S\ref{photboth}).  
First, however, we discuss a possible sub-mm source identification.

\subsubsection{\eleven\ Field Possible Sub-mm Counterpart}\label{phot1126submm}
%
\eleven~(384) was chosen for SED fitting since
it is a possible counterpart of sub-mm source \eleven~(SM1) (\S\ref{submm}).
The best fit (\chinu=2.78) is an instantaneous burst with 
\zph~$=1.50_{-0.15}^{+0.30}$ and \ebv~$=0.50_{-0.14}^{+0.20}$
(Figure \ref{fig_photz384.1126}); 
the constant SFR fit also requires a large \ebv~$=0.55_{-0.11}^{+0.20}$.
These large $E(B-V)$ values support the idea that Q~1126+101~(384) 
is the optical counterpart of the SCUBA source in this field.
The galaxy has $K=17.7$, making it $\sim$0\fm4 brighter than the magnitude
of a brightest cluster galaxy at the quasar redshift as estimated by HG98, but
it is within the observed scatter in such objects' magnitudes at $z\simeq1$
\citep{tas00}.  As with \oate~(458), this possible luminous sub-mm-detected
starburst at the quasar redshift clearly warrants spectroscopic followup.

\subsubsection{\eleven\ Field $J-K$ selected EROs} \label{phot1126jdrops}
%
Four of the brightest $J-K$ selected EROs in the \eleven\ field had data
sufficient for SED fitting.
Figure~\ref{fig_photz447.1126}
shows the best fit of these, 
\eleven~(447) and \eleven~(424).
\eleven~(447) is best fit as a dusty instantaneous burst 
consistent with the
quasar redshift (\zph~$=1.9_{-0.5}^{+0.4}$ and \ebv~$=0.75_{-0.45}^{+0.55}$).
\eleven~(424) is best fit as a less dusty, constant SFR model
background to the quasar at $>$99.9\% confidence
(\zph~$=3.00_{-0.40}^{+0.95}$ and \ebv~$=0.46_{-0.10}^{+0.24}$).
This agrees with our suggestion in \S7.3 of HG98 that it is an example of a 
dusty star-forming galaxy at $z\gtrsim2.5$, with the 4000~\AA\ break 
in $J$ or beyond.  

The other two $J-K$ selected EROs we fit were \eleven~(381), which has very red
$z-J>3.4$, and \eleven~(425), the brightest $J$-band dropout in the field.  In
HG98 we suggested that these were examples of dusty star-forming galaxies at the
quasar redshift $z_q=1.5173$, which puts the 4000~\AA\ break between 
$z$ and $J$.  Figure~\ref{fig_photz425.1126} shows that both of these objects
are best fit with highly
reddened constant SFR models (though for \eleven~(381) an instantaneous burst
is almost as good) background to the quasar at almost exactly 90\% confidence.
\eleven~(381) has \zph~$=1.95_{-0.35}^{+0.45}$ and \ebv~$=0.55_{-0.11}^{+0.40}$.
\eleven~(425) has \zph~$=2.35_{-0.25}^{+0.60}$ and \ebv~$=0.55_{-0.07}^{+0.20}$.
This \zph\ for \eleven~(425) is slightly lower than the value of 
\zph~$=3.5\pm0.5$ reported in \citet{hal99tp3a} based on earlier modelling,
but the \ebv\ is the same within the uncertainties.
The fits for both objects have large \chinu\ values because they underpredict
the flux in the $r$ band,
and the same is true to a lesser extent for \eleven~(424).  This may
represent a limitation of our simple models with uniform \ebv.  The addition
of a small amount of unreddened light from a young stellar population could
bring the fits into agreement with the $r$-band data while not destroying
their agreement with the observed fluxes at longer wavelengths (cf. HG98).

As mentioned in HG98, the large \ebv\ values for these objects are consistent
with their \jk\ colors being comparable to, or redder than, the prototypical
dusty ERO HR10, which requires dust reddening to fit its SED at its known 
$z=1.44$ in any reasonable cosmology \citep{hr94,gd96,cim97}.
However, the prediction of HG98 that \eleven~(381) and \eleven~(425) would be
at the quasar redshift seems erroneous based on their photometric redshifts,
though the quasar redshift is excluded at only 90\% confidence.
The \zph\ of these objects are plausible in the sense that their 
$K$ magnitudes 
would be fainter than those of powerful radio galaxies 
(typically the brightest galaxies at any redshift) at the same redshifts
\citep{vb98}.
Their being background to the quasar is also consistent with the lack of a
strong concentration of excess galaxies around the quasar on the sky in this
field (\S\ref{wfdisc}).
We also fitted solar metallicity models to these four objects to see if their
\ebv\ values or redshifts depended strongly on metallicity.
The fits were of comparable but slightly worse \chinu\ on average, and none of
the best-fit redshifts or \ebv\ values changed enough to affect our conclusions.

In summary, fits to the SEDs of these four $J-K$ selected EROs all require
strong reddening, with \ebv\ values of $\sim0.6\pm0.3$. 
Two are better fit by constant SFR models, while two have nearly identical
\chinu\ for either constant SFR or instantaneous burst models.  The best-fit
photometric redshifts are $1.9\leq z_{ph} \leq3$, and only \eleven~(447)
is consistent with being at the quasar redshift at 90\% confidence.
Small amounts of unreddened light from a young stellar population are needed
to match the $r$-band data for all three of the $z_{ph}>z_q$ objects.
All these conclusions hold for both solar and 20\% solar metallicity templates.
These \jk\ selected EROs may be representative of a population of dusty and
star-forming galaxies missing from optically selected samples.

\subsection{Very Red Objects and Extremely Red Objects}		\label{photboth}
Photometric redshifts were also computed for all objects with \rk~$>5.4$ and
$K<19.6$ in both fields.  Such objects could be old stellar populations
at high redshift or dusty starbursting galaxies, both populations of great
interest to the study of galaxy evolution.
Objects with $r-K>5.4$ but $r-K<6.322$ are not red enough to meet our 
definition of an ERO, but objects with $r-K>5.322$ are called EROs by 
\citet{cea99} and \citet{dad00},
and objects with $r-K>5.622$ are called EROs by 
\citet{pm00} and \citet{mar00}.
To avoid confusion with our definition of EROs we refer to objects with
$5.4<r-K<6.322$ as Very Red Objects \citep[VROs; cf.][]{coh99a}.
There are 12 such objects in the \oate\ field, one of which 
was not successfully fit.
Excluding two objects with the colors of late M or L dwarf stars 
\citep[cf.][]{thesis}, 
there are 12 VROs in the \eleven\ field with sufficient photometric data 
for fitting,
one of which was not successfully fit.
Overall, in both fields, we successfully fit 
22 \rk~$>5.4$ objects with $K<19.6$.\footnote{The two unsuccessfully fit
objects are \oate~(618) and \eleven~(50).
\oate~(618) has \zph~$<0.9$ at 90\% confidence, but its age and \ebv\ are
unconstrained since the SED can be fit either as an extremely red 
$z<0.5$ galaxy and an unreddened $z\sim1.5$ galaxy.
We suspect \eleven~(50) could not be fit because of 
its red $r-z$ and blue $z-J$ colors, which can probably be
simultaneously fit only by a solar metallicity model at $z\sim1$.}
These 22 objects include four from 
the spatially compact group discussed in \S\ref{phot0835rp} and three 
of the four \jk\ selected EROs discussed in \S\ref{phot1126jdrops}.
One of these 22, \eleven~(91), is best fit as a $\lesssim$0.2~Gyr old 
instantaneous burst with \ebv~$=0.6\pm0.2$ at $z_{ph}=5.0\pm0.3$.
If this is correct, with $K=19.5$ it would be one magnitude brighter than the
$z=5.19$ radio galaxy TN~J0924-2201 \citep{vb99}, but consistent with the
scatter seen in the $K-z$ relation at $1<z<3$.  Thus this $z_{ph}$ is not
impossible, though it is unlikely.  
In this section we discuss only the remaining 21 of 22 very or extremely red 
objects, which all have $1 \lesssim z_{ph} \lesssim 2$.

Figure~\ref{fig_photbothzebv}a plots \ebv\ vs. \zph\ for these 21
objects, and Figure~\ref{fig_photbothzebv}b plots age vs. \zph.
The distribution of \ebv\ values, and to a lesser extent the distribution of
ages, are both skewed to slightly larger values than in 
Figure~\ref{fig_photz0835zebv},
with median \ebv~$\sim0.4$ instead of $\sim0.25$ 
and median age $\sim0.7$~Gyr instead of $\sim0.3$~Gyr.  Both trends
are understandable since the objects in Figure~\ref{fig_photz0835zebv}
were chosen only for their proximity to the quasar, while those in
Figure~\ref{fig_photbothzebv} were chosen on the basis of their red colors.
Figure~\ref{fig_photbothzebv} also shows that 
there are more objects consistent (at the 90\% confidence level)
with the quasar redshift in the \oate\ field (open squares) than in the
\eleven\ field (filled squares).  This is expected, since many of the
\eleven\ field objects are $J-K$ selected EROs, and at least some of that
population is not expected to be at the quasar redshift 
(\S\ref{wfdisc} and \S\ref{phot1126jdrops}).

Following \citet{pm00}, 
we plot these 21 objects 
on an $r-K$ vs. $J-K$ color-color diagram to study the separation
of unreddened and reddened galaxies at $1 \lesssim z_{ph} \lesssim 2$
\citep[see also][]{mar00}.
Figure~\ref{fig_photbothjkrkebv} shows most objects' fits require dust
reddening of \ebv~$>0.3$ 
(filled symbols, with symbol size scaling linearly with \ebv). The separation
of reddened and unreddened galaxies (open symbols) is not as clean as
predicted by the modelling of \citet{pm00} for objects at $1<z<2$.
Our modelling is very similar to that of \citet{pm00}, but we fit a wider range
of ages and \ebv\ values and use information from more filters than just $RJK$
or $IJK$.  In particular, they do not consider reddened instantaneous burst
models, which we find are the best fits for many of our objects.
Larger samples plus spectroscopic redshifts and spectral types
are clearly desirable to calibrate this classification system.
For now, we note that the general trend of objects with larger \ebv\ being
redder in $J-K$ is valid, but that this trend is not accurately predictive on
an object by object basis.  \ebv~$=0$ objects may be an exception, since they 
seem to occupy a distinct region of the diagram with little contamination.  

We also note that our results may differ from those of \cite{mcd00}, who 
assembled {\em HST} images of a heterogeneous sample of VROs and EROs 
in both blank and candidate high-redshift cluster fields.
They found that VROs and EROs with irregular morphologies, which are likely
to include a higher percentage of star-forming galaxies than objects with
regular morphologies, tend to have redder $R-K$ or $I-K$ colors.  
We may be able to compare these irregular EROs to objects in our sample which 
have large best-fit \ebv\ values, which also should include a higher percentage
of star-forming galaxies than galaxies with low reddenings.  As seen in
Figure~\ref{fig_photbothjkrkebv} and discussed above, we find that objects with
large reddening do not tend to have redder $r-K$ colors, although they do tend
to have redder $J-K$ colors.
Our heavily reddened EROs and the irregular EROs of \cite{mcd00} may be 
different populations, of course, but it would be very interesting if they
were in fact comparable.  In that case this result would indicate
a breakdown in the relationship between morphology and star formation
properties at $z \gtrsim 1$; for example, if we are seeing irregular
galaxies with relatively old or quiescent stellar populations, and young
or actively star forming galaxies with nonetheless regular morphologies.

To test the sensitivity of our fits to metallicity, we refit 16 
objects using solar metallicity models instead of 20\% solar ones.  These
solar fits had comparable \chinu, with a tendency toward worse fits on average.
Best-fit redshifts and \ebv\ values were unaffected by the different
metallicity, with no discrepancies at $>$90\% confidence.
Best-fit ages were found to be more sensitive to the metallicity, but even so,
only in five cases where the solar metallicity fits had lower \chinu\ did the
best-fit ages change (decrease) at $\gtrsim$90\% confidence.
All five cases are best-fit instantaneous bursts, and four 
are \ebv~$=0$ objects.  This is consistent with our \ebv\ values being 
essentially the same for solar and 20\% solar metallicity fits;
the majority of our objects require dust for both metallicities,
while a minority do not require it for either metallicity.

Finally, we attempted to fit four objects with 
$z-J>2.9$, suggestive of a spectral break near 1\micron,
which have neither $r-K>5.4$ nor $J-K>2.5$.
One object --- \eleven~(253) --- was not successfully fit due to its detection
only in $J$ and $K$.  The three successfully fit objects are denoted `red $z-J$'
in the Notes column of Table~\ref{t_photz}, which summarizes the fit results.
\oate~(343) and (379) have fits similar to those of VROs and EROs.
The $r-K>5.4$ definition of a VRO is somewhat arbitrary,
and these two red $z-J$ objects have $r-K \simeq 5$,
so the similarity of their fits to those of VROs is not surprising.
\eleven~(269) also has a best fit similar to those of VROs or EROs,
but it has a large \chinu\ value.  We believe this is because it has only a 
moderately red color $r-K=4.4$ which our simple stellar population models 
have trouble fitting simultaneously with this object's very red $z-J>3.7$.
A higher metallicity model with a slight contribution from an unreddened 
young population is probably needed to improve the \chinu\ for this 
object.\footnote{This object is one of seven in Table~\ref{t_photz} 
with 
\chinu~$\gtrsim 10$.  
Two others were discussed in \S\ref{phot1126jdrops}; they also require
a contribution from an unreddened young population to fit the observed SED.
Of the remaining four, \oate~(307) probably requires both a metallicity
$>0.2 Z_\odot$ and a young spectral component while \oate~(280) and (410) 
probably require only metallicities $>0.2 Z_\odot$ and \oate~(66) is 
definitely better fitted by a solar metallicity model.
It is encouraging that all seven objects' high \chinu\ can be understood
as due to limitations of our modelling.
}

%
In summary, we have performed SED fits to
a sample of red objects with a high {\em a priori}
probability of having strong dust reddening or old stellar populations.
Many of these objects' SED fits yield photometric redshifts 
consistent with the quasar redshifts.
We find that significant dust is required to fit most of these objects,
including about half of those which also require old ages.
Our fits have \ebv\ values ranging from 0 to 1.1, with median \ebv~$\sim$0.4.
For comparison, \citet{lin01} find an average \ebv~$\simeq0.1$ from fitting
the luminosity density evolution of $z<0.65$ field galaxies from the CNOC2
redshift survey,
\citet{tws00} find a distribution \ebv~$\leq0.6$, albeit peaked at
\ebv~$\leq0.1$, for $z<2$ galaxies in the deep NICMOS image of the HDF-N,
and \cite{hea01} find \ebv~$\leq1$, peaked at \ebv~$\sim0.5$, for a sample
of luminous compact galaxies at $z \sim 0.65$.
Our \ebv\ values do not change significantly when we fit solar metallicity 
models instead of 20\% solar models.
However, the best-fit ages for the four objects with best-fit \ebv~$=0$
do decrease significantly for solar metallicity fits.

Thus the very red objects in our fields appear to consist of two populations.
The minority population has no dust and is red due to age and/or high
metallicity, while the majority population is red due to dust, 
albeit with old age contributing to the red colors in some cases.
This result may differ from those of \cite{st00} and \cite{mcd00}, who find
that the fraction of dusty starbursts among VROs and EROs is $\lesssim$30\%
and $\sim$20$-$50\%, respectively, based on their morphologies as seen by
{\em HST}.  As discussed above, our heavily reddened objects and the irregular
and exponential disk objects of \cite{st00} and \cite{mcd00} may be different
populations, but it would be very interesting if this discrepancy was real and
indicated a breakdown in the relationship between morphology and star formation
properties at $z \gtrsim 1$.

\section{Conclusions} \label{concl}

In previous work we have identified an excess population of predominantly red
galaxies around a sample of 31 radio-loud quasars at $1<z<2$.  In this paper
we have presented new multiwavelength data and analyses on the fields of
four of these quasars at $z\sim1.54$, obtained to build more detailed pictures
of the environments of these quasars and the galaxies within them.
Our conclusions are as follows.

1.  These fields have a surface density of extremely red objects
(EROs, with $R-K>6$) 2.7 times higher than the general field.  
Assuming these EROs are passively evolved galaxies at the quasar redshifts,
we find that they have characteristic luminosities of only $\sim L^*$.
Thus these \rk\ selected EROs are consistent with being drawn from
the bright end of the luminosity function of early-type galaxies.
The higher density of such EROs in RLQ fields can be easily understood
if the RLQs are located in overdense regions.  
However, evolution in the dust content of these EROs could affect 
the luminosity estimates in up to $\sim$40\% of the population.
We also show that only one of four RLQ fields has an excess of $J-K$ selected
EROs with $J-K>2.5$.
The majority of these EROs are therefore probably unrelated to the quasars.

2. Wide-field $J$ and \ks\ data show that the galaxy excess around \oate\ 
extends to 140\arcsec\ at $2.35\sigma$ significance, with a richness of
$N_{0.5}$=27$\pm$11 (corresponding to Abell richness 2$\pm$1).
The galaxy excess around \eleven\ extends to only 50\arcsec, suggesting 
that this quasar is embedded in a small-scale overdensity even though 
the overall counts in the field are higher than the literature average,

3. In three fields totaling 10.156~arcmin$^2$ we present the deepest 
narrow-band redshifted H$\alpha$ observations published to date.
We detect five candidate galaxies at the quasar redshifts, a surface density
2.5 times higher than in the only existing random-field survey of similar depth.
However, photometric SED fitting of one candidate suggests it is background to
the quasar, and that \OIII\,$\lambda$5007 instead of H$\alpha$ was detected in
the narrow-band imaging.

4. Submillimeter observations of three fields with SCUBA detect two of the 
quasars and two galaxies whose SEDs are best fit as
highly reddened galaxies (\ebv~$\simeq0.55$) at the quasar redshifts.
While many galaxies whose SEDs indicate the presence of considerable dust are
not detected, the SCUBA limits are only sufficient to rule out the hypothesis 
that these galaxies are hyperluminous infrared galaxies.

5.  $H$-band adaptive optics imaging is used to estimate structural redshifts
for two moderately red bulge-dominated galaxies in the \oate\ field using 
the Kormendy relation between central surface brightness and half-light radius.
Both objects have structural redshifts consistent with ellipticals 
foreground to the quasar at $z\lesssim0.2$ or $1 \lesssim z \lesssim 1.35$.
We have calculated photometric redshifts for these two objects for comparison
to the structural redshifts.  One object is not well fit by any of the simple 
models we consider, whereas the other is only fit by a moderately reddened
and young galaxy at $z_{ph}=4.50\pm0.15$.  It seems likely that our optical 
photometry for these objects is corrupted by scattered light from the nearby 
bright star used for adaptive optics, resulting in inaccurate SED fits.
Optical adaptive optics or {\em HST} imaging may be necessary 
to obtain accurate photometry for future study of these objects.

6. Quantitative SED fits and the resultant photometric redshifts are presented
and discussed for numerous galaxies from specific populations of interest.
Thirteen of eighteen objects in the spatially compact group around \oate\ are
consistent with being at the quasar redshift at $\gtrsim$90\% confidence.
One of the thirteen is a candidate for a very old galaxy with no ongoing star
formation; the others appear to have some ongoing star formation, as indicated
by a best-fit constant star formation model or a young best-fit age.
Fits to four $J-K$ selected EROs in the \eleven\ field all require large 
reddenings of $E(B-V)$~$\simeq0.6\pm0.3$.  One is consistent with the quasar 
redshift at 90\% confidence, while the remaining three have best-fit photometric
redshifts of $1.9<z_{ph}<3$.  These objects may be indicative of a population
of dusty, star-forming high-redshift 
galaxies potentially underrepresented in optically selected samples.
Fits to 21 very or extremely red objects show that many of them have 
photometric redshifts consistent with the quasar redshifts.
Significant dust is required to fit most of these objects, including
about half of those which also require relatively old stellar populations.

Overall, our observations support the hypothesis that 
radio-loud quasars at $z>1$ can be found in galaxy overdensities.
Among all but the very reddest galaxies in these overdensities, ongoing or
recent star formation with moderate amounts of dust seems to be common.
A similar result has been obtained by \cite{yam97} for the galaxy excess around
the RLQ 1335.8+2834 at $z=1.1$, and a variation in recent star formation 
histories has also been suggested for the large scale structures at $z\sim1.2$
in the field of 3C~324 \citep{kaj00a}.
These results suggest that the $z>1$ redshift range is at least approaching the 
redshift range in which the majority of early-type cluster galaxies are 
undergoing significant bursts of star formation.  
However, our finding that most very or extremely red objects require significant
dust reddening may conflict with {\em HST} imaging results that show only 
$\sim$20$-$50\% of such objects have disklike or irregular morphologies
suggestive of recent or ongoing star formation.  This raises the intriguing
possibility of a breakdown in the relationship between morphology and star 
formation properties at $z \gtrsim 1$.
Spectroscopy will be needed to confirm these speculations, of course, 
and is needed in any case to confirm or calibrate our 
SED fitting results.

\acknowledgements

We thank all our telescope operators and support astronomers,
D. Thompson and N. Drory for providing and discussing results prior to
publication, P. Smith for IRTF observing help, and 
T. Pickering, 
T. Webb, A. Barger, 
M. Dickinson, 
and C. Kulesa 
for assorted help.
Data were obtained (in part) using the 2.4m Hiltner Telescope of the MDM
Observatory.  TIFKAM was funded by the Ohio State University, the MDM
consortium, MIT, and NSF grant AST-9605012.  NOAO and USNO paid for the
development of the ALADDIN arrays and contributed the array in use in TIFKAM
during our observations.  
MS acknowledges support from NSF grant AST-9618686 and from the Natural
Sciences and Engineering Research Council (NSERC) of Canada.
RAF acknowledges support from a UA/NASA Spacegrant Fellowship, NSF
grant AST-9623788, and NASA GSRP training grant NGT5-50283.
HL acknowledges support provided by NASA through Hubble Fellowship grant
\#HF-01110.01-98A awarded by the Space Telescope Science Institute, which
is operated by the Association of Universities for Research in Astronomy,
Inc., for NASA under contract NAS 5-26555.  ASE was supported by RF9736D.

\footnotesize 

%
\begin{deluxetable}{ccccc}
\tabletypesize{\scriptsize}
\tablecaption{Observations\label{t_obs}}
\rotate
\tablewidth{570.00000pt}
\tablehead{
\colhead{Quasar}	& \colhead{Q~0835+580}	& \colhead{Q~1126+101}	& \colhead{Q~2149+212}	& \colhead{Q~2345+061}
}
\startdata
Right Ascension (J2000)	& 08:39:06.459		& 11:29:14.19		& 21:51:45.874		& 23:48:31.836 \\
Declination (J2000)	& +57:54:17.12		& +09:51:59.6		& +21:30:13.51		& +06:24:59.25 \\
Redshift		& 1.5358$\pm$0.0006	& 1.5173$\pm$0.0010	& 1.5385$\pm$0.0008	& 1.5396$\pm$0.0012 \\

\cutinhead{Wide-Field Observations}
Telescope+Instrument	& SO~2.3m+PISCES	& MDM~2.4m+TIFKAM	& \nodata		& \nodata \\
$J$ Date		& 1999 Jan 6		& 1999 Apr 6		& \nodata		& \nodata \\
$J$ Exposure		& 10560 		& 10560			& \nodata		& \nodata \\
$J$ $3\sigma$ Limit	& 22.51			& 22.37			& \nodata		& \nodata \\
$K_s$ Date	& 1998 Sep 30, 1999 Apr 27-28	& 1999 Apr 7		& \nodata		& \nodata \\
$K_s$ Exposure		& 8980			& 12360			& \nodata		& \nodata \\
$K_s$ $3\sigma$ Limit	& 20.63			& 21.03			& \nodata		& \nodata \\

\cutinhead{Narrow-band $H\alpha$ Observations}
Telescope+Instrument	& IRTF+NSFCAM		& \nodata		& CFHT+REDEYE		& IRTF+NSFCAM ; CFHT+REDEYE \\
Date		& 1997 Mar 19, Oct 26-30, Nov 1	& \nodata		& 1993 Sep 9		& 1997 Oct 26-30, Nov 1 ; 1993 Sep 9 \\
Filter $\lambda$ 	& 1.6549--1.6734\micron\ & \nodata		& 1.6631--1.6693\micron\	& 1.6574--1.6759\micron\ ; 1.6631--1.6693\micron\ \\
$H$-band Exposure 	& 7920 			& \nodata		& 810 			& 5940 ; 570 \\
$H\alpha$ Exposure 	& 34620			& \nodata		& 4200 			& 37440 ; 7200 \\
Area 			& 0.926			& \nodata		& 3.982			& 0.697 ; 4.551 \\
f$_{H\alpha}$ $3\sigma$ limit\tablenotemark{a}	& 1.71	& \nodata	& 0.78			& 1.92 ; 0.87 \\


\cutinhead{JCMT + SCUBA Observations}
Date			& 1999 Mar 26		& 1999 Mar 25-26	& \nodata		& 1998 July 23 \\
Exposure		& 7040 			& 15360			& \nodata		& 10240 \\
$\tau_{CSO}$		& 0.02--0.08		& 0.02--0.07		& \nodata		& 0.055 \\
RMS Noise @ 850\micron\	& 2.21 mJy/beam		& 1.83 mJy/beam		& \nodata		& 4.2 mJy/beam \\ 

\cutinhead{CFHT $Pue'o$ + KIR $H$-band Adaptive Optics Observations}
Date			& 1999 Dec 30-31	& \nodata		& \nodata		& \nodata \\
Exposure		& 10350			& \nodata		& \nodata		& \nodata \\

\enddata
\tablenotetext{a}{Units of 10$^{-17}$~ergs~cm$^{-2}$~s$^{-1}$.}
\tablecomments{
Redshifts are from the detailed study of Tytler \& Fan (1992), 
except for Q~1126+101 which was computed using their method.
All observation dates are UT.
All exposure times are in seconds.
All areas are in arcmin$^2$.
}
\end{deluxetable}

\clearpage

\begin{deluxetable}{ccccccrcrr}
\tablecaption{Candidate H$\alpha$ Emitters\label{t_ha_objs}}
\rotate
\tablewidth{625.93623pt}
\tablehead{
\colhead{ID} &
\colhead{$H$} &
\colhead{($H$$-$$H\alpha$)$_{AB}$} &
\colhead{$H\alpha$ REW\tablenotemark{a}} &
\colhead{f$_{H\alpha}$\tablenotemark{b}} &
\colhead{SFR\tablenotemark{c}} &
\colhead{$z_{H\alpha}$\tablenotemark{d}} &
\colhead{ID\#\tablenotemark{e}} &
\colhead{$\Delta$$\alpha$} &
\colhead{$\Delta$$\delta$} }
\startdata
Q~0835+580~(H$\alpha$1) & 20.13$\pm$0.17 & 1.56$\pm$0.25 &   316$\pm$79   & 13.9$\pm$1.3 & 		14.7$\pm$1.4 & 1.5358$\pm$0.0141 & 398 & 4.12 & 22.31\\
Q~2149+212~(H$\alpha$1) & 20.21$\pm$0.18 & 1.17$\pm$0.25 &  50.2$\pm$18.9 & 1.5$\pm$0.2 & 		1.6$\pm$0.2 & 1.5389$\pm$0.0047 & 171 & 12.48 & --21.66\\
Q~2345+061~(H$\alpha$1) & 18.27$\pm$0.01 & 0.54$\pm$0.07 &  51.6$\pm$3.8  & 4.6$\pm$0.2 & 		4.9$\pm$0.2 & 1.5396$\pm$0.0141\tablenotemark{f} & 195 & --13.86 & 28.06\\
Q~2345+061~(H$\alpha$2) & $>$20.63 ($2\sigma$) & 1.78$^{+\infty}_{-0.53}$ &  114$^{+\infty}_{-45}$  & 	1.7$^{+0.6}_{-0.4}$ & 1.8$\pm$0.5 & 1.5389$\pm$0.0047 & 133 & 8.96 & --1.08\\
Q~2345+061~(H$\alpha$3) & $>$20.63 ($2\sigma$) & 1.81$^{+\infty}_{-0.52}$ &  118$^{+\infty}_{-44}$  & 1.8$^{+0.7}_{-0.4}$ & 1.9$^{+0.7}_{-0.4}$ & 1.5389$\pm$0.0047 & \nodata & 22.79 & 21.47\\
\enddata
\tablenotetext{a}{Rest Equivalent Width in units of \AA,
calculated from Eq.~5 of Bunker {\it et~al.} (1995).  
}
\tablenotetext{b}{Units of 10$^{-17}$~ergs~cm$^{-2}$~s$^{-1}$,
calculated from Eq.~6 of Bunker {\it et~al.} (1995) 
using the appropriate zeropoint.}
\tablenotetext{c}{Star Formation Rate in units of $M_{\odot}~{\rm yr}^{-1}$
for $H_{\rm 0}$=75 km~s$^{-1}$~Mpc$^{-1}$, $q_{\rm 0}$=0.1,
derived from the relation of Kennicutt (1983).  
}
\tablenotetext{d}{Possible redshift range determined from the 
bandpass of the filter in which the excess was detected.}
\tablenotetext{e}{ID number in our catalog of the field
(available on request from the first author; cf. Table~4.)}
\tablenotetext{f}{Excluding the range 1.5389$\pm$0.0047 (see text).}
\tablecomments{$\Delta$$\alpha$ and $\Delta$$\delta$ are offsets in arcseconds 
from the quasar in that field, with positive offsets to the North and East.}
\end{deluxetable}

\clearpage

\begin{deluxetable}{lrrrrrcl}
\tablecaption{Probable and Possible Sub-mm Detections\label{t_scuba_objs}}
\rotate
\tablewidth{482.22331pt}
\tablehead{
\colhead{Source} &
\colhead{$\Delta$$\alpha$} &
\colhead{$\Delta$$\delta$} &
\colhead{S/N} &
\colhead{$S_{850}$, mJy} &
\colhead{CC} &
\colhead{Catalog ID} &
\colhead{Notes}}
\startdata
Q~2345+061& 		0&	0& 	2.8& 	11.7&	\nodata& 157&	\nodata \\ 
Q~0835+580~(SM1)&	8.5&	--44.8&	4.6&	10.1&	0.37&	117&	tentative \\
Q~0835+580~(SM2)&	14.7&	36.1&	3.4&	7.5&	0.29&	458&	tentative \\
Q~1126+101&		--2.0&  --0.7& 	3.4&	6.2&	0.49&	350&	\nodata \\ 
Q~1126+101~(SM1)&	--14.7&	14.7&	3.0&	5.5& 	0.46& 	384&    \nodata \\ 
Q~1126+101~(SM2)&	--45.7&	--13.7& 4.4&	8.0&	0.71&	\nodata& \nodata \\ 
Q~1126+101~(SM3)&	--47.4&	--70.6& 6.7&	12.6&	0.69&	\nodata&tentative; edge \\ 
\enddata
\tablecomments{
$\Delta$$\alpha$ and $\Delta$$\delta$ are offsets in arcseconds relative to
the optical position of the quasar.
CC is the correlation coefficient for each source.
Catalog ID is the ID number of the possible optical/near-IR
counterpart(s) in our catalogs of these fields
(available on request from the first author; cf. Table~4).
}
\end{deluxetable}

\clearpage

\begin{deluxetable}{lrrcccccrcccl} 
\tabletypesize{\scriptsize}
\tablecaption{SED Fitting Results\label{t_photz}}
\rotate
\tablewidth{600.00000pt}
\tablehead{
\colhead{ID} &
\colhead{$\Delta$$\alpha$} &
\colhead{$\Delta$$\delta$} &
\colhead{$K$} &
\colhead{$r-K$} &
\colhead{$J-K$} &
\colhead{$z-J$} &
\colhead{Model} &
\colhead{$\chi_{\nu}^2$} &
\colhead{$z_{ph}$} &
\colhead{$E(B-V)$} &
\colhead{Age} &
\colhead{Notes} }
\startdata
\cutinhead{Q~0835+580 Field Objects}
65~ & ~~5.5 & -58.7 & 19.52$\pm$0.08 & 5.59$\pm$0.21 & 1.92$\pm$0.08 & 2.80$\pm$0.19 & Burst & ~1.23 & 1.70$_{-0.30}^{+0.40}$ & 0.28$_{-0.00}^{+0.62}$ & 0.36$_{-0.36}^{+0.36}$ & VRO \\
106 & -37.1 & -67.9 & 17.73$\pm$0.02 & 5.05$\pm$0.04 & 2.10$\pm$0.03 & 1.81$\pm$0.03 & Burst & ~5.83 & 4.50$_{-0.15}^{+0.15}$ & 0.23$_{-0.18}^{+0.25}$ & 0.055$_{-0.052}^{+0.126}$ & AO Target \\
112 & -49.6 & -67.2 & 15.32$\pm$0.02 & 3.64$\pm$0.02 & 1.46$\pm$0.02 & 1.60$\pm$0.01 & Const & 19.60 & 3.10$_{-0.05}^{+0.00}$ & 0.00$_{-0.00}^{+0.02}$ & 0.806$_{-0.166}^{+0.472}$ & AO Target \\
117 & ~11.4 & -46.1 & 19.83$\pm$0.13 & 3.91$\pm$0.11 & 1.16$\pm$0.10 & 2.16$\pm$0.10 & Burst & ~2.19 & 1.60$_{-0.15}^{+0.15}$ & 0.00$_{-0.00}^{+0.12}$ & 0.286$_{-0.125}^{+0.740}$ & Q~0835+580 (SM1) \\
248 & ~-8.0 & -14.9 & 20.88$\pm$0.20 & 5.08$\pm$0.33 & 1.75$\pm$0.17 & $>2.66$~~~~~~ & Const & ~1.33 & 1.55$_{-0.40}^{+0.45}$ & 0.28$_{-0.10}^{+0.62}$ & 19.50$_{-19.49}^{+0.50}$ & SCG \\
280 & ~63.8 & ~-8.6 & 18.96$\pm$0.07 & 5.48$\pm$0.18 & 1.85$\pm$0.06 & 3.07$\pm$0.23 & Burst & 17.00 & 1.85$_{-0.25}^{+0.20}$ & 0.46$_{-0.20}^{+0.19}$ & 0.64$_{-0.44}^{+0.64}$ & VRO \\
307 & -18.9 & ~~3.5 & 20.95$\pm$0.22 & 4.49$\pm$0.28 & 1.30$\pm$0.19 & 2.24$\pm$0.22 & Const & 11.50 & 1.15$_{-0.30}^{+0.35}$ & 0.20$_{-0.12}^{+0.22}$ & 19.50$_{-17.10}^{+0.50}$ & SCG \\
319 & ~16.0 & ~~5.0 & 19.27$\pm$0.07 & 3.30$\pm$0.07 & 1.55$\pm$0.08 & 1.68$\pm$0.07 & Burst & ~1.56 & 2.20$_{-0.35}^{+0.15}$ & 0.24$_{-0.12}^{+0.14}$ & 0.026$_{-0.019}^{+0.054}$ & SCG \\
320 & ~15.3 & ~~1.9 & 20.23$\pm$0.11 & 4.99$\pm$0.21 & 1.94$\pm$0.13 & 2.65$\pm$0.29 & Const & ~2.63 & 1.75$_{-0.30}^{+0.50}$ & 0.30$_{-0.08}^{+0.35}$ & 20.00$_{-19.71}^{+0.00}$ & SCG \\
329 & -19.8 & ~~7.6 & 19.98$\pm$0.11 & 3.98$\pm$0.14 & 1.54$\pm$0.13 & 1.81$\pm$0.12 & Const & ~1.49 & 1.45$_{-0.25}^{+0.40}$ & 0.12$_{-0.10}^{+0.16}$ & 4.75$_{-4.35}^{+15.25}$ & SCG \\
333 & ~14.9 & ~10.5 & 20.00$\pm$0.11 & 3.96$\pm$0.09 & 1.34$\pm$0.08 & 2.27$\pm$0.09 & Burst & ~4.11 & 1.95$_{-0.35}^{+0.20}$ & 0.00$_{-0.00}^{+0.55}$ & 0.29$_{-0.29}^{+0.11}$ & SCG \\
339 & ~-6.0 & ~~9.5 & 18.64$\pm$0.04 & $>7.03$~~~~~~ & 1.99$\pm$0.04 & 3.55$\pm$0.20 & Burst & ~5.57 & 1.65$_{-0.30}^{+0.20}$ & 0.01$_{-0.01}^{+0.19}$ & 19.75$_{-13.00}^{+0.25}$ & SCG, ERO \\
340 & ~-5.7 & ~11.8 & 19.91$\pm$0.08 & 5.69$\pm$0.25 & 1.98$\pm$0.08 & $>2.78$~~~~~~ & Burst & ~1.70 & 1.70$_{-0.20}^{+0.40}$ & 0.42$_{-0.28}^{+0.43}$ & 0.23$_{-0.23}^{+0.67}$ & SCG \\
343 & -12.2 & ~-0.1 & 19.06$\pm$0.06 & 4.95$\pm$0.10 & 1.59$\pm$0.05 & 2.93$\pm$0.12 & Const & ~2.92 & 1.60$_{-0.25}^{+0.25}$ & 0.32$_{-0.14}^{+0.12}$ & 2.75$_{-2.24}^{+17.25}$ & SCG, red $z-J$ \\
345 & ~~3.7 & ~11.0 & 21.06$\pm$0.30 & 2.79$\pm$0.21 & 0.80$\pm$0.22 & 2.05$\pm$0.14 & Burst & ~1.22 & 2.20$_{-2.20}^{+0.30}$ & 0.00$_{-0.00}^{+0.71}$ & 0.18$_{-0.18}^{+0.46}$ & SCG \\
347 & ~-3.8 & ~-9.9 & 19.53$\pm$0.07 & $>6.79$~~~~~~ & 1.86$\pm$0.06 & $>3.89$~~~~~~ & Burst & ~7.96 & 1.75$_{-0.15}^{+0.30}$ & 0.46$_{-0.22}^{+0.19}$ & 0.72$_{-0.52}^{+0.89}$ & SCG, ERO, radio hotspot \\
348 & -11.2 & ~-7.3 & 20.51$\pm$0.19 & 2.82$\pm$0.13 & 1.25$\pm$0.14 & 2.03$\pm$0.13 & Burst & ~2.02 & 2.20$_{-2.00}^{+0.30}$ & 0.01$_{-0.01}^{+0.37}$ & 0.09$_{-0.09}^{+0.07}$ & SCG \\
349 & ~-1.2 & ~~9.6 & 19.18$\pm$0.06 & 5.73$\pm$0.18 & 1.69$\pm$0.05 & 2.97$\pm$0.14 & Burst & ~2.94 & 1.65$_{-0.20}^{+0.30}$ & 0.32$_{-0.24}^{+0.18}$ & 0.29$_{-0.21}^{+0.73}$ & SCG, VRO \\
352 & ~~9.3 & ~~0.3 & 18.00$\pm$0.02 & 3.25$\pm$0.02 & 1.32$\pm$0.02 & 1.83$\pm$0.02 & Burst & ~5.40 & 0.35$_{-0.05}^{+0.20}$ & 0.16$_{-0.06}^{+0.06}$ & 0.11$_{-0.10}^{+0.05}$ & SCG \\
353 & ~~9.7 & ~~3.2 & 20.85$\pm$0.15 & 3.03$\pm$0.11 & 1.45$\pm$0.13 & 1.59$\pm$0.10 & Const & ~0.02 & 2.30$_{-2.20}^{+0.30}$ & 0.26$_{-0.26}^{+0.29}$ & 0.064$_{-0.064}^{+19.936}$ & SCG \\
354 & ~~3.1 & ~~6.3 & 18.90$\pm$0.05 & 3.28$\pm$0.04 & 1.27$\pm$0.04 & 1.67$\pm$0.03 & Burst & ~0.67 & 1.82$_{-1.72}^{+0.43}$ & 0.60$_{-0.38}^{+0.05}$ & 0.001$_{-0.001}^{+0.160}$ & SCG \\
355 & ~-7.8 & ~-5.5 & 18.74$\pm$0.03 & 5.26$\pm$0.07 & 1.71$\pm$0.03 & 2.74$\pm$0.07 & Burst & ~1.05 & 1.70$_{-0.15}^{+0.20}$ & 0.34$_{-0.19}^{+0.16}$ & 0.20$_{-0.13}^{+0.37}$ & SCG \\
356 & ~-3.9 & ~-5.1 & 19.37$\pm$0.04 & 5.80$\pm$0.13 & 1.88$\pm$0.04 & 2.73$\pm$0.09 & Const & ~2.23 & 1.45$_{-0.20}^{+0.25}$ & 0.32$_{-0.04}^{+0.18}$ & 19.00$_{-17.57}^{+1.00}$ & SCG, VRO \\
379 & ~62.0 & ~19.0 & 19.47$\pm$0.08 & 4.91$\pm$0.16 & 1.53$\pm$0.07 & 3.02$\pm$0.23 & Burst & ~1.03 & 1.85$_{-0.30}^{+0.25}$ & 0.12$_{-0.12}^{+0.22}$ & 0.45$_{-0.29}^{+0.35}$ & red $z-J$ \\
398 & ~~4.1 & ~22.3 & 19.20$\pm$0.08 & 3.20$\pm$0.06 & 1.38$\pm$0.06 & 1.70$\pm$0.05 & Burst & ~0.32 & 2.20$_{-1.95}^{+0.25}$ & 0.34$_{-0.34}^{+0.16}$ & 0.003$_{-0.000}^{+0.140}$ & Q~0835+580 (H$\alpha$1) \\
440 & ~41.1 & ~35.2 & 18.59$\pm$0.04 & 5.65$\pm$0.11 & 1.72$\pm$0.04 & 2.27$\pm$0.06 & Burst & 17.10 & 1.05$_{-0.10}^{+0.25}$ & 0.34$_{-0.26}^{+0.21}$ & 0.72$_{-0.72}^{+0.56}$ & VRO \\
458 & ~26.2 & ~38.9 & 17.72$\pm$0.02 & 5.35$\pm$0.04 & 2.03$\pm$0.02 & 2.24$\pm$0.03 & Burst & ~0.23 & 1.20$_{-0.10}^{+0.20}$ & 0.55$_{-0.05}^{+0.15}$ & 0.035$_{-0.026}^{+0.018}$ & Q~0835+580 (SM2) \\
461 & ~45.9 & ~39.8 & 18.60$\pm$0.06 & 5.63$\pm$0.15 & 2.01$\pm$0.06 & 2.65$\pm$0.14 & Const & ~4.55 & 1.45$_{-0.40}^{+0.35}$ & 0.36$_{-0.06}^{+0.29}$ & 19.50$_{-18.99}^{+0.50}$ & VRO \\
478 & ~72.0 & ~41.7 & 18.27$\pm$0.04 & 5.63$\pm$0.14 & 2.00$\pm$0.05 & 2.72$\pm$0.12 & Burst & ~4.27 & 1.65$_{-0.90}^{+0.35}$ & 0.80$_{-0.60}^{+0.50}$ & 0.004$_{-0.004}^{+0.720}$ & VRO \\
479 & ~69.4 & ~44.8 & 19.07$\pm$0.06 & 5.53$\pm$0.21 & 1.99$\pm$0.07 & 2.36$\pm$0.14 & Const & ~2.85 & 1.15$_{-0.35}^{+0.40}$ & 0.32$_{-0.08}^{+0.53}$ & 19.25$_{-19.21}^{+0.75}$ & VRO \\
483 & ~36.9 & ~43.8 & 18.18$\pm$0.04 & 5.88$\pm$0.12 & 1.82$\pm$0.03 & 2.61$\pm$0.07 & Burst & ~0.43 & 1.60$_{-0.15}^{+0.35}$ & 0.24$_{-0.10}^{+0.14}$ & 0.72$_{-0.46}^{+0.30}$ & VRO \\
\\ \\ \\ \\ \\ \\ \\ \\ \\ \\ \\ \\ 
\\ \\ \\ 
\cutinhead{Q~1126+101 Field Objects}
66~ & -19.9 & -63.7 & 19.50$\pm$0.13 & 6.09$\pm$0.31 & 1.45$\pm$0.10 & 3.53$\pm$0.21 & Burst & 18.40 & 1.40$_{-0.10}^{+0.25}$ & 0.03$_{-0.03}^{+0.05}$ & 17.25$_{-12.75}^{+2.75}$ & VRO \\
91~ & -10.3 & -57.5 & 19.49$\pm$0.12 & 5.70$\pm$0.24 & 1.84$\pm$0.11 & 2.74$\pm$0.17 & Burst & ~2.68 & 5.00$_{-0.35}^{+0.00}$ & 0.60$_{-0.36}^{+0.15}$ & 0.005$_{-0.005}^{+0.220}$ & VRO \\
269 & -35.7 & -10.6 & 20.46$\pm$0.24 & 4.38$\pm$0.32 & 0.88$\pm$0.25 & $>$3.694~~~~~ & Burst & ~9.98 & 1.90$_{-0.25}^{+0.35}$ & 0.20$_{-0.20}^{+0.30}$ & 0.72$_{-0.54}^{+0.55}$ & red $z-J$ \\
298 & -28.0 & ~-4.1 & 19.32$\pm$0.09 & 5.71$\pm$0.22 & 1.73$\pm$0.09 & 3.29$\pm$0.23 & Burst & ~0.21 & 1.90$_{-0.35}^{+0.30}$ & 0.22$_{-0.22}^{+0.48}$ & 0.45$_{-0.45}^{+0.98}$ & VRO \\
346 & ~~1.0 & ~12.4 & 19.34$\pm$0.08 & 6.15$\pm$0.24 & 1.69$\pm$0.07 & 3.10$\pm$0.14 & Burst & ~6.59 & 1.40$_{-0.10}^{+0.30}$ & 0.00$_{-0.00}^{+0.12}$ & 20.00$_{-15.50}^{+0.00}$ & VRO \\
370 & ~49.1 & ~17.4 & 18.64$\pm$0.05 & 5.96$\pm$0.16 & 2.31$\pm$0.07 & 2.54$\pm$0.13 & Const & ~0.03 & 1.40$_{-1.20}^{+0.65}$ & 0.44$_{-0.10}^{+1.26}$ & 15.75$_{-15.75}^{+4.25}$ & VRO \\
381 & -10.9 & ~15.5 & 19.36$\pm$0.07 & 5.65$\pm$0.17 & 2.69$\pm$0.14 & $>$3.36~~~~~~ & Const & 12.10 & 1.95$_{-0.35}^{+0.45}$ & 0.55$_{-0.11}^{+0.40}$ & 17.25$_{-17.07}^{+2.75}$ & VRO \\
384 & -14.8 & ~15.8 & 17.67$\pm$0.03 & 4.34$\pm$0.03 & 1.91$\pm$0.03 & 1.95$\pm$0.03 & Burst & ~2.78 & 1.50$_{-0.15}^{+0.30}$ & 0.50$_{-0.14}^{+0.20}$ & 0.010$_{-0.010}^{+0.027}$ & Q~1126+101 (SM1) \\
387 & -43.1 & ~19.3 & 19.52$\pm$0.08 & 5.57$\pm$0.21 & 2.18$\pm$0.12 & $>$3.46~~~~~~ & Const & ~5.54 & 1.85$_{-0.40}^{+0.35}$ & 0.44$_{-0.22}^{+0.41}$ & 2.75$_{-2.74}^{+17.25}$ & VRO \\
424 & -32.3 & ~30.3 & 19.83$\pm$0.10 & 5.54$\pm$0.22 & $>$3.68~~~~~~ & \nodata       & Const & ~5.88 & 3.00$_{-0.40}^{+0.95}$ & 0.46$_{-0.10}^{+0.24}$ & 19.75$_{-18.61}^{+0.25}$ & VRO \\
425 & -35.3 & ~28.0 & 18.93$\pm$0.05 & 6.18$\pm$0.22 & 3.40$\pm$0.20 & $>$2.72~~~~~~ & Const & 14.60 & 2.35$_{-0.25}^{+0.60}$ & 0.55$_{-0.07}^{+0.20}$ & 19.75$_{-17.45}^{+0.25}$ & VRO \\
447 & -26.1 & ~33.5 & 19.38$\pm$0.09 & $>6.95$~~~~~~ & 2.57$\pm$0.13 & 2.88$\pm$0.31 & Burst & ~4.93 & 1.90$_{-0.50}^{+0.40}$ & 0.75$_{-0.45}^{+0.55}$ & 0.14$_{-0.14}^{+1.14}$ & ERO \\
520 & -33.4 & ~55.8 & 19.30$\pm$0.08 & 6.22$\pm$0.30 & 2.18$\pm$0.10 & 2.72$\pm$0.19 & Burst & ~2.50 & 0.70$_{-0.50}^{+1.25}$ & 1.10$_{-0.45}^{+0.90}$ & 0.32$_{-0.32}^{+19.68}$ & VRO \\
539 & -56.0 & ~60.4 & 19.21$\pm$0.09 & $>7.04$~~~~~~ & 1.69$\pm$0.08 & 2.87$\pm$0.19 & Burst & ~6.21 & 1.40$_{-0.20}^{+0.35}$ & 0.00$_{-0.00}^{+0.12}$ & 20.00$_{-16.00}^{+0.00}$ & ERO \\
\enddata
\tablecomments{
Catalog ID is the ID number of the possible optical/near-IR
counterpart(s) in our catalogs of these fields
(available on request from the first author).
$\Delta$$\alpha$ and $\Delta$$\delta$ are offsets in arcseconds 
from the quasar, with positive offsets to the North and East.
Magnitudes are FOCAS total magnitudes; colors are FOCAS isophotal colors.
All lower limits to colors are $3\sigma$ limits.
The Model column gives the best-fit of the two star formation histories we
considered: an instantaneous burst (Burst) or a constant star formation rate
(Const), both with 20\% solar metallicity.
Ages are given in Gyr.
In the Notes column, 
ERO denotes $r-K\geq6.322$ and VRO denotes $5.4<r-K<6.322$ (\S7.3),
red $z-J$ denotes an object with $z-J>2.9$ but $r-K<6.322$ and $J-K<2.5$ 
(\S7.3),
and SCG denotes a member of the spatially compact group around Q~0835+580 
(\S7.1.1).
}
\end{deluxetable}

\clearpage

\begin{figure} 
\caption[]{
\singlespace 
CFHT REDEYE narrow-band H$\alpha$ images, each 128$''$$\times$128$''$.
North is up and east is left.  Bright objects produce elongated crosstalk
``images'' at the same position in the other three quadrants.
(a) Q~2345+061 field.  The candidate H$\alpha$ emitters are
marked, including one selected as a candidate only in IRTF data.
(b) Q~2149+212 field.  The quasar is the brightest object in the SE quadrant.
The candidate H$\alpha$ emitter is marked.
}\label{fig_ha_img}
\end{figure}

\begin{figure}
\epsscale{1.0}
\caption[]{
\singlespace 
\small 
$K$ image of the Q~0835+580 field (data from HGC98)
overlaid with 850\micron\ SCUBA contours 
(this paper, 7.04 ksec exposure, $1 \sigma$ RMS noise 2.21~mJy/beam).
The sub-mm and optical images have been aligned using the default JCMT pointing.
Light (green) contours are positive (2.5, 3.0, 3.5... $\sigma$)
and dark (red) contours are negative (-2.5, -3.0, -3.5... $\sigma$),
with the outermost dark (red) contour marking the edge of the full SCUBA maps.
The quasar is marked by Q, and sub-mm sources discussed in the text by SM\#.
Positive features marked by a letter (or not marked at all) do not correspond
to peaks in the cross-correlation map and so are not considered real.
}\label{fig_scuba_img1}
\end{figure}

\begin{figure}
\epsscale{1.0}
\caption[]{
\singlespace 
\small 
$K$ image of the Q~1126+101 field (data from HGC98)
overlaid with 850\micron\ SCUBA contours 
(this paper, 15.36 ksec exposure, $1 \sigma$ RMS noise 1.83~mJy/beam).
Feature A, although strong, was seen only on night 2 and is probably noise.
See Figure \ref{fig_scuba_img1} for further details.
}\label{fig_scuba_img2}
\end{figure}

\begin{figure} 
\epsscale{1.0}
\plotone{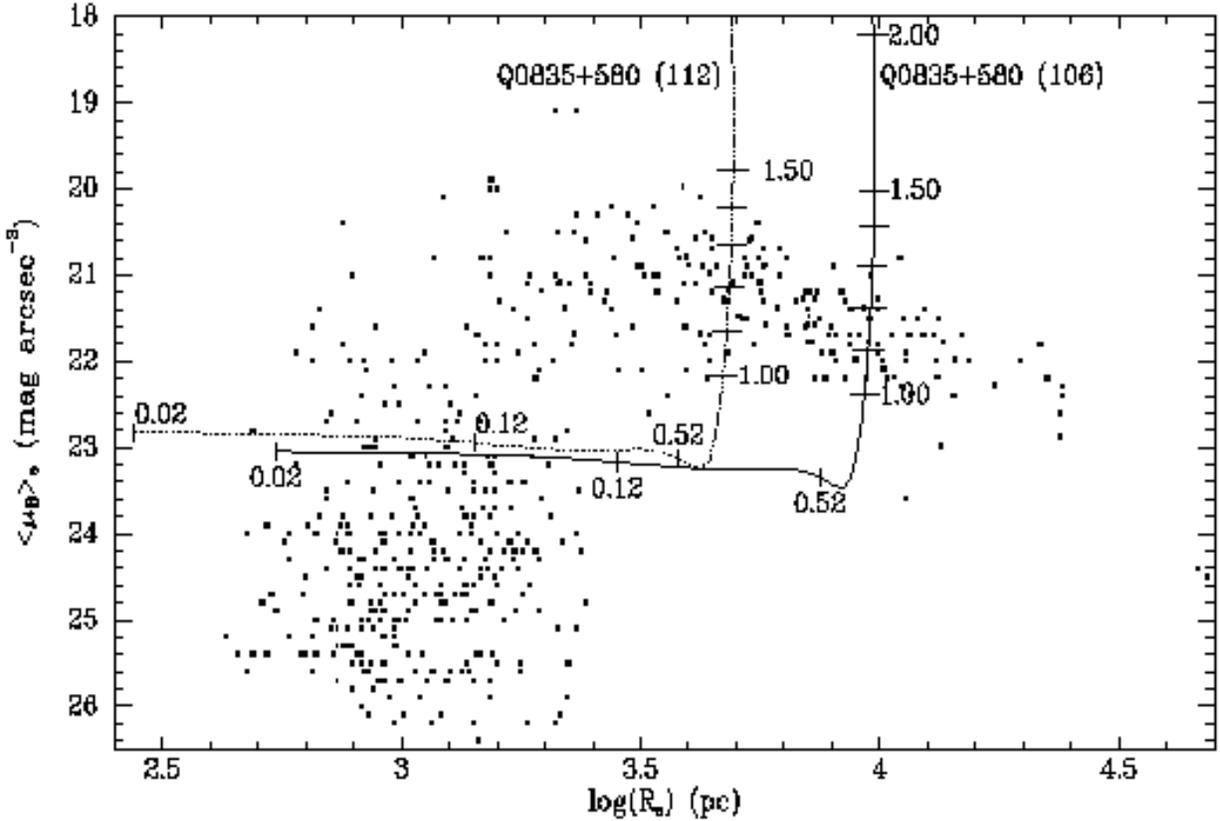}	
\caption[]{
\singlespace 
Redshift estimates for adaptive optics targets Q~0835+580 (106) (solid line) and
Q~0835+580 (112) (dotted line) using the average rest frame $B$-band surface
brightness-effective radius relation for giant elliptical galaxies.  
Labelled tick marks show the positions for various assumed redshifts,
including every $\Delta$z=0.1 from z=1 to z=1.5.
Points are local ellipticals from Sandage \& Perelmuter (1990).
Both curves assume $q_0$=0.225.  Assuming $q_0$=0.0 (0.5) would make
little difference at z$\leq$0.52 but would shift the vertical curves rightward
(leftward) by $\sim$0.1 in the log by z=2.
}\label{fig_remue}
\end{figure}

\begin{figure}
\epsscale{2.00}
\plottwo{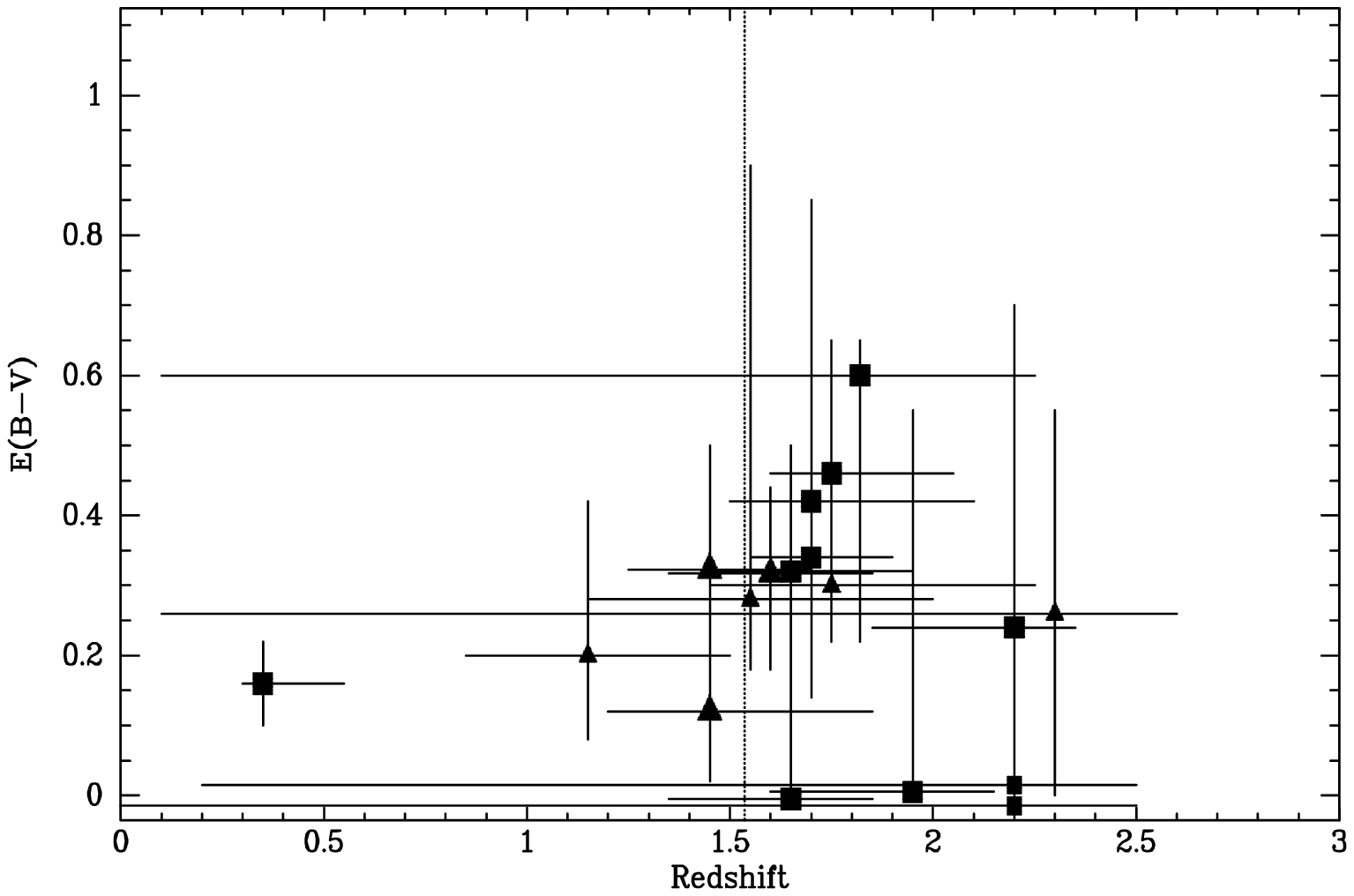}{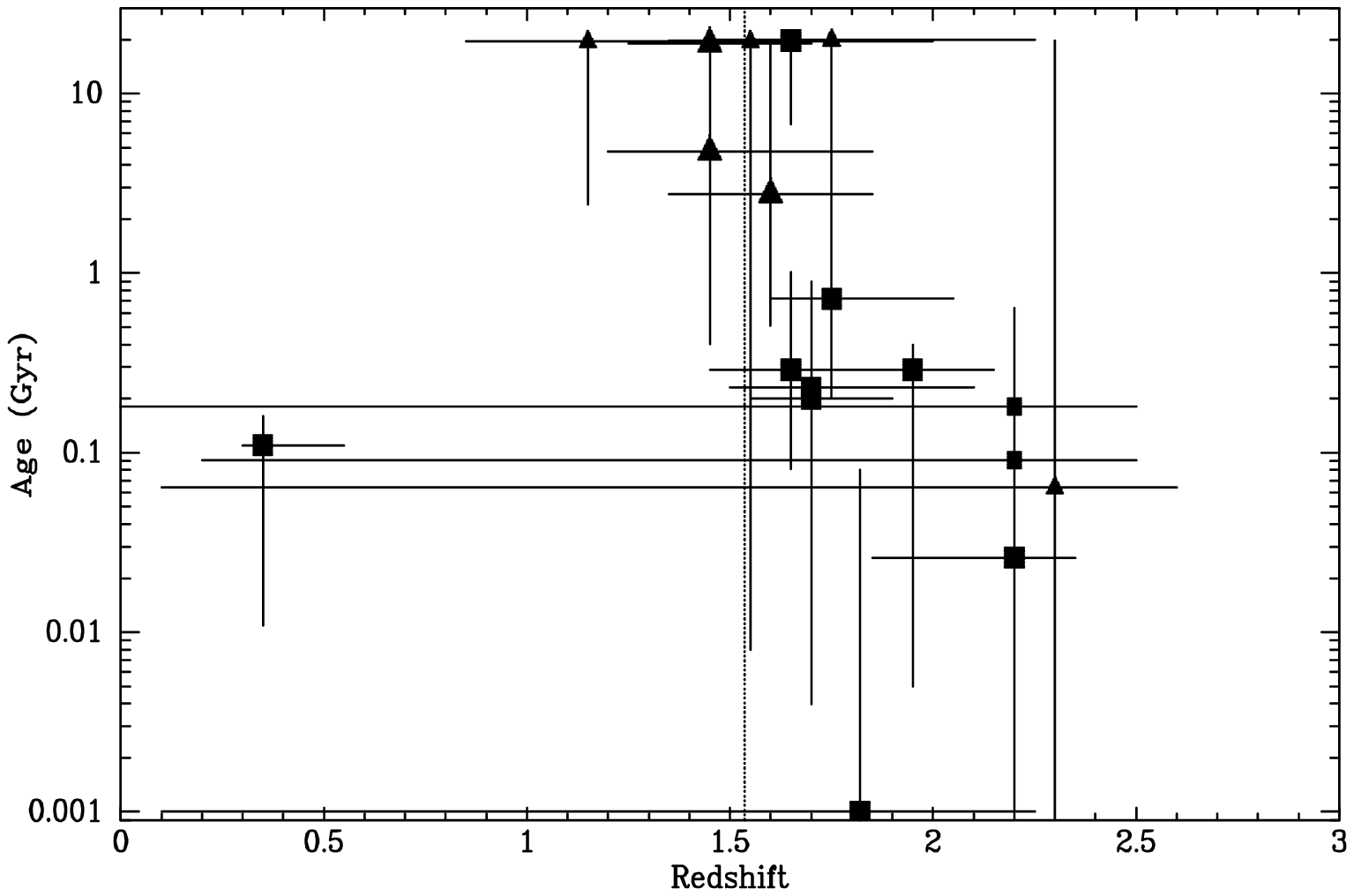}
\caption[]{
\singlespace 
(a) $E(B-V)$ vs. $z_{ph}$ for the best fit model of twelve objects with 
$K\leq20$ (large symbols) and six with $K>20$ (small symbols) within 21\farcs5 
of Q~0835+580.  Instantaneous bursts (shown as squares) are preferred over
constant star formation histories (triangles) 
in 9 of 12 cases at $K\leq20$ but only 2 of 6 cases at $K>20$.
The vertical dotted line shows the quasar redshift.
The error bars are projected 90\% confidence limits (see text).
Objects with $E(B-V)$=0 have been arbitrarily offset by small amounts in 
$E(B-V)$ to show the error bars.
(b) Age vs. $z_{ph}$ for the same objects.
}\label{fig_photz0835zebv}
\end{figure}

\begin{figure}
\epsscale{1.11}
\plotone{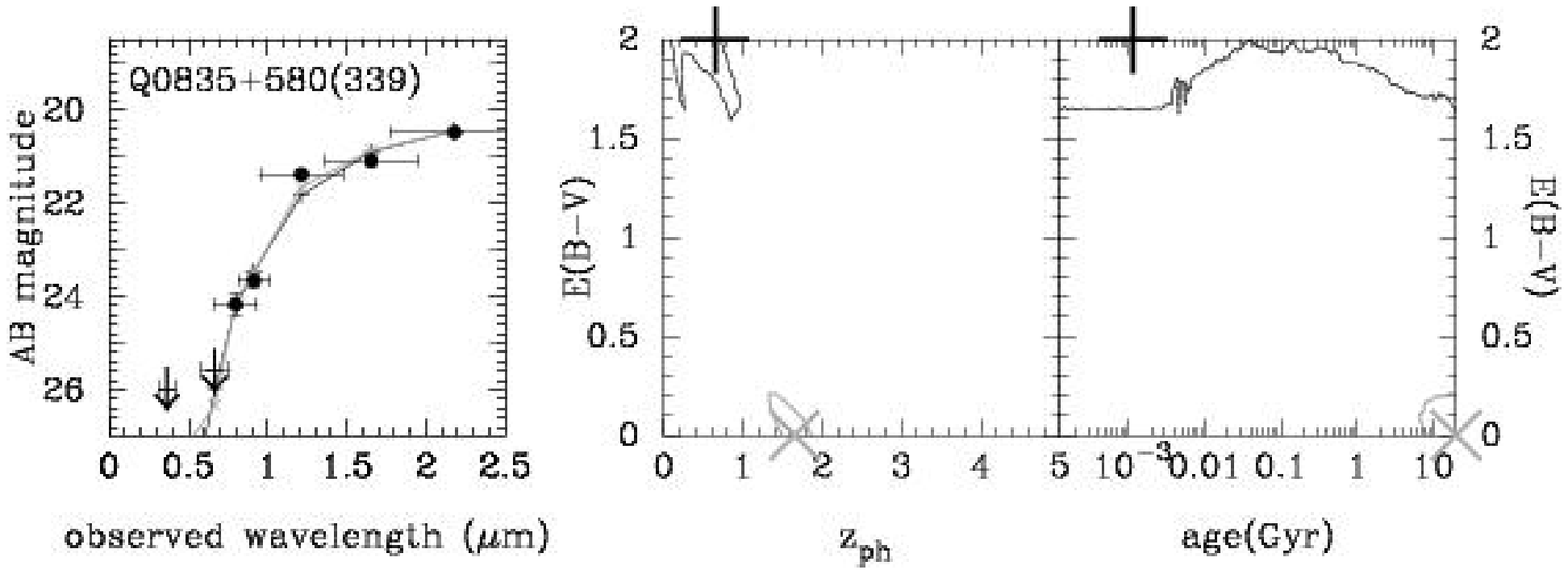}
\plotone{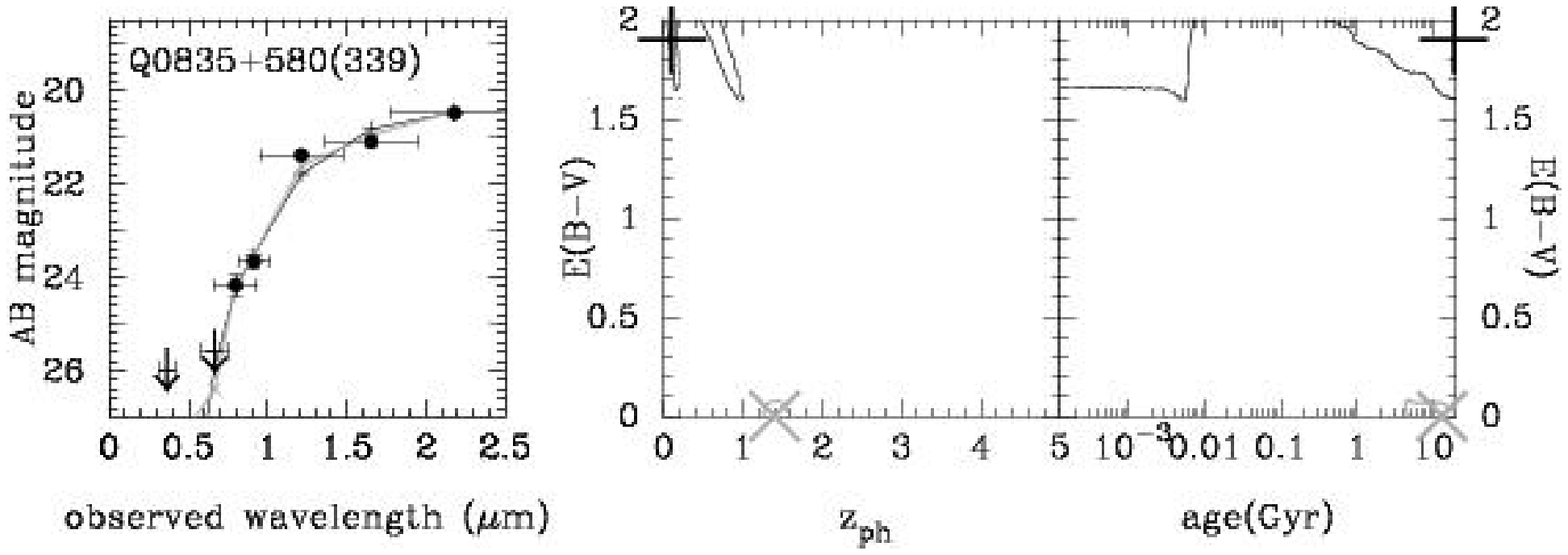}
\caption[]{
\singlespace 
SED fitting results for ERO Q~0835+580 (339).
(a) 20\% solar metallicity fit.
(b) solar metallicity fit.
In each plot, the left panel shows the observed flux points 
and the best-fit SED points connected by straight line segments.
In all panels, the two different star formation histories we modelled
are denoted by lines of different colors:  gray for an instantaneous burst
and black for a constant star formation rate.
The middle and right panels are two perpendicular projections 
of the three-dimensional space of $E(B-V)$, $z$ and age.
In each of these two panels, the cross and the plus sign mark the best-fit
values for instantaneous burst and constant star formation models, respectively,
while the contours show the projection of the 90\% confidence volume.
}\label{fig_photz0835.339}
\end{figure}

\begin{figure}
\epsscale{0.75}
\plotone{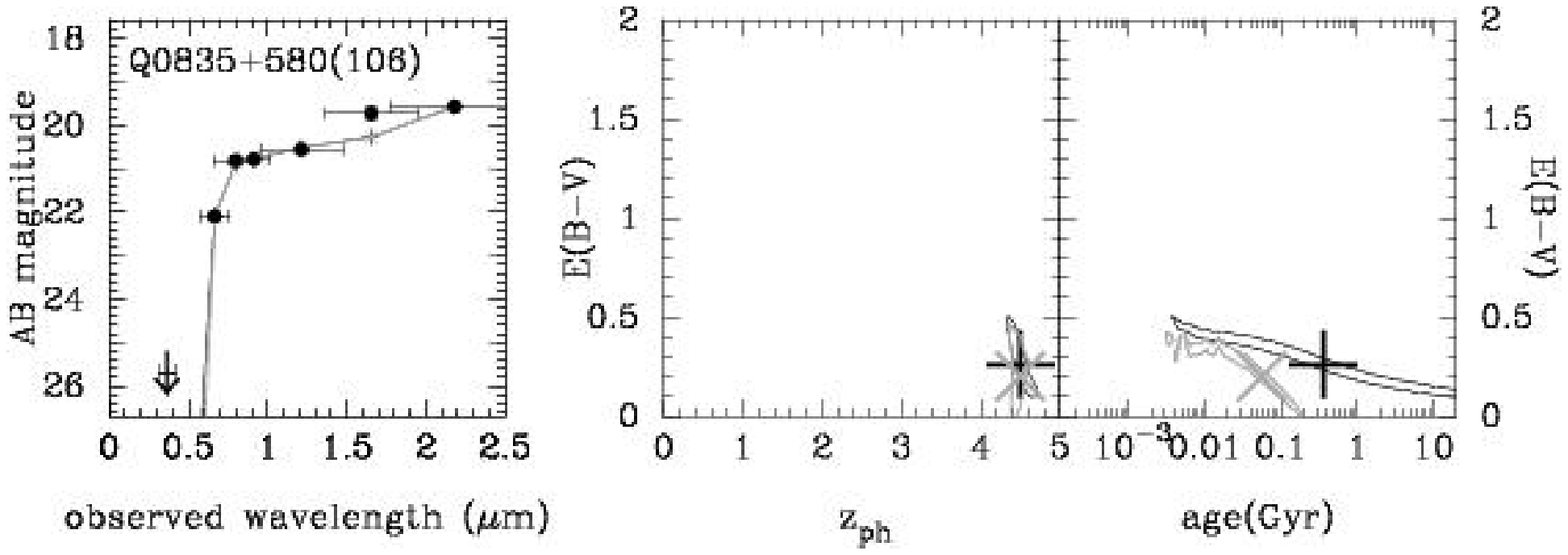}
\plotone{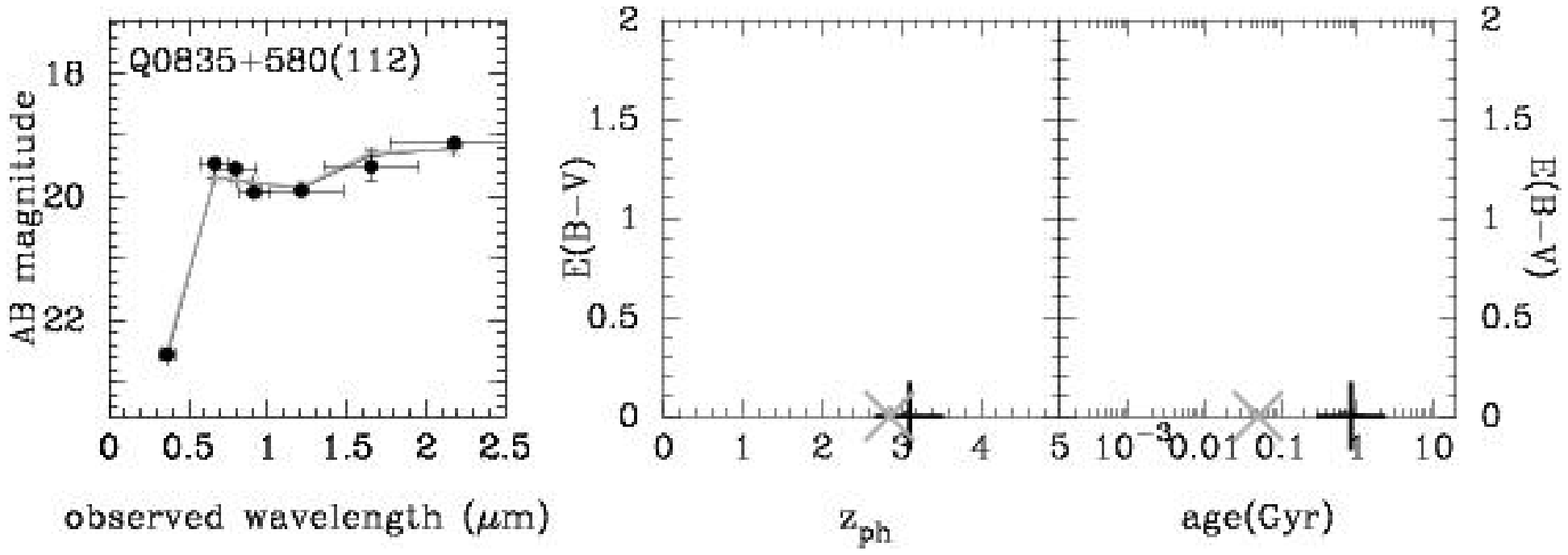}
\caption[]{
\singlespace 
SED fitting results for adaptive optics targets.
(a) Q~0835+580 (106).
(b) Q~0835+580 (112).
See Figure~\ref{fig_photz0835.339} for details.
}\label{fig_photz106.0835}
\end{figure}

\begin{figure}
\epsscale{0.75}
\plotone{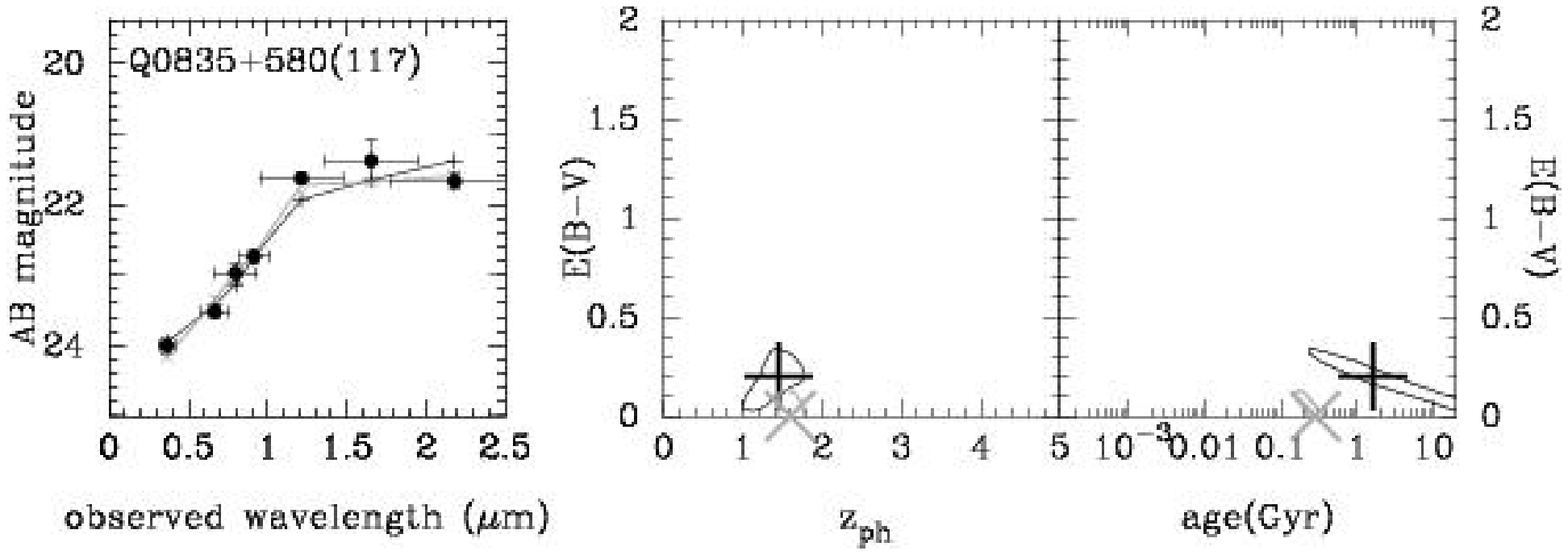}
\plotone{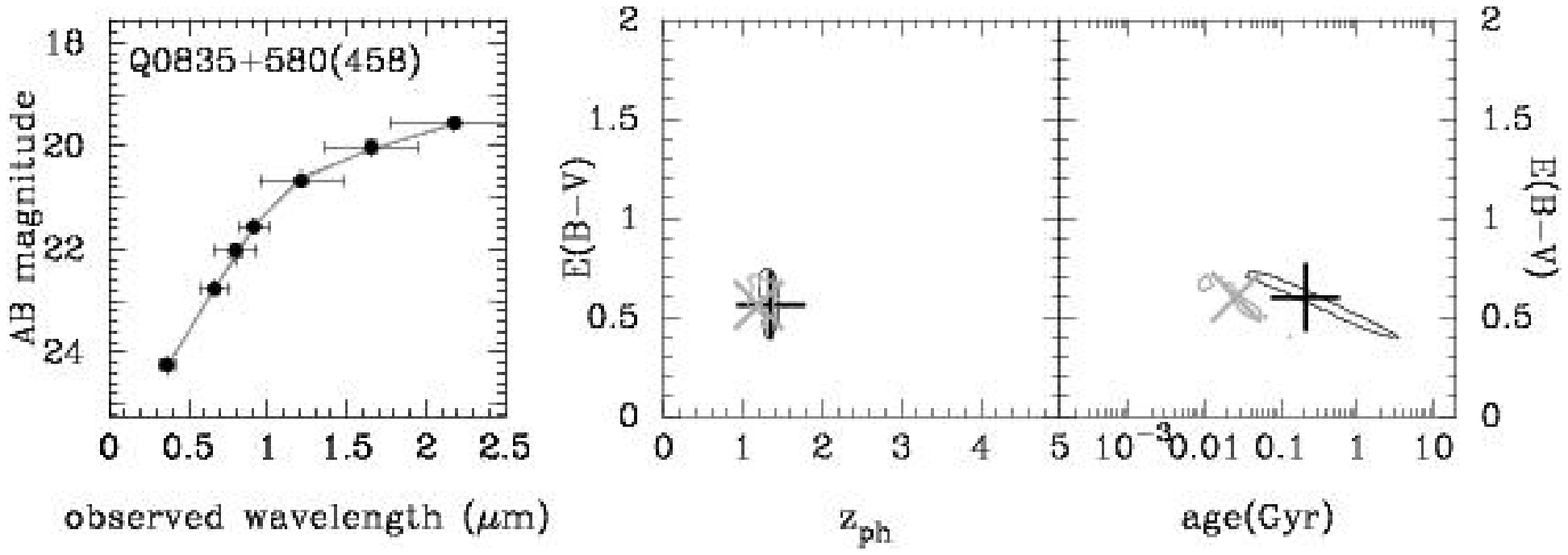}
\caption[]{
\singlespace 
SED fitting results for candidate sub-mm source 
counterparts in the Q~0835+580 field.
(a) Q~0835+580 (117).
(b) Q~0835+580 (458).
See Figure~\ref{fig_photz0835.339} for details.
}\label{fig_photz117.0835}
\end{figure}

\begin{figure}
\epsscale{1.00}
\plotone{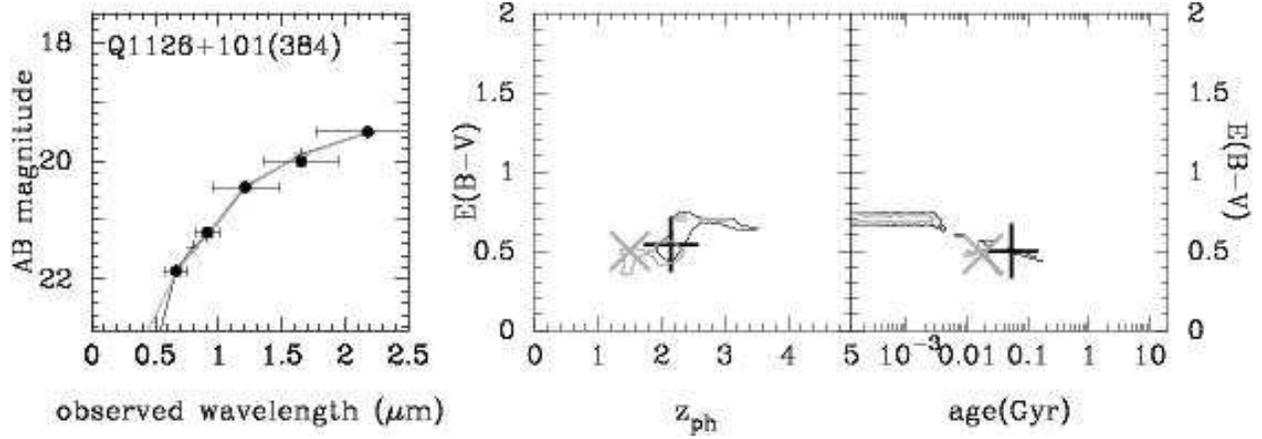}
\caption[]{
\singlespace 
SED fitting results for candidate sub-mm source
counterpart Q~1126+101 (384).
See Figure~\ref{fig_photz0835.339} for details.
}\label{fig_photz384.1126}
\end{figure}

\begin{figure}
\epsscale{1.00}
\plotone{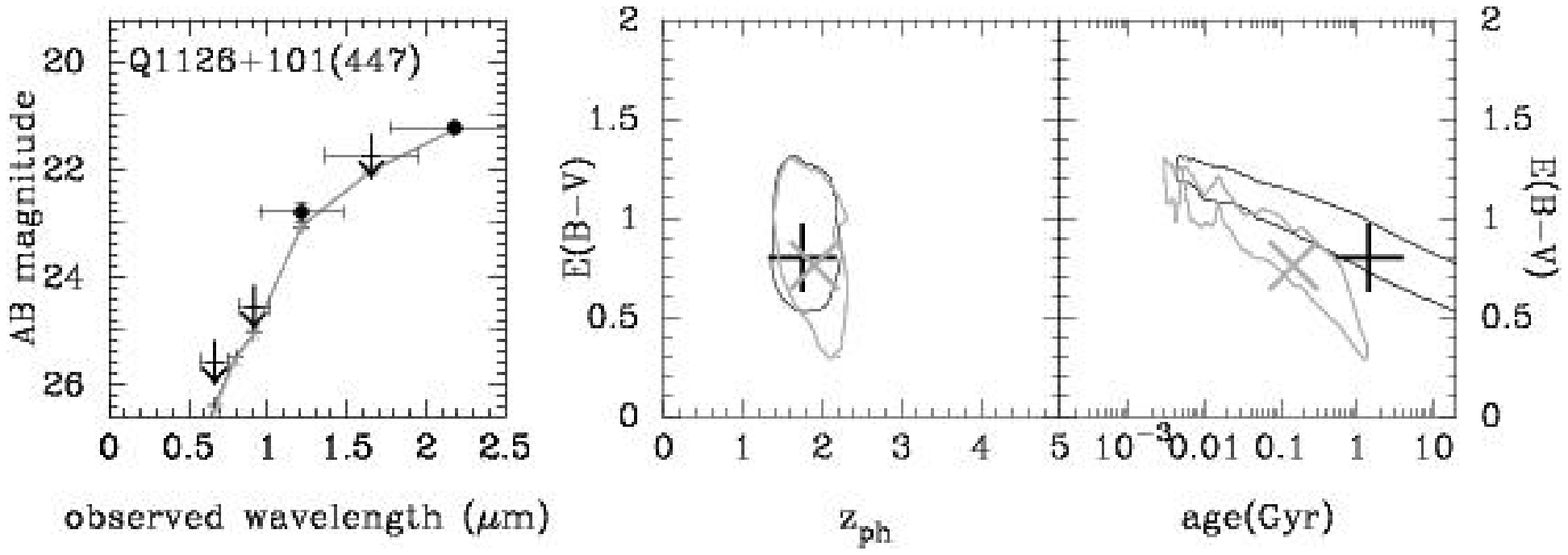}
\plotone{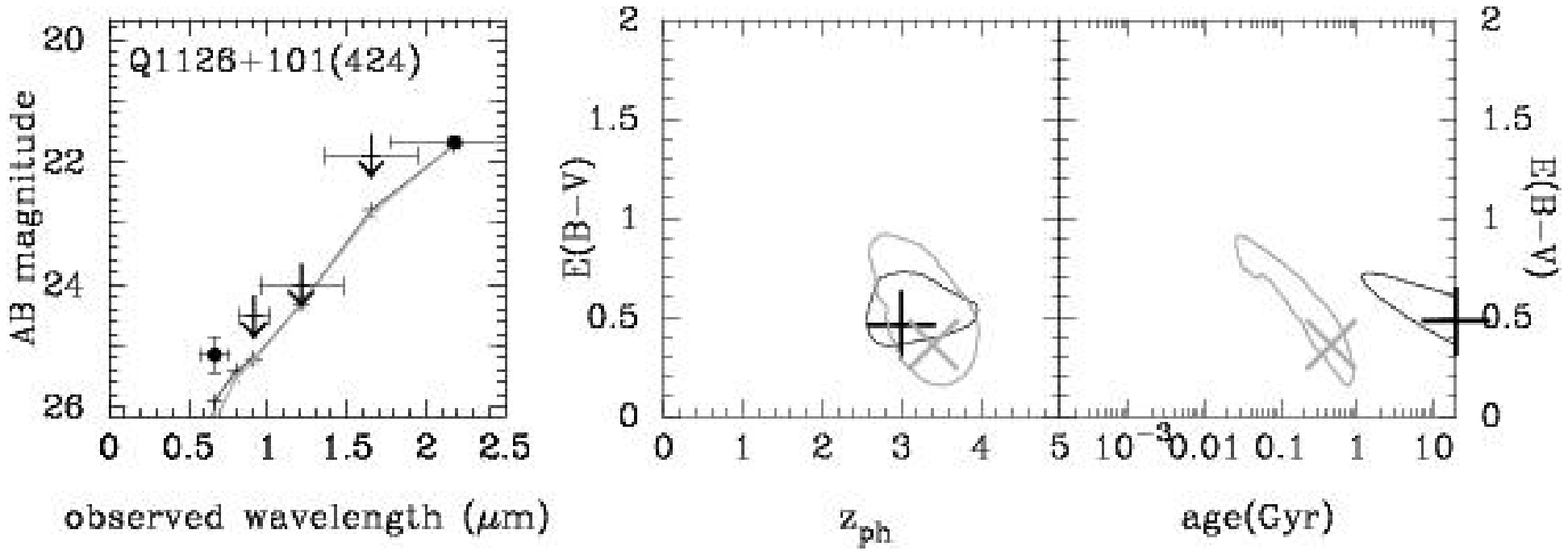}
\caption[]{
\singlespace 
SED fitting results for $J-K$ selected EROs.
(a) Q~1126+101 (447).
(b) Q~1126+101 (424).
See Figure~\ref{fig_photz0835.339} for details.
}\label{fig_photz447.1126}
\end{figure}

\begin{figure}
\epsscale{1.11}
\plotone{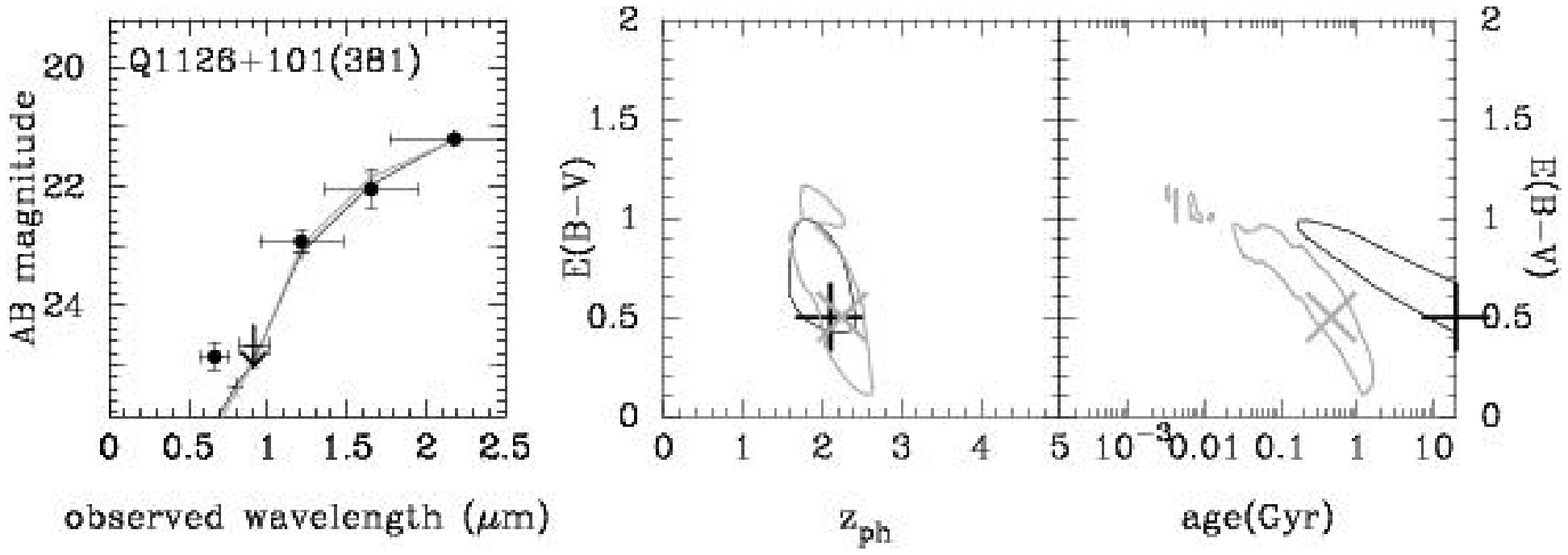}
\plotone{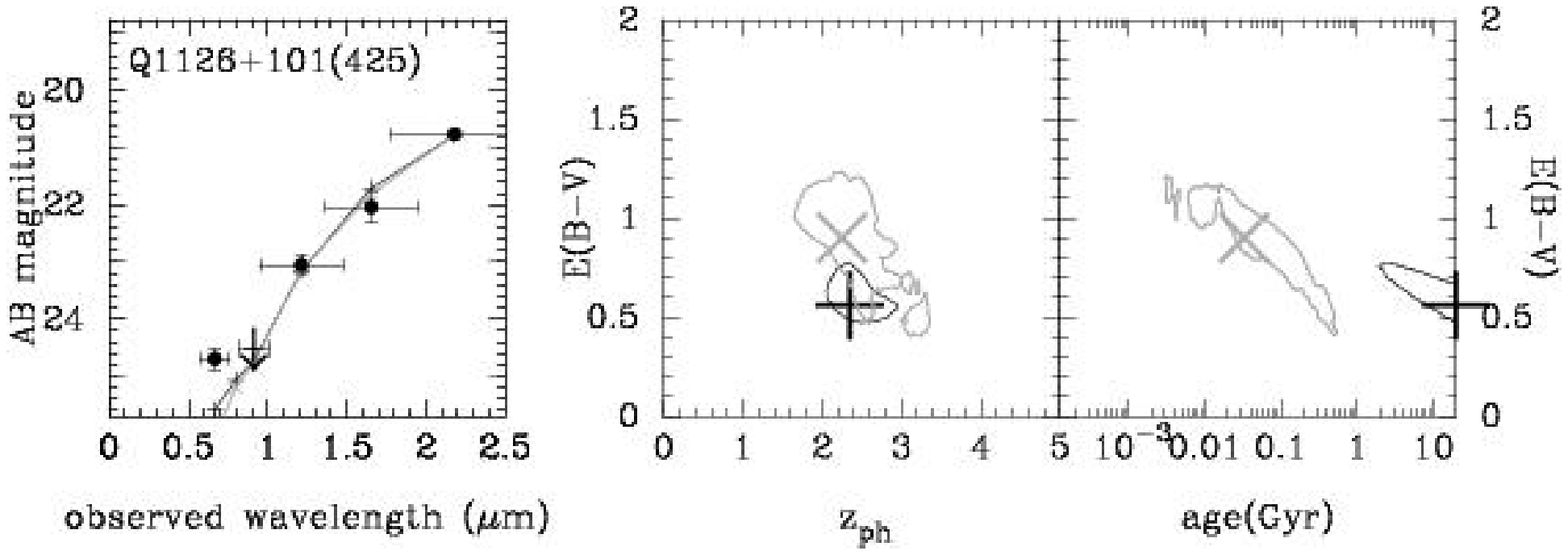}
\caption[]{
\singlespace 
SED fitting results for $J-K$ selected EROs.
(a) Q~1126+101 (381).
(b) Q~1126+101 (425).
See Figure~\ref{fig_photz0835.339} for details.
}\label{fig_photz425.1126}
\end{figure}

\clearpage

\begin{figure}
\epsscale{2.00}
\plottwo{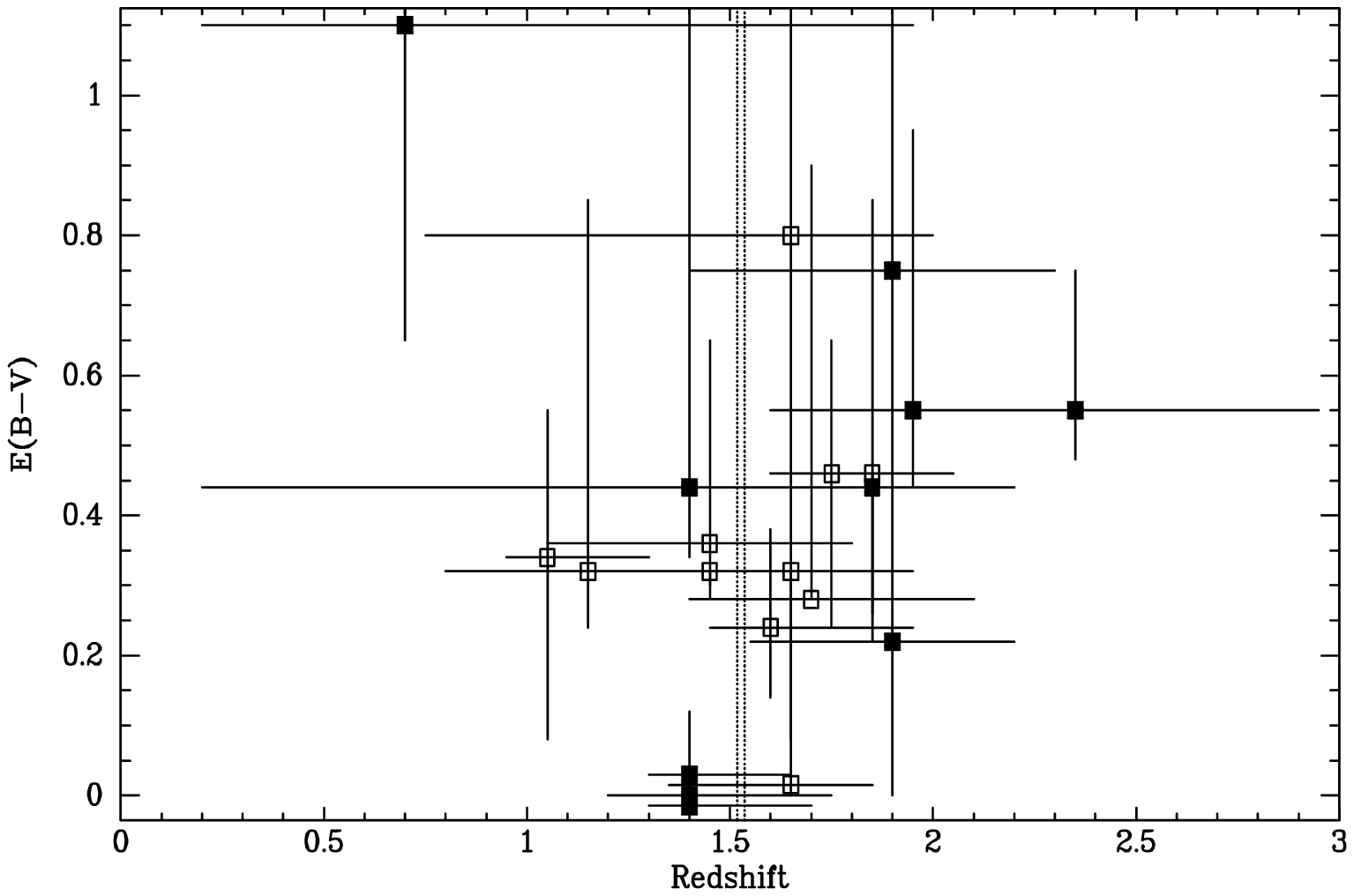}{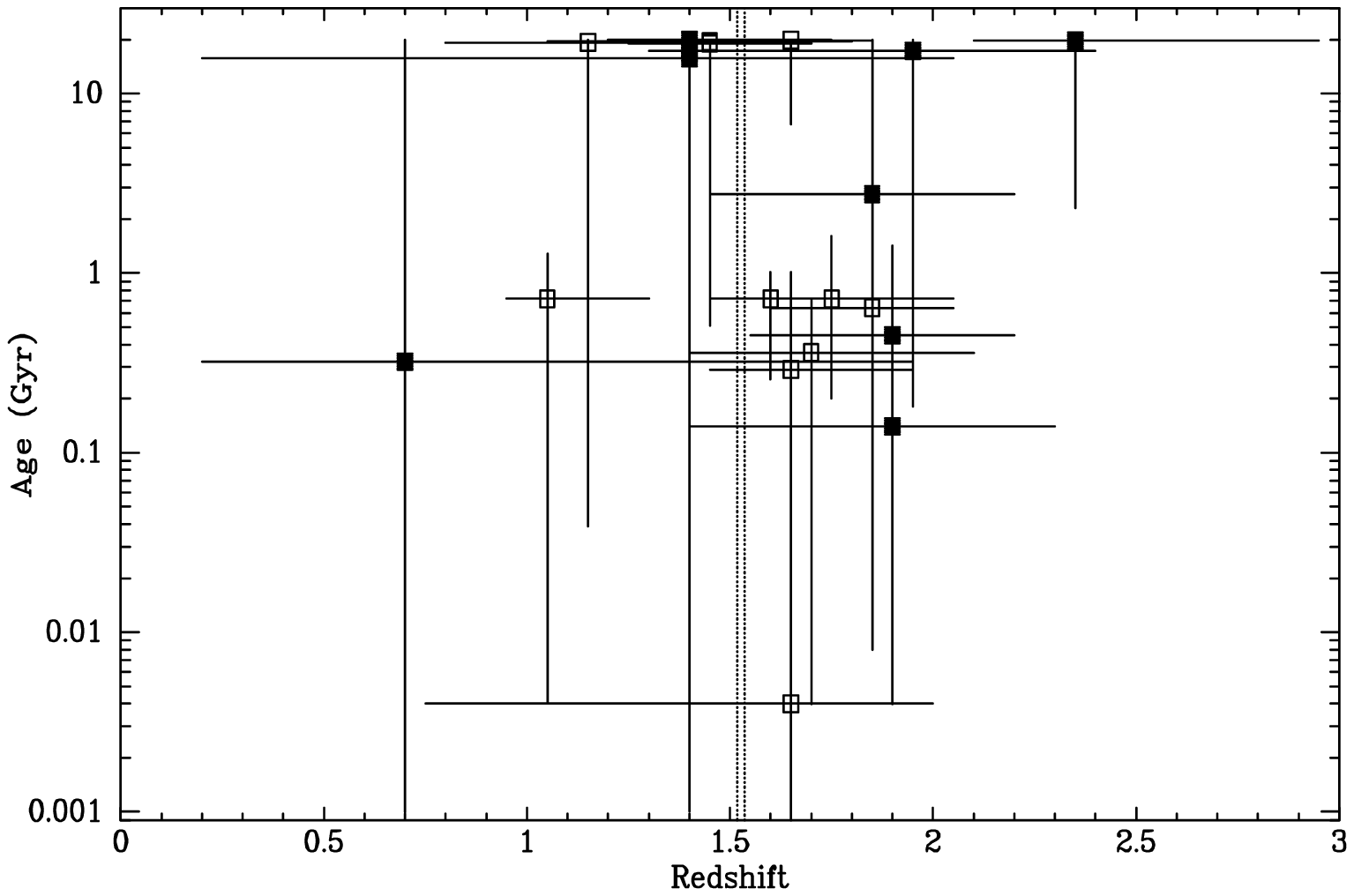}
\caption[]{
\singlespace 
(a) $E(B-V)$ vs. $z_{ph}$ for all successfully fit objects with $r-K>5.4$ and 
$K<19.6$ in the fields of Q~0835+580 (open squares) and Q~1126+101 
(filled squares).  The dotted vertical lines show the quasar redshifts.
The error bars are projected 90\% confidence limits (see text).
(b) Age vs. $z_{ph}$ for the same objects.
}\label{fig_photbothzebv}
\end{figure}

\begin{figure}
\epsscale{1.00}
\plotone{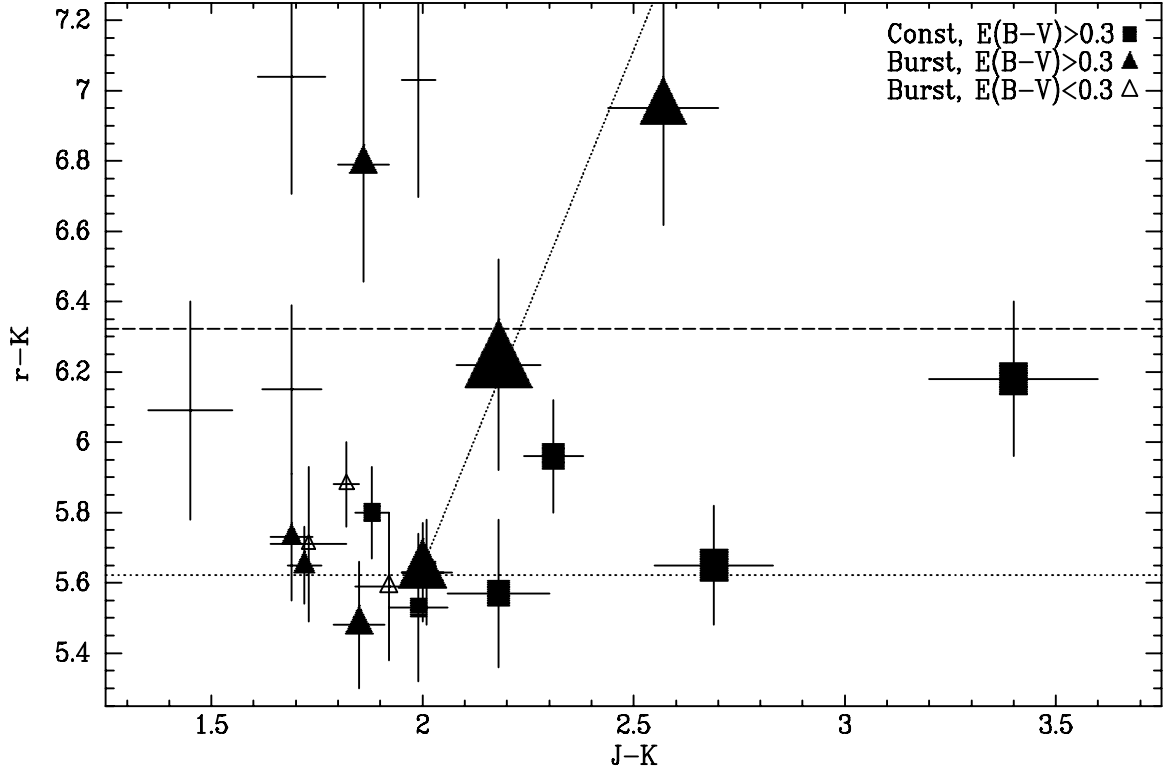}
\caption[]{
\singlespace 
$r-K$ vs. $J-K$ for all successfully fit objects with $r-K>5.4$ and $K<19.6$
in the fields of Q~0835+580 and Q~1126+101 (see Table 4).
Triangles denote objects for which instantaneous burst models are the best fits,
and squares objects for which constant SFR models are best.
Filled symbols denote $E(B-V)>0.3$, and open symbols best-fit $E(B-V)<0.3$.
Note that all constant SFR models have best-fit $E(B-V)>0.3$.
The symbol sizes scale linearly with $E(B-V)$, with the largest best-fit value
being 1.1.  Objects without symbols are all best fit as $E(B-V)=0$ bursts.
Error bars extending off the plot indicate $3\sigma$ lower limits on the $r-K$
or $J-K$ colors.  Objects above the dashed line are EROs by our definition
($R-K>6$, using $r=R+0.322$).  The dotted lines show the elliptical 
(upper left quadrant) and dusty starburst (upper right quadrant) regions 
defined by Pozzetti \& Mannucci (2000), again using $r=R+0.322$.  
Our SED fitting results indicate that their criteria are not as robust 
as might be hoped.
}\label{fig_photbothjkrkebv}
\end{figure}


\begin{thebibliography}{}


\bibitem[Barger \etal(1999)]{bar99} \reference{} 
Barger, A.~J., Cowie, L.~L., Smail, I., Ivison, R.~J., Blain, A.~W., 
and Kneib, J.-P.  1999, \aj, 117, 2656 

\bibitem[Barkhouse \& Hall(2001)]{bh00}
Barkhouse, W. A., and Hall, P. B. 2001, \aj, in press





\bibitem[Ben\'\i tez \etal(1999)]{ben99} \reference{}
Ben\'\i tez, N., Broadhurst, T., Bouwens, R., Silk, J. and Rosati, P.  1999,
\apjl, 515, L65



\bibitem[Bruzual \& Charlot(1993)]{bc93} \reference{} 
Bruzual A., G., and Charlot, S. 1993, \apj, 405, 538

\bibitem[Bruzual \& Charlot(1996)]{bc96} \reference{}    
Bruzual A., G., and Charlot, S.  1996, in ``A Data Base for Galaxy Evolution
Modeling,'' eds. C. Leitherer \etal, \pasp, 108, 996

\bibitem[Bunker \etal(1995)]{bun95} \reference{}
Bunker, A.~J., Warren, S.~J., Hewett, P.~C., and Clements, D.~L.  1995,
\mnras, 273, 513

\bibitem[Burbidge \etal(1990)]{bur90} \reference{}        
Burbidge, G., Hewitt, A., Narlikar, J.~V., and Das Gupta, P. 1990, \apjs, 74, 675

\bibitem[Calzetti(1997)]{cal97} \reference{}
Calzetti, D.  1997, in ``The Ultraviolet Universe at Low and High Redshift,"
eds. W.~H. Waller, M.~N. Fanelli, and A.~C. Danks (AIP: New York), 403

\bibitem[Casali \& Hawarden(1992)]{ch92} \reference{}
Casali, M., and Hawarden, T.  1992, JCMT-UKIRT Newsletter, 4, 33

\bibitem[Chapman, McCarthy \& Persson(2000)]{cmp00} \reference{}
Chapman, S.~C., McCarthy, P.., and Persson, S.~E.  2000, 
\aj, 120, 1612 

\bibitem[Cimatti \etal(1997)]{cim97} \reference{}       
Cimatti, A., Bianchi, S., Ferrara, A., and Giovanardi, C.  1997, \mnras, 290, L43

\bibitem[Cimatti \etal(1999)]{cea99} \reference{}       
Cimatti, A., Daddi, E., di Serego Alighieri, S., Pozzetti, L., Mannucci, F., 
Renzini, A., Oliva, E., Zamorani, G., Andreani, P. and R\"ottgering, H.\ J.\ A.
1999, \aap, 352, L45

\bibitem[Cimatti \etal(2000)]{cim00} \reference{}       
Cimatti, A., Villani, D., Pozzetti, L., and di Serego Alighieri, S.  2000,
\mnras, 318, 453 

%

\bibitem[Cohen \etal(1999)]{coh99a} \reference{}
Cohen, J.~G., Hogg, D.~W., Pahre, M.~A., Blandford, R., and Shopbell, P.~L. 
1999, \apjs, 120, 171

\bibitem[Cohen \etal(1999)]{coh99b} \reference{}
Cohen, J.~G., Blandford, R., Hogg, D.~W., Pahre, M.~A., and Shopbell, P.~L. 
1999, \apj, 512, 30

\bibitem[Coleman, Wu, \& Weedman(1980)]{cww80} \reference{} 
Coleman, G. D., Wu, C.-C., and Weedman, D. W. 1980, \apjs, 43, 393 


\bibitem[Cutri \etal(2000)]{cut00}
Cutri, R. M., Nelson, B. O., Kirkpatrick, J. D., Huchra, J. P. and Smith, P. S.
2000, to appear in The New Era of Wide-Field Astronomy, ed. R. Clowes
(San Francisco: ASP)

\bibitem[Daddi \etal(2000)]{dad00} \reference{} 
Daddi, E., Cimatti, A., Pozzetti, L., Hoekstra, H., Roettgering, H. J. A.,
Renzini, A., Zamorani, G., and Mannucci, F.  2000, \aap, 361, 535


\bibitem[De Propris \etal(1998)]{dep98} \reference{}
De Propris, R., Eisenhardt, P. R., Stanford, S. A., and Dickinson, M.  1999,
\apjl, 503, L45

\bibitem[De Propris \etal(1999)]{dep99} \reference{}
De Propris, R., Stanford, S. A., Eisenhardt, P. R., Dickinson, M., and Elston,
R.  1999, \aj, 118, 719 

\bibitem[Dey, Spinrad \& Dickinson(1995)]{dsd95} \reference{}
Dey, A., Spinrad, H., and Dickinson, M.  1995, \apj, 440, 515

\bibitem[Dey \etal(1999)]{dea99} \reference{}
Dey, A., Graham, J.~R., Ivison, R.~J., Smail, I., Wright, G.~S., and
Liu, M.~C.  1999, \apj, 519, 610 (astro-ph/9902044)

\bibitem[Drory \etal(1999)]{dro99} \reference{}	
Drory, N., Hopp, U., Bender, R., Feulner, G., Snigula, J., Mendes de Oliveira,
C., and Hill, G.  1999, in {\it Clustering at High Redshift}, 
eds. A. Mazure, O. Le Fevre, and V. LeBrun (ASP: San Francisco), 91 (D99)

\bibitem[Dwek \etal(1998)]{dwe98} \reference{}	
Dwek, E., \etal\ 1998, \apj, 508, 106

\bibitem[Eisenhardt \etal(1996)]{eis96} \reference{}
Eisenhardt, P.~R., Armus, L., Hogg, D.~W., Soifer, B.~T., Neugebauer, G.,
and Werner, M.~W.  1996, \apj, 461, 72

\bibitem[Eisenhardt \etal(2000)]{eis00} \reference{}
Eisenhardt, P.~R., Elston, R., Stanford, S.~A., Dickinson, M., Spinrad, H.,
Stern, D. and Dey, A.  2000, to appear in proceedings of the Xth Rencontres
de Blois on ``The Birth of Galaxies", ed. B. Guiderdoni \etal\ (astro-ph/0002468)






\bibitem[Fioc \& Rocca-Volmerange(1997)]{frv97} \reference{}     
Fioc, M., and Rocca-Volmerange, B.  1997, \aap, 326, 950 



\bibitem[Fukugita, Shimasaku \& Ichikawa(1995)]{fsi95}\reference{}       
Fukugita, M., Shimasaku, K., and Ichikawa, T.  1995, \pasp, 107, 945

\bibitem[Gardner \etal(1997)]{gar97} \reference{}         
Gardner, J.~P., Sharples, R.~M., Frenk, C.~S., and Carrasco B.~E.  1997,
\apjl, 480, L99

\bibitem[Gehrels(1986)]{geh86}\reference{}
Gehrels, N.  1986, \apj, 303, 336

\bibitem[Girard \etal(1989)]{gir89} \reference{}
Girard, T.~M., Grundy, W.~M., L\'opez, C.~E., and van Altena, W.~F.  1989,
\aj, 98, 227



\bibitem[Graham \& Dey(1996)]{gd96}\reference{}
Graham, J.~R., and Dey, A.  1996, \apj, 471, 720 


\bibitem[Hall(1998)]{thesis} \reference{}
Hall, P.~B.  1998, PhD Thesis, University of Arizona

\bibitem[Hall, Green \& Cohen(1998)]{hgc98} \reference{}
Hall, P.~B., Green, R.~F., and Cohen, M.  1998, \apjs, 119, 1 (HGC98)

\bibitem[Hall \& Green(1998)]{hg98} \reference{}
Hall, P.~B., and Green, R.~F.  1998, \apj, 507, 558 (HG98)

\bibitem[Hall \etal(1999)]{hal99tp3a} \reference{}
Hall, P.~B., Sawicki, M., Pritchet, C.~J., Hartwick, F.~D.~A., and Evans, A.  
1999, in ``The Hy-Redshift Universe: Galaxy
Formation and Evolution at High Redshift," eds. A. J. Bunker \&
W. J. M. van Breugel (ASP: San Francisco), 415 

\bibitem[Hall, Sawicki \& Lin(2000)]{hal99tp3b} \reference{}
Hall, P.~B., Sawicki, M., and Lin, H.  2000, in ``Clustering at High
Redshift," eds. A. Mazure, O. Le Fevre, and V. Le Brun (ASP: San Francisco), 205

\bibitem[Hammer \etal(2001)]{hea01}\reference{}
Hammer, F., Gruel, N., Thuan, T., Flores, H., and Infante, L.  2001,
\apj, in press (astro-ph/0011218)

\bibitem[Hill \& Lilly(1991)]{hl91}\reference{}
Hill, G., and Lilly, S. 1991, \apj, 367, 1

\bibitem[Holland \etal(1998)]{hea98} \reference{}
Holland, W.~S., Cunningham, C.~R., Gear, W.~K., Jenness, T., Laidlaw, K., 
Lightfoot, J.~F., and Robson, E.~I.  1998, \procspie, 3357, 305


\bibitem[Hu \& Ridgway(1994)]{hr94} \reference{}
Hu, E.~M., and Ridgway, S.~E.  1994, \aj, 107, 1303


\bibitem[Hunt \etal(1998)]{hun98} \reference{} 
Hunt, L.~K., Mannucci, F., Testi, L., Migliorini, S., Stanga, R.~M.,
Baffa, C., Lisi, F., and Vanzi, L.  1998, \aj, 115, 2594



\bibitem[Jenness \& Lightfoot(1998)]{jl98} \reference{}
Jenness, T., and Lightfoot, J.~F.  1998, Starlink User Note 216


\bibitem[Kajisawa \etal(2000)]{kaj00a} \reference{}
Kajisawa, M., \etal\  2000, \pasj, 52, 53

\bibitem[Kajisawa \etal(2000)]{kaj00b} \reference{}
Kajisawa, M., \etal\  2000, \pasj, 52, 61


\bibitem[Kennicutt(1983)]{ken83} \reference{}
Kennicutt, R.~C.  1983, \apj, 272, 54


\bibitem[Kennicutt(1998)]{ken98} \reference{}
Kennicutt, R.~C.  1998, \araa, 36, 189



\bibitem[Kormendy \& Djorgovski(1989)]{kd89} \reference{}
Kormendy, J., and Djorgovski, G.  1989, \araa, 27, 235

\bibitem[Krisciunas \etal(1987)]{kri87}\reference{}
Krisciunas, K., Sinton, W., Tholen, D., Tokunaga, A., Golisch, W., Griep, D.,
Kaminski, C., Impey, C., and Christian, C.  1987, \pasp, 99, 887
 
\bibitem[Leggett \& Deanult(1996)]{ld96} \reference{}
Leggett, S., and Denault, T.  1996, NSFCAM 256x256 InSb Infrared Array Camera
User's Guide, NASA Infrared Telescope Facility
 

\bibitem[Liu \etal(2000)]{liu00} \reference{}
Liu, M.~C., Dey, A., Graham, J.~R., Bundy, K.~A., Steidel, C.~C.,
Adelberger, K., and Dickinson, M.~E.  2000, \aj, 119, 2556 

\bibitem[Lin \etal(2001)]{lin01} \reference{}
Lin, H., \etal\  2001, in preparation






\bibitem[Martini(2001)]{mar00} \reference{}
Martini, P.  2001, \aj, in press (astro-ph/0009287)


\bibitem[McCarthy \etal(1998)]{mcc98} \reference{}
McCarthy, D.~W., Ge, J., Hinz, J.~L., Finn, R.~A., Low, F.~J.,
Cheselka, M., and Salvestrini, K.  1998, \baas, 193, 11.09

\bibitem[McCarthy \etal(2000)]{mea00} \reference{}
McCarthy, D.~W., Ge, J., Hinz, J.~L., Finn, R.~A. and de Jong, R. S. 2000,
\pasp, submitted

\bibitem[McCarthy \etal(1999)]{mea99} \reference{}
McCarthy, P.~J., \etal\  1999, \apj, 520, 548

\bibitem[Mendes de Oliveira \etal(1998)]{mdo98} \reference{}
Mendes de Oliveira, C., Drory, N., Hopp, U., Bender, R., and Saglia, 
R.~P.  1998, to appear in ``Wide Field Surveys in Cosmology'', eds. Y. Mellier
and S. Colombi (Editions Frontieres: Gif-sur-Yvette) (astro-ph/9809406)



\bibitem[Moriondo, Cimatti \& Daddi(2000)]{mcd00} \reference{}
Moriondo, G. and Cimatti, A. and Daddi, E.  2000, \aap, in press 
(astro-ph/0010335)

\bibitem[Pahre \& Djorgovski(1995)]{pd95} \reference{}
Pahre, M.~A., and Djorgovski, S.~G.  1995, \apj, 449, L1


\bibitem[Persson \etal(1998)]{per98}\reference{1998AJ....116.2475P} 
Persson, S. E., Murphy, D. C., Krzeminski, W., Roth, M. \& Rieke, M. J. 1998,
\aj, 116, 2475


\bibitem[Pogge \etal(1998)]{pog98} \reference{}
Pogge, R.W., DePoy, D.L., Atwood, B., O'Brien, T.P., Byard, P.L., Martini, P.,
Stephens, A., Gatley, I., Merril, K.M., Vrba, F.J., and Henden, A.A.  1998,
SPIE, 3354, 414

\bibitem[Poggianti(1997)]{pog97} \reference{}
Poggianti, B.~M.  1997, \aaps, 122, 399

\bibitem[Pozzetti \& Mannucci(2000)]{pm00} \reference{}
Pozzetti, L. \& Mannucci, F.  2000, \mnras, 317, L17 




\bibitem[S{\'a}nchez \& Gonz{\'a}lez-Serrano(1999)]{sgs99} \reference{}
S{\'a}nchez, S.\ F. and Gonz{\'a}lez-Serrano, J.\ I.  1999, \aap, 352, 383

\bibitem[Sandage \& Perelmuter(1990)]{sp90} \reference{}
Sandage, A., and Perelmuter, J.-M.  1990, \apj, 361, 1

\bibitem[Saracco \etal(1999)]{sar99} \reference{}
Saracco, P., D'Odorico, S., Moorwood, A., Buzzoni, A., Cuby, J.-G., and Lidman,
C.  1999, \aap, 349, 751 

\bibitem[Sawicki, Lin \& Yee(1997)]{sly97} \reference{} 
Sawicki, M., Lin, H., and Yee, H.~K.~C.  1997, \aj, 113, 1

\bibitem[Sawicki \& Yee(1998)]{sy98} \reference{} 
Sawicki, M., and Yee, H. K. C. 1998, \aj, 115, 1329

\bibitem[Scodeggio \& Silva(2000)]{ss00} \reference{} 
Scodeggio, M., and Silva, D.~R.  2000, \aap, 359, 953 

\bibitem[Smail \etal(1998)]{sma98} \reference{}	
Smail, I., Ivison, R., Blain, A., and Kneib, J.-P.  1998,
in "After The Dark Ages: When Galaxies Were Young",
eds. S.~S. Holt and E.~P. Smith (AIP: New York), 312 

\bibitem[Smail \etal(1999)]{sma99} \reference{}	
Smail, I., \etal\ 1999, \mnras, 308, 1061 

\bibitem[Soifer \etal(1999)]{soi99} \reference{}	
Soifer, B.~T., Matthews, K., Neugebauer, G., Armus, L., Cohen, J.~G., and
Persson, S.~E.  1999, \aj, 118, 2065 


\bibitem[Stanford \etal(1997)]{sea97} \reference{}	
Stanford, S.\ A., Elston, R., Eisenhardt, P.\ R., Spinrad, H., Stern, D. and 
Dey, A.  1997, \aj, 114, 2232

\bibitem[Steidel \etal(1996)]{ste96} \reference{} 
Steidel, C.~C., Giavalisco, M., Pettini, M., Dickinson, M., and Adelberger,
K.~L.  1996, \apjl, 462, L17


\bibitem[Stiavelli \& Treu(2000)]{st00}\reference{}
Stiavelli, M., and Treu, T.  2000, to appear in ``Galaxy Disks and Disk 
Galaxies", eds. J. G. Funes and E. M. Corsini (astro-ph/0010100)



\bibitem[Teplitz, McLean \& Malkan(1999)]{tmm99} \reference{} 
Teplitz, H., McLean, I., and Malkan, M.  1999, \apj, 520, 469 



\bibitem[Thompson \etal(1999a)]{tea99} \reference{}
Thompson, D., \etal\  1999, \apj, 523, 100 (T99)

\bibitem[Thompson \etal(1999b)]{teb99} \reference{}
Thompson, D., Kelly, A.~E., Sawicki, M., Soifer, B.~T., and Matthews, K.  1999,
in "Photometric Redshifts and High Redshift Galaxies", eds. R. Weymann, 
L. Storrie-Lombardi, M. Sawicki \& R. Brunner (San Francisco: ASP), 291

\bibitem[Thompson, Aftreth \& Soifer(2000)]{tas00} \reference{}
Thompson, D., Aftreth, O., and Soifer, B.~T.  2000, \aj, 120, 2331 

\bibitem[Thompson, Weymann \& Storrie-Lombardi(2000)]{tws00} \reference{}
Thompson, R., Weymann, R., and Storrie-Lombardi, L.  2000, \apj, 546, 694


\bibitem[Tytler \& Fan(1992)]{tf92} \reference{}
Tytler, D., and Fan, X.-M.  1992, \apjs, 79, 1

\bibitem[Valdes(1982)]{val82a}\reference{}
Valdes, F.  1982, FOCAS User's Manual,
    Kitt Peak National Observatory, Central Computer Services, Tucson AZ

\bibitem[van Breugel \etal(1998)]{vb98} \reference{}
van Breugel, W.~J.~M., Stanford, S.~A., Spinrad, H., Stern, D., and Graham,
J.~R.  1998, \apj, 502, 614

\bibitem[van Breugel \etal(1999)]{vb99} \reference{}
van Breugel, W.~J.~M., De Breuck, C., Stanford, S.~A., Stern, D., 
R{\"o}ttgering, H., and Miley, G.  1999, \apjl, 518, L61

\bibitem[van der Werf(1997)]{vdw97} \reference{}
van der Werf, P.~P.  1997, to appear in ``Extragalactic Astronomy in the 
Infrared," eds. G.~A. Mamon, T.~X. Thuan, and J. Tran Thanh Van (Editions
Frontieres: Gif-sur-Yvette) (astro-ph/9706130)

\bibitem[Yamada \etal(1997)]{yam97} \reference{} 
Yamada, T., Tanaka, I., Aragon-Salamanca, A., Kodama, T., Ohta, K. and Arimoto,
N.  1997, \apjl, 487, L125



\bibitem[Yee \& Ellingson(1993)]{ye93}\reference{}
Yee, H.~K.~C., and Ellingson, E.  1993, \apj, 411, 43 

\bibitem[Yee(1998)]{yee98} \reference{}
Yee, H.~K.~C.  1998, to appear in the proceedings of
the Xth Recontres de Blois: The Birth of Galaxies (astro-ph/9809347)

\end{thebibliography}
\end{document}